# Quantifying the Role of Higher-Lying Excited States in Organic Emitters via Multistate *Ab Initio* Kinetic Modeling

Yue He, Daniel Escudero*

Quantum Chemistry and Physical Chemistry Division, Department of Chemistry, KU Leuven, Celestijnenlaan 200F, 3001 Leuven, Belgium

*E-mail: Daniel.Escudero@kuleuven.be

**Abstract:** Higher-lying excited states beyond $S_1$ and $T_1$ are widely recognized in many photophysical systems, including thermally activated delayed fluorescence (TADF). However, their explicit and quantitative impact on photophysical observables such as photoluminescence quantum yields (PLQY) and lifetimes is difficult to be attained experimentally and it has not been systematically assessed within a fully *ab initio* kinetic modeling framework. To address this gap, we developed KinLuv, a multistate excited state kinetic model that includes higher-lying excited states ($S_2$, $T_2$) and all possible monomolecular interconversion processes between all the electronic states, whose rate constants were computed using Fermi's golden rule explicitly including the Herzberg-Teller (HT) vibronic coupling effect. We applied KinLuv to prototypical multi-resonance TADF (MR-TADF) emitters and their derivatives, as well as other representative organic chromophores, demonstrating its broad applicability across diverse photophysical playgrounds beyond TADF. The resulting simulations quantitatively reproduce key experimental observables, including PLQY and prompt/delayed fluorescence lifetimes. Beyond its predictive power, the present results establish clear criteria for identifying when higher-lying excited states influence the excited-state decay and when simplified models remain adequate. This framework enables rational selection of minimal kinetic models that balance physical insight with numerical robustness, with direct implications for the *in silico* design of high-performance organic emitters.

## 1. INTRODUCTION

Understanding the excited-state dynamics of organic chromophores has traditionally relied on a simplified three-state picture involving the ground state ($S_0$) and the lowest singlet ($S_1$) and triplet ($T_1$) excited states. This approximation is generally valid when $S_1$ and $T_1$ are well separated in energy from higher-lying excited states ($S_{n>1}$ and $T_{n>1}$). However, advances in quantum-chemical methods and ultrafast spectroscopies have increasingly demonstrated the protagonist role of higher-lying excited states in a broad variety of photophysical and photochemical scenarios. Representative examples include emission[1,2] and photochemical reactions[3] originating from higher-lying excited states (i.e., anti-Kasha photochemistry), triplet formation through $S_1 \rightarrow T_{n>1}$ intersystem crossing (ISC) processes[4–7], and internal conversion (IC) mediated by higher-lying excited states that enable access to pathways inaccessible from $S_1$[8,9]. In the above scenarios, a conventional three-state model becomes inadequate to model the excited-state dynamics of organic chromophores, as the explicit inclusion of higher-lying states may qualitatively and quantitatively alter the photophysical outcomes. From an experimental perspective, unambiguously establishing their involvement remains challenging because it requires excitation-wavelength-dependent time-resolved spectroscopy combined with global kinetic analysis across multiple excited-state models and timescales[9–11]. Consequently, such tremendous efforts are rarely reported in the literature.

The above considerations are particularly relevant for thermally activated delayed fluorescence (TADF) materials[12], for which a three-state picture is often assumed. However, increasing evidence indicates that more complex excited-state kinetics beyond this picture are not uncommon in TADF emitters[13–16]. In these systems, delayed fluorescence arises from reverse intersystem crossing (rISC) that upconverts triplet excited states ($T_n$) to singlet excited states ($S_n$), rendering observables such as photoluminescence quantum yield (PLQY) and prompt/delayed fluorescence lifetimes highly sensitive to the underlying excited-state decay pathways[17]. More recently, multi-resonant TADF (MR-TADF) emitters[18] have attracted considerable attention owing to their narrowband emission and efficient rISC enabled by short-range charge-transfer (SRCT) character and small singlet-triplet energy gaps[19–21]. As mentioned previously, increasing experimental and theoretical studies have suggested that higher-lying excited states may play a key mechanistic role in mediating ISC and rISC processes in these systems. Specifically, spin-vibronic coupling involving higher-lying triplet states has been shown to strongly enhance ISC and rISC processes especially when direct $S_1 \rightarrow T_1$ or $T_1 \rightarrow S_1$ coupling is symmetry-forbidden or intrinsically weak[22–25]. Computational studies by Olivier et al.[13,14] and Marian et al.[15] have further demonstrated that "dark" triplet states with distinct orbital characters can act as intermediate spin-mixing channels, facilitating efficient population transfer between $S_1$ and $T_1$.

Despite these mechanistic insights, most quantitative kinetic analyses of TADF photophysics continue to rely predominantly on simplified three-state models. Experimentally, the transient photoluminescence measurements of TADF emitters are typically fitted by a biexponential decay profile, where the fast component is assigned to the prompt fluorescence and the slow one to the delayed fluorescence. Early kinetic models, such as those by Adachi et al.[26] and Dias et al.[17], estimated rISC rates under simplifying assumptions, e.g., neglecting non-radiative decay or phosphorescence, whereas later approaches by Monkman et al.[27] and Tsuchiya et al.[28] have progressively relaxed these constraints and enabled direct extraction of all relevant kinetic parameters. Nevertheless, these models generally assume that ISC and rISC occur solely between $S_1$ and $T_1$. From a mathematical perspective, the limited number of experimentally measurable quantities (i.e., PLQY and fluorescence lifetimes) is insufficient to uniquely determine all unknowns in the excited-state decay kinetics. Consequently, the roles of higher-lying states cannot be unambiguously proved from experiment alone, making computational simulations indispensable for achieving a complete and quantitative description of the excited-state dynamics. Unlike the above-mentioned kinetic models that extract rate constants by fitting photoluminescence transient data, several theoretical studies have instead attempted to first compute rate constants and then use kinetic models to predict PLQY and fluorescence lifetimes[29]. Shizu et al.[30,31] have made significant contributions to the quantitative modeling of excited-state kinetics. They applied their method to molecules like DABNA-1, BNOO, BNSS, and BNSeSe. As with any theoretical framework, their methodology involves certain approximations that define its range of applicability. For example, they neglect Herzberg-Teller (HT) vibronic coupling effects in the rate constant calculations, leading to an underestimation of (r)ISC rate constants. This deficiency is evident from the calculated rISC $T_1 \rightarrow S_1$ rate constants in DABNA-1 (2.5 $s^{-1}$ versus an experimental value of ~$9.9\times10^3$ $s^{-1}$) and BNOO (8.5 $s^{-1}$ versus ~$4.3\times10^4$ $s^{-1}$).

As a result, a fundamental gap remains in our understanding of the excited-state dynamics of organic chromophores, including TADF emitters: although higher-lying excited states are frequently invoked qualitatively, their explicit and quantitative impact on PLQY and excited-state lifetimes has not been systematically assessed within a fully *ab initio* kinetic modelling framework.

In this study, we address this gap by developing multistate kinetic models explicitly including higher-lying singlet and triplet states ($S_2$ and $T_2$) and incorporating HT vibronic coupling in the rate constant calculations. Our framework provides the community with a quantitative basis for determining when explicit inclusion of higher-lying excited states becomes essential and when a simplified kinetic model (i.e., three-state picture) remains adequate. We first present a computational protocol that enables the calculation of all radiative and non-radiative rate constants involved in the interconversion between the electronically excited states. We then introduce KinLuv, a Python-based kinetic modeling program that computes prompt and delayed fluorescence lifetimes and PLQY from the calculated rate constants. To validate the proposed framework, we adopt a hierarchical strategy. DOBNA and DiKTa serve as prototypical MR-TADF systems, while DABNA-1 (a DOBNA derivative) and DQAO (a DiKTa derivative) assess the applicability of our approach to related derivatives. To further evaluate the generality of our methods, we additionally consider two non-TADF organic chromophores, namely dibenzothiophene (DBT) and bis-benzo-fused pyrano[3,2-b]pyran-2,6-dione (PPDs-1). DBT[32] and PPDs-1[33] were chosen as representative organic luminophores whose photophysical properties are strongly influenced by higher-lying excited states. In DBT, triplet formation through $S_1 \rightarrow T_{n>1}$ ISC processes has been proved[34,35], while preliminary investigations in our group show IC mediated by higher-lying excited states in PPDs-1.By systematically comparing multistate kinetic models of increasing complexity, this unified framework enables us to identify when higher-lying excited states play a significant role in their excited-state dynamics, to delineate the limitations of simplified kinetic models, and to quantify the trade-off between mechanistic completeness (inclusion of all physically accessible excited states and pathways) and quantitative necessity (inclusion required to reproduce experimentally observable quantities) with a fully *ab initio* kinetic modeling framework. This framework offers a principled strategy for constructing minimal yet robust kinetic models, thereby guiding the *in silico* design of high-performance organic emitters.

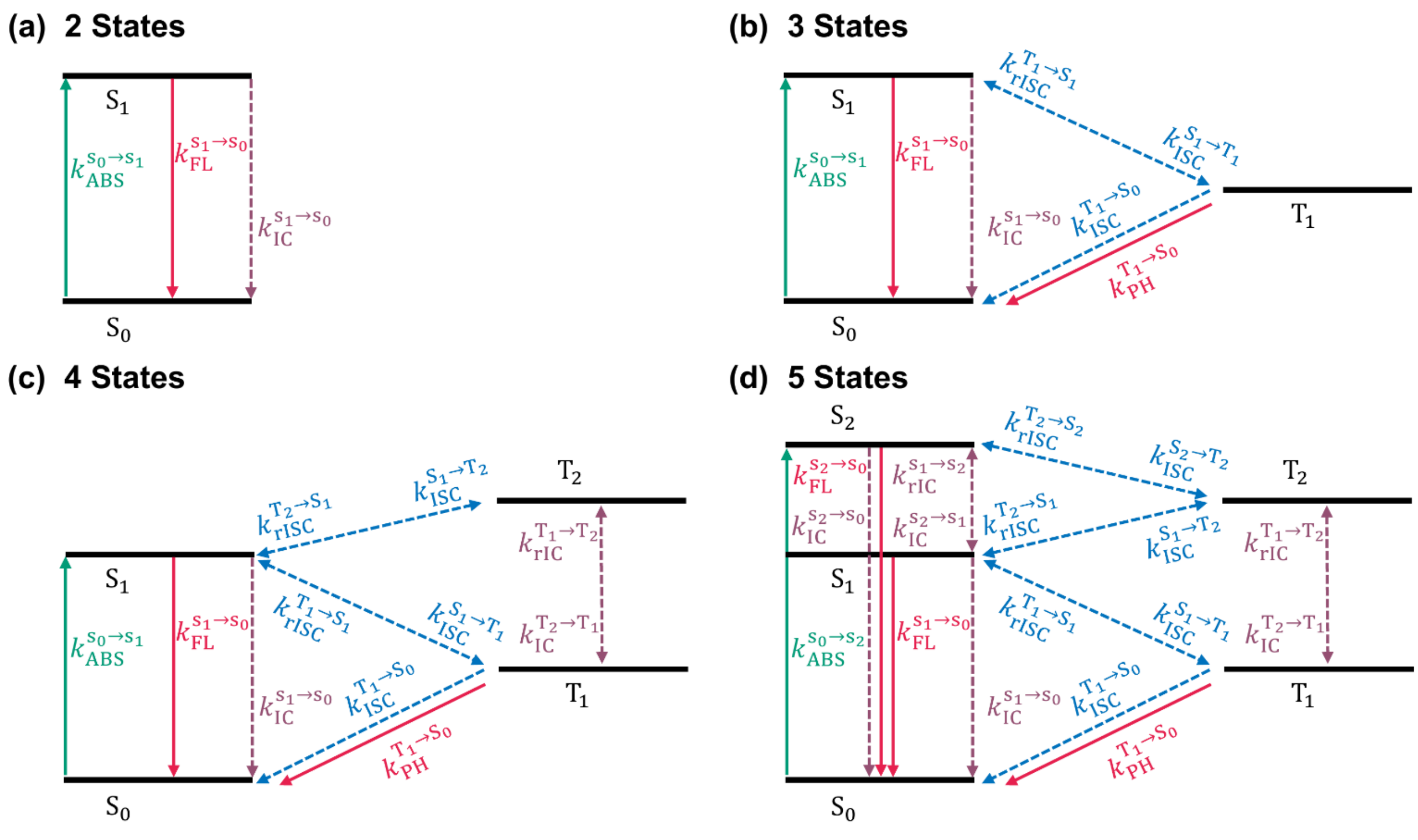

**Figure 1.** Schematic diagram of the examined excited state kinetic models for the investigated chromophores: (a) two-state ($S_0$, $S_1$), (b) three-state ($S_0$, $S_1$, $T_1$), (c) four-state ($S_0$, $S_1$, $T_1$, $T_2$), (d) five-state ($S_0$, $S_1$, $S_2$, $T_1$, $T_2$) model. The considered interconversion processes between the involved electronic states are highlighted for each model: absorption (ABS), fluorescence (FL), phosphorescence (PH), internal conversion (IC), reverse internal conversion (rIC), intersystem crossing (ISC), and reverse intersystem crossing (rISC).

## 2. RESULTS AND DISCUSSION

**2.1. Excited State Kinetic Models.** If a molecule in its excited state is regarded as a distinct chemical species from that in its ground state, the interconversion between its involved electronic states can be formulated as a chemical reaction chain and analyzed with conventional kinetic methods. Within this framework, we systematically examined a hierarchy of excited state kinetic models with increasing complexity to evaluate the mechanistic relevance of higher-lying excited states in the quantitative calculation of PLQY and excited state lifetimes for both TADF and non-TADF organic chromophores.

Figure 1 schematically summarizes the four kinetic models considered in this study, which differ in the number of electronic states and the possible interconversion processes between them. The two-state model considers the ground state ($S_0$) and the first singlet excited state ($S_1$) (Figure 1a) and is therefore applicable only to model prompt fluorescence in the simplest dyes. The three-state model shown in Figure 1b additionally includes the first triplet excited state ($T_1$), allowing for an approximate description of TADF. However, as mentioned above, higher-lying excited states cannot be ruled out in many molecular systems[16], necessitating more complex excited-state kinetic model. This is illustrated here by the four- and five-state models, which include a higher triplet excited state ($T_2$) and concomitantly a higher singlet excited state ($S_2$), respectively. To the best of our knowledge, the latter two models have not yet been widely applied to quantitatively study the excited-state dynamics of functional chromophores. Here, they are deliberately employed to assess the impact of higher-lying excited states on PLQY and prompt/delayed fluorescence lifetimes, across a diverse set of molecular playgrounds, including TADF emitters.

As shown in Figure 1, all possible interconversion processes between adjacent electronic states, including absorption, fluorescence (both Kasha and anti-Kasha), phosphorescence, (r)IC and (r)ISC, are explicitly considered. By applying the law of mass action and the rate equation to each pathway, a set of ODEs is obtained in a general form (eq 1):

$$R_i * N_i(t) = \frac{\mathrm{d}}{\mathrm{d}t} N_i(t) \quad (1)$$

where $R_i$ is the rate-constant matrix encompassing all possible pathways in the $i$-state model, $N_i(t)$ is the time-dependent population vector of $i$ states. Detailed ODEs for each model are provided in Section 2-3 of the supporting information (SI). Once the rate constants are determined, the ODEs can be solved analytically or numerically to yield the population evolution of ground and excited states, from which the prompt and delayed fluorescence lifetime can be extracted.

While the time-dependent solutions of the above ODEs allow for simulating excited-state lifetime, PLQY is conventionally measured by steady-state conditions. In this limit, the system approaches a dynamic equilibrium in which all states populations become time independent. Applying this steady state approximation (SSA), the ODEs are reduced to ordinary algebraic equations (eq 2), from which the PLQY is directly derived from its definition (see Section 4 of the SI).

$$R_i * N_i(t) = 0 \quad (2)$$

All relevant rate constants were calculated using our computational protocols (see Section 4 below and Section 1 of SI), and the corresponding ODEs were automatically solved using our Python-based code package, KinLuv. The resulting PLQY and prompt/delayed fluorescence lifetimes of the studied emitters (Figure 2) were subsequently compared with experimental data (Table 1). For the studied systems, anti-Kasha emission from the higher-lying singlet state $S_2$ is not considered, as no experimental evidence for such anomalous emission has been reported. In addition, rISC processes are not considered for DBT and PPDs-1, as delayed fluorescence has not been experimentally observed for these non-TADF compounds. This workflow enables a quantitative assessment of the validity and predictive capability of fully *ab initio* multistate kinetics models for functional organic chromophores, with particular emphasis on TADF systems, where both PLQY and prompt/delayed fluorescence lifetimes are of primary interest.

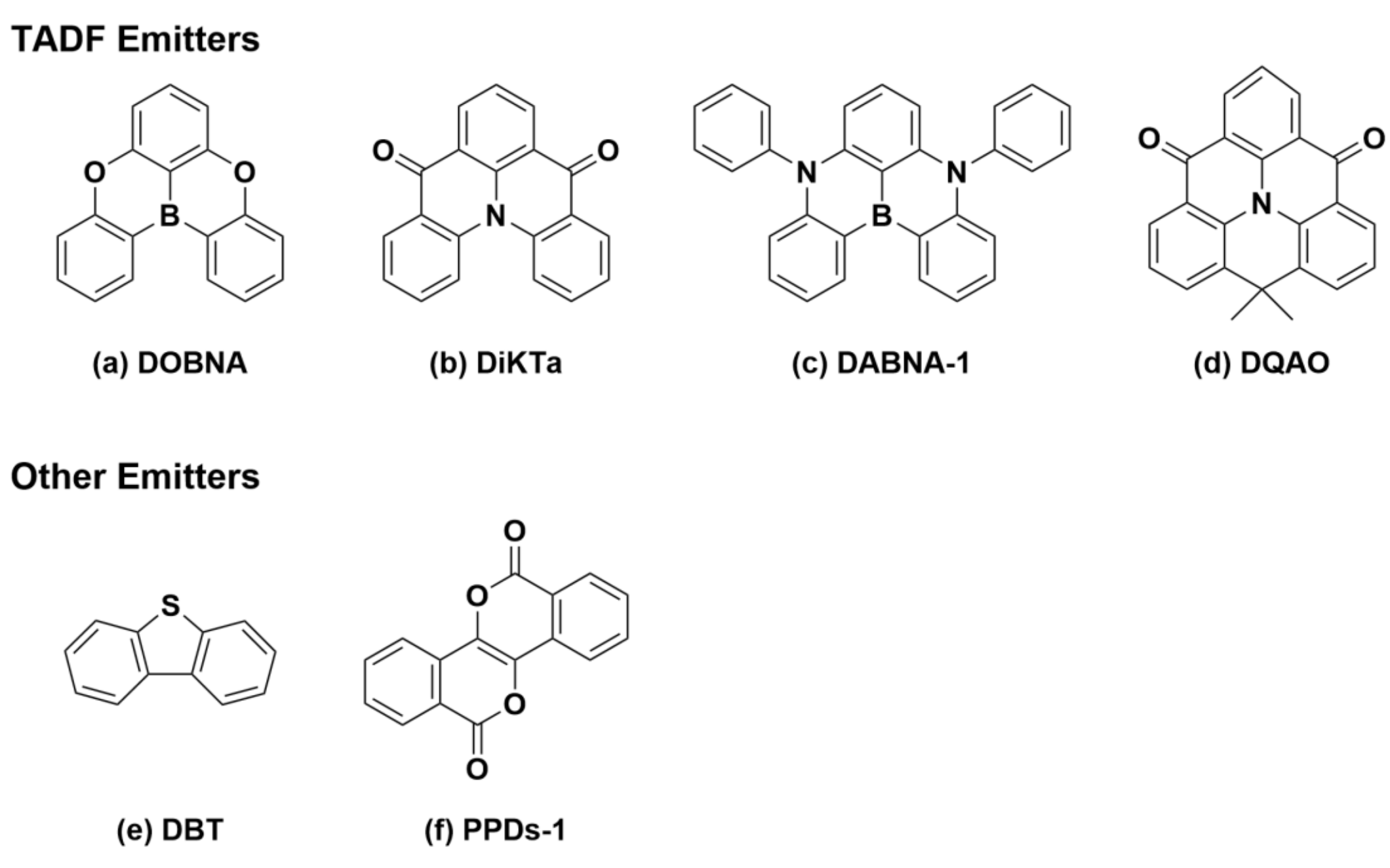

**Figure 2.** Molecular structure of DOBNA (a), DiKTa (b), DABNA-1 (c), DQAO(d), DBT(e) and PPDs-1(f).

**2.2. Excited-State Energetics and Decay Rate Constants.** Six functional chromophores were deliberately selected as representative systems to examine the putative role of higher-lying excited states (Figure 2). The selection follows a hierarchical design. We first focus on two prototypical MR-TADF emitters, DOBNA (Figure 2a) and DiKTa (Figure 2b), which serve as stringent benchmark systems for in-depth analysis. DOBNA features a central boron atom as the electron acceptor and oxygen atoms as electron donors, while DiKTa uses carbonyl groups as electron acceptors instead of boron. To assess the generality and robustness of the proposed methodology, DABNA-1 (Figure 2c) and DQAO (Figure 2d) were subsequently selected as structurally related derivatives of these prototype MR-TADF emitters. These systems allow us to examine whether the conclusions drawn for DOBNA and DiKTa persist upon molecular modification, and to further evaluate the transferability of our approach. Finally, DBT (Figure 2e) and PPDs-1 (Figure 2f) were included as representative non-TADF organic chromophores, as previously mentioned to be inadequately described by simplified three-state models. Their incorporation enables the assessment of our approach beyond TADF and provides a critical control for assessing the role of higher-lying excited states. Table 1 summarizes the previously reported experimental results for all the investigated compounds[19,32,21,36,37,33].

**Table 1. Experimental photophysical properties of selected molecules from previous studies**

| Molecule | $\Phi_F$ (%) | $\tau_{PF}$ (ns) [g)] | $\tau_{DF}$ (μs) [h)] | $\Delta E_{ST}$ ($S_1$-$T_1$) [i)] |
|---|---|---|---|---|
| DOBNA [a)] | 57 | 11.6 | 66.1 | 0.18 |
| DiKTa [b)] | 26 | 5.1 | 23 | 0.18 |
| DABNA-1 [c)] | 88 | 8.8 | 93.7 | 0.2 |
| DQAO [d)] | 59.3 | 10.09 | 110.58 | 0.19 |
| DBT [e)] | 9 | 3 | - | - |
| PPDs-1 [f)] | 0.9 | 0.066 | - | - |

[a)] Dispersed in PMMA (1 wt%) at 300 K; [b)] In degassed dilute toluene at 300 K; [c)] in mCBP films (1 wt%) at 300 K; [d)] in mCP films (8 wt%) at 298 K; [e)] in deaerated diluted cyclohexane at 298 K; [f)] in dichloromethane (DCM) at 298 K; [g)]Prompt fluorescence lifetime; [h)] Delayed fluorescence lifetime; [i)] Energy gap between $S_1$ and $T_1$ measured from emission spectra at 77 K.

The optimized ground- and excited-state geometries of all systems are provided in Section 6 of SI: DOBNA and DiKTa (Figure S3), DABNA-1 and DQAO (Figure S4), and DBT and PPDs-1 (Figure S5). In all cases, the structural changes associated with excitations are minimal (Root-Mean-Square Deviation well below 0.2 Å), thereby satisfying one of the prerequisites to apply the harmonic approximation and Fermi's golden rule (FGR) to calculate the excited state decay rate constants. The natural transition orbitals (NTOs) calculated with TD(A)-CAMB3LYP involved in the excited states of selected compounds, i.e., DOBNA and DiKTa, are shown in Section 5 of the SI. These NTOs were compared with those obtained with SCS-ADC(2). As shown in Figure S1 for DOBNA, the NTOs calculated using the CAM-B3LYP functional closely resemble those obtained from SCS-ADC(2). Similarly, for DiKTa (Figure S2), the CAM-B3LYP NTOs show good agreement with those from SCS-ADC(2) for most of the excited states, except for one hole NTO in the $T_2$ state.

While geometries, Hessian matrix, and associated properties at the TD(A)-DFT level have been reported to be sufficiently accurate for excited state decay rate constant calculations, excitation energies are often insufficiently accurate[38–40]. To obtain refined adiabatic excitation energies and more accurate estimations of adiabatic energy gaps, we performed single-point SCS-ADC(2) calculations at the TD(A)-DFT optimized geometries. For instance, for the MR-TADF emitters, the adiabatic TD(A)-DFT excitation energies differ from the SCS-ADC(2) results by 0.1-0.8 eV (Figure S6). Specifically, TD(A)-DFT tends to yield larger excitation energies for singlet states ($S_n$) and smaller energies for triplet states ($T_n$) compared to SCS-ADC(2), resulting in larger singlet-triplet energy gaps at the TD(A)-DFT level. Notably, this discrepancy is particularly pronounced for the $T_2$ state of DOBNA and DABNA-1, highlighting the necessity of using a higher-level method such as SCS-ADC(2) to refine the excited-state energy gaps. In the case of the non-TADF systems DBT and PPDs-1, deviations between TD(A)-DFT and SCS-ADC(2) excitation energies remain below 0.5 eV. Table 2 lists the adiabatic excitation energies calculated with SCS-ADC(2) for all the compounds, obtained by subtracting the ground-state energy at its optimized geometry from the excited-state energies at their respective minima. For DOBNA, the calculated $S_1$-$T_1$ energy gap, i.e., $\Delta E_{ST}(S_1\text{-}T_1)$, is 0.18 eV, which shows excellent agreement with the experimental value (Table 1). DiKTa exhibits a slightly larger calculated $\Delta E_{ST}(S_1\text{-}T_1)$ of 0.23 eV, which is within 0.05 eV of the experimental value, and comparable accuracy is obtained for DABNA-1 and DQAO. These results highlight the capability of the SCS-ADC(2) method to accurately determine the $\Delta E_{ST}(S_1\text{-}T_1)$ values of MR-TADF emitters. In contrast, the calculated $\Delta E_{ST}(S_1\text{-}T_1)$ values for the non-TADF emitters are sufficiently large (>0.7 eV), indicating a slowed down rISC process and consistent with the experimental observed absence of delayed fluorescence in these two compounds. Moreover, the $\Delta E_{ST}(S_1\text{-}T_2)$ values for DiKTa, DQAO, DBT and PPDs-1, in contrast to those of DOBNA and DABNA-1, are sufficiently small (within ±0.25 eV) to qualitatively suggest that triplet formation via higher-lying triplet excited states may be energetically feasible in these systems. However, this qualitative analysis is based solely on energetic considerations. In addition, a definitive quantitative assessment requires the approach developed here.

**Table 2. Adiabatic excitation energies (eV) of selected molecules from SCS-ADC(2) calculations**

| Molecule | $S_1$ | $S_2$ | $T_1$ | $T_2$ | $\Delta E_{ST}$ ($S_1$-$T_1$) | $\Delta E_{ST}$ ($S_1$-$T_2$) |
|---|---|---|---|---|---|---|
| DOBNA | 3.55 | 4.31 | 3.37 | 3.99 | 0.18 | -0.44 |
| DiKTa | 3.27 | 3.45 | 3.04 | 3.40 | 0.23 | -0.13 |
| DABNA-1 | 3.15 | 3.98 | 3.02 | 3.69 | 0.13 | -0.54 |
| DQAO | 3.26 | 4.02 | 3.04 | 3.41 | 0.23 | -0.14 |
| DBT | 4.11 | 4.76 | 3.42 | 3.87 | 0.69 | 0.25 |
| PPDs-1 | 3.66 | 3.77 | 2.51 | 3.68 | 1.15 | -0.02 |

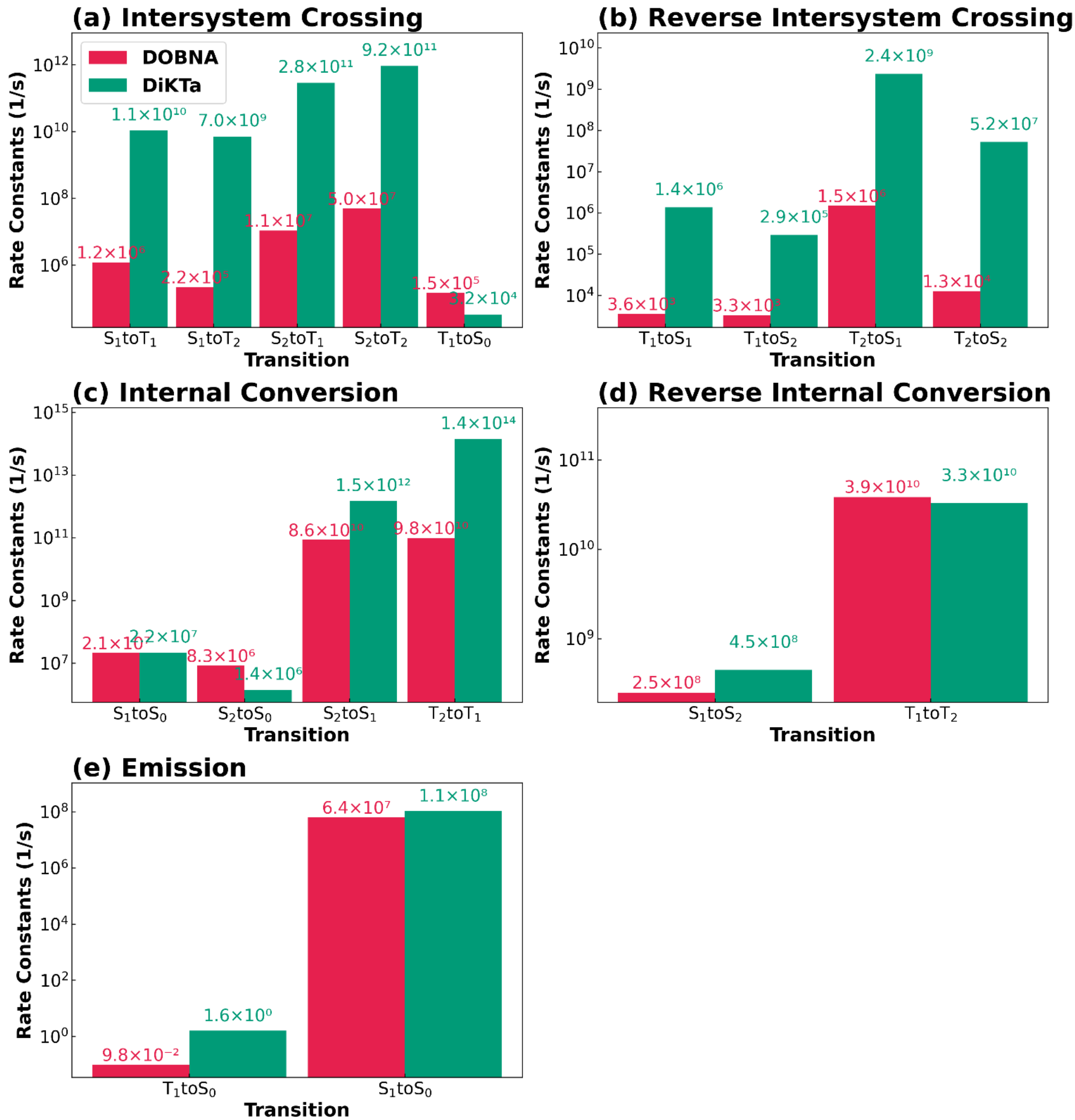

**Figure 3.** Calculated rate constants of (a) intersystem crossing (ISC), (b) reverse intersystem crossing (rISC), (c) internal conversion (IC), (d) reverse internal conversion (rIC) and (e) emission for molecule DOBNA (red) and DiKTa (green). Calculations at SCS-ADC(2)/def2-TZVP//TDA-CAM-B3LYP/6-311G(d,p) level with the adiabatic Hessian, Franck-Condon/Herzberg-Teller (FC/HT) vibronic model and a 10 $cm^{-1}$ Lorentzian broadening.

Thus, in the following section we focus on a detailed analysis of the calculated rate constants for DOBNA and DiKTa, which serve as representative prototype examples of MR-TADF emitters with expected different behaviors. For completeness, the corresponding data for the other two MR-TADF derivatives, i.e., DABNA-1 and DQAO, as well as for the two non-TADF emitters, are provided in Section 9 of SI. Figures 3a-b present the calculated average rate constants of (r)ISC between singlet states ($S_n$) and triplet states ($T_n$) for DOBNA (red) and DiKTa (green), respectively. The calculated ISC rate constants for the individual sublevels are provided in Tables S1-S2 of the SI. As seen in Figures 3a-b, DiKTa exhibits larger (r)ISC rate constants than DOBNA for most transitions, with the exception of ISC $T_1 \rightarrow S_0$. In particular, the larger $S_1 \rightarrow T_1$ ISC rate constant calculated for DiKTa than for DOBNA cannot be explained by energetic considerations alone, as DOBNA has a slightly smaller $\Delta E_{ST}(S_1\text{-}T_1)$ value of 0.18 eV compared to 0.23 eV for DiKTa (Table 2). According to Marcus theory for ISC rate constants, two main factors account for this behavior: differences in spin-orbit coupling matrix elements (SOCMEs) or variations in the Marcus regime governing the ISC process (i.e., normal or inverted regime). As shown in Tables S4-S5, both compounds feature a very small reorganization energy (~0.02 eV) compared to their singlet-triplet gaps, indicating that the $S_1 \rightarrow T_1$ ISC occurs within the same Marcus regime for both compounds. In contrast, the calculated SOCMEs at the reference FC geometry (Tables S6-S7) indicate that DiKTa exhibits larger $S_1 \rightarrow T_1$ SOCMEs than DOBNA. These results identify enhanced SOCMEs as the primary origin of the faster $S_1 \rightarrow T_1$ ISC rate in DiKTa. Still, the calculated ISC rate constants for both DiKTa and DOBNA are exceptionally large, roughly ranging from $10^5$ $s^{-1}$ to $10^{11}$ $s^{-1}$. Such high values cannot be rationalized solely based on calculated SOCMEs at FC geometry, nor by invoking simple empirical rules such as the heavy-atom effect[41] or El-Sayed rule[42]. The latter rule predicts faster ISC processes between states of different orbital characters (e.g., $n\pi^* \rightarrow \pi\pi^*$). However, analysis of the calculated NTOs (Figures S1-S2) reveals that all relevant transitions are of $\pi\pi^*$ character, with negligible $n\pi^*$ contributions. Instead, the large (r)ISC rate constants stem from the vibronic coupling, as the HT effect is explicitly considered in the rate constant calculations. As shown in Tables S8-S9, the calculated (r)ISC rate constants are dominated by HT contributions for nearly all transitions. Unlike the FC approximation, the HT effect enables electronic transitions to couple with vibrational motions via a breakdown of the Condon approximation[43]. The vibrational motions of nuclei effectively increase the SOCMEs, thereby facilitating rapid ISC processes.

Figures S7-S8 present the Huang-Rhys factors and reorganization energies per vibrational normal mode for key ISC processes of DOBNA and DiKTa, revealing a strong mode dependence. For DOBNA, in-plane vibrations contribute the most to r(ISC) processes, as they have both the largest Huang-Rhys factors and the largest reorganization energies. Conversely, out-of-plane modes exhibit the largest Huang-Rhys factors for DiKTa, but in-plane modes still contribute the most to the reorganization energies. Overall, these results highlight that out-of-plane vibrational modes have more relevance for DiKTa than for DOBNA. To further elucidate the origins of the HT effects, Figures S9-S10 show the SOCMEs calculated at the reference FC geometry and at displaced geometries along specific vibrational modes corresponding to the HT regime for the key (r)ISC channels of DOBNA and DiKTa, respectively. Notably, the SOCMEs in the HT regime are significantly enhanced compared to those in the FC regime particularly for the $S_1 \rightarrow T_1$ (Figure S9a and S10a) and $T_1 \rightarrow S_1$ (Figure S9e and S10e) (r)ISC processes. Structurally, all atoms in DOBNA lie in a strictly planar configuration, leading to predominantly in-plane vibrational modes. In contrast, DiKTa exhibits a slight torsion between its molecular fragments, allowing out-of-plane vibrations that more effectively break molecular symmetry and enhance vibronic coupling[44]. This structural distinction likely contributes to the stronger HT effect observed in DiKTa (Figures S9-S10). This analysis demonstrates that vibronic couplings effectively enhance SOCMEs via the HT effect.

Figure 3c-d show the rate constants for IC and rIC; respectively, for DOBNA and DiKTa. The $S_1 \rightarrow S_0$ IC rate constants are on the order of $10^7$ $s^{-1}$. Notably, in DOBNA (Figure 3d), the calculated rIC rate constant for $T_1 \rightarrow T_2$ is comparable in magnitude to that of the IC $T_2 \rightarrow T_1$ process and is significantly larger than those of $T_2 \rightarrow S_1$ and $T_1 \rightarrow S_1$ processes, indicating an efficient and nearly reversible population exchange between the two triplet states prior to population transfer back to the singlet manifold. Figure 3e collects the values of the calculated radiative rate constants, which are ca. $10^8$ $s^{-1}$ for fluorescence and ca. 1 $s^{-1}$ for phosphorescence, aligned to typical values for these compounds.

The calculated excited state decay rate constants for MR-TADF derivatives DABNA-1 and DQAO, as well as for the non-TADF systems (DBT, PPDs-1), are briefly summarized below. Calculated values are provided in Section 9 of the SI. DABNA-1, being closely related to DOBNA shows similar trends in the computed rate constants. Both compounds possess smaller (r)ISC rate constants than DQAO, which is closely related to DiKTa and, in turn, also exhibits analogous rate constant trends (Tables S10-S11). Importantly, by explicitly including HT contributions in the rate constant calculations, the rISC $T_1 \rightarrow S_1$ rate constant of DABNA-1 (~$2.41 \times 10^4$ $s^{-1}$) is obtained with a more reasonable magnitude compared to the smaller value reported by Shizu et al.[30] (2.5 $s^{-1}$). This comparison highlights the indispensable role of vibronic coupling in accurately describing (r)ISC processes. Interestingly, the calculated rIC rate constant for $T_1 \rightarrow T_2$ in DABNA-1 (Table S10) is comparable in magnitude to the IC $T_2 \rightarrow T_1$ rate constant. This trend is also found in DOBNA as previously mentioned. These rate constant behaviors observed for the prototype emitters are consistently reproduced in their derivatives, indicating the generality of the underlying trends. In addition, the magnitudes of the ISC rate constants are also substantial for non-TADF compounds DBT and PPDs-1, where the large values arise primarily from HT contributions (Table S15). Both of them show a larger ISC $S_1 \rightarrow T_2$ rate constant than that of $S_1 \rightarrow T_1$, indicating the non-negligible role of $T_2$ in mediating ISC. At the same time, PPDs-1 shows a comparably large rIC $S_1 \rightarrow S_2$ rate constant compared to IC $S_2 \rightarrow S_1$, indicating a nearly reversible population exchange between these two singlet states.

### 2.3. Excited-State Kinetics Simulation of MR-TADF.

The following discussion focuses on excited-state decay kinetics and the quantitative determination of PLQY and lifetimes. The discussion first focuses on DOBNA and DiKTa, as prototype MR-TADF emitters, and then on their structurally related derivatives (DABNA-1 and DQAO). The remaining systems, PPDs-1 and DBT, where higher-lying excited states are also relevant, are discussed thereafter. The complete excitation and decay profiles directly generated by KinLuv for all compounds are provided in Section 12 of SI. Figures 4 and 5 collect the simulated decay kinetics of DOBNA and DiKTa using kinetic models of increasing complexity, respectively. A logarithmic time scale is employed to simultaneously capture both prompt and delayed emission regimes within a single plot. Additionally, KinLuv can directly output the PLQY as well as the prompt and delayed fluorescence lifetimes. For the two- and three-state models, the decay of $S_1$ population can be described by analytical expressions, whereas numerical solutions are employed for the four- and five-state models to ensure computational efficiency. The calculated PLQY, prompt and delayed lifetimes for the four investigated MR-TADF emitters are summarized in Figure 6.

DOBNA is first discussed as a representative example (Figure 4). In the two-state model (Figure 4a), after initial population of $S_1$ (green), radiative and nonradiative decay to $S_0$ (red) occur in the nanosecond regime, giving rise exclusively to prompt fluorescence. Incorporation of the $T_1$ state in the three-state model (Figure 4b) introduces an ISC pathway from $S_1$ to $T_1$ (blue), followed by rISC back to $S_1$. Consequently, two distinct regimes showing a marked decrease in $S_1$ population are observed in Figure 4b. The first regime corresponds to prompt fluorescence (as in the two-state model) while the second one (ranging between ~$10^2$ ns and ~$10^2$ µs) stems from the delayed fluorescence via the $T_1 \rightarrow S_1 \rightarrow S_0$ pathway. However, only a minute fraction of $T_1$ population is accumulated because the ISC $S_1 \rightarrow T_1$ rate constant (~$1.2\times10^6$ $s^{-1}$) is much smaller than the competing fluorescence (~$6.4\times10^7$ $s^{-1}$) and IC (~$2.1\times10^7$ $s^{-1}$) rate constants. Accordingly, the delayed decay of the $S_1$ population originates from rISC of the $T_1$ population and is characterized by a small onset amplitude of ~$10^{-6}$ (Figure 4b). In the four-state model, inclusion of the higher-lying triplet state $T_2$ leads to nearly overlapping $T_1$ (blue) and $T_2$ (purple) population curves (Figures 4c and 4d), indicating synchronized population dynamics. This behavior arises from the comparable rIC rate constant for $T_1 \rightarrow T_2$ (~$3.9\times10^{10}$ $s^{-1}$) and IC rate constant for $T_2 \rightarrow T_1$ rate constant (~$9.8\times10^{10}$ $s^{-1}$), resulting in rapid thermal equilibration between these two states. Although $T_2$ state introduces an alternative delayed fluorescence pathway ($T_1 \rightarrow T_2 \rightarrow S_1 \rightarrow S_0$), the limited population accumulated in $T_1$ restricts the overall impact of this pathway, increasing the onset amplitude of delayed $S_1$ population only modestly, from ~$10^{-6}$ (in the three-state model) to ~$10^{-4}$. In the five-state model, the initial population of the higher-lying singlet state $S_2$ is followed by ultrafast IC to $S_1$ (~$8.6\times10^{10}$ $s^{-1}$), which leads to negligible effect on the global excited-state dynamics of the emitter.

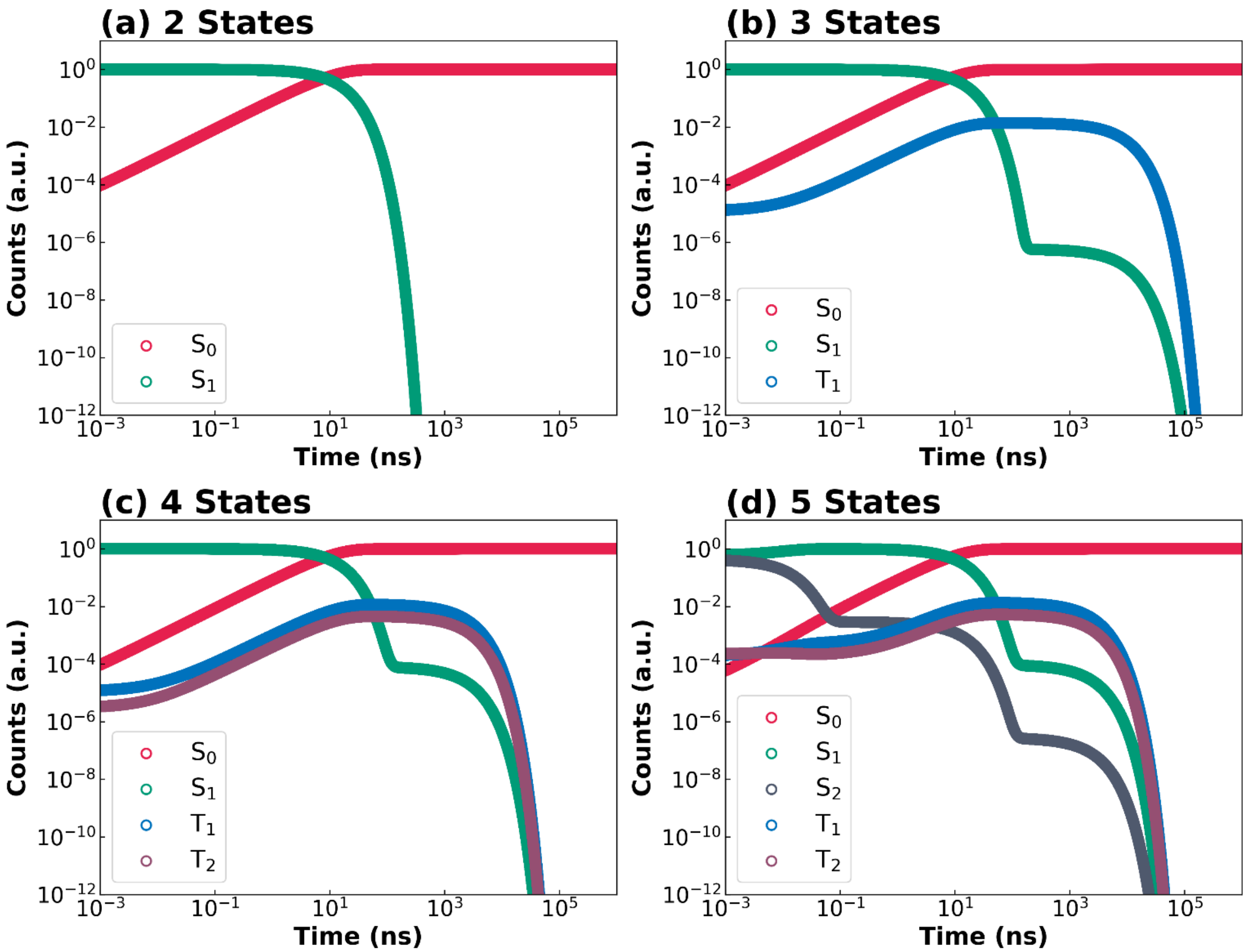

**Figure 4.** Simulated population decay kinetics of DOBNA over a 1 ms timescale based on (a) two-state ($S_0$, $S_1$), (b) three-state ($S_0$, $S_1$, $T_1$), (c) four-state ($S_0$, $S_1$, $T_1$, $T_2$), and (d) five-state ($S_0$, $S_1$, $S_2$, $T_1$, $T_2$) kinetic models. The number of absorbed photons is set to 1, with the excitation pulse width of 10 ps. Both time (x-axis) and population (y-axis) are plotted on logarithmic scales.

Despite the mechanistic differences between the three-, four- and five-state models, the calculated PLQY remains nearly constant at ~74% across all models (see red bars in Figure 6). The calculated prompt fluorescence lifetime reproduces the experimental value of 11.6 ns in the three-state model and shows only minor changes in the four- and five- state model. These results indicate that for DOBNA while the inclusion of higher-lying excited states is mechanistically important to provide a more complete description of the TADF pathways, it is not essential for accurately predicting experimentally observables such as PLQY, prompt and delayed fluorescence lifetimes. This can be attributed to the relatively slower timescales of (r)ISC processes compared to fluorescence and IC, which limit population accumulation in higher-lying triplets and thereby minimize their impact on the overall decay dynamics.

As a derivative of DOBNA, DABNA-1 exhibits excited-state dynamics that are very similar to those of DOBNA (Figure S11). The calculated PLQY (~82% versus experimental ~88%) and prompt fluorescence lifetime (8.4 ns versus ~8.8 ns) remain essentially unchanged across the three-, four-, and five-state models (see blue bars in Figure 6). By contrast, the delayed fluorescence lifetime decreases markedly from ~41 µs in the three-state model to ~2.3 µs in the four and five state model. Consistent with the behavior observed for DOBNA, higher-lying states may impact the TADF pathways mechanistically, yet the slow (r)ISC rates ensure a negligible effect on PLQY, prompt and delayed fluorescence. The consistency between DOBNA and its derivative DABNA-1 demonstrates that the conclusions drawn from the prototype system extend beyond a single molecule, thereby underscoring the robustness of our approach. Taken together, these results demonstrate that the adequacy of simplified three-state kinetic models is governed by both the magnitude of (r)ISC rate constants relative to fluorescence and IC and whether higher-lying excited states provide kinetically competitive rISC pathways. In systems such as DOBNA and DABNA-1, where (r)ISC processes are slow and the population of higher-lying states remains kinetically limited, their contribution to rISC is secondary, and three-state models remain sufficient to reproduce experimentally observable quantities, including PLQY and fluorescence lifetimes.

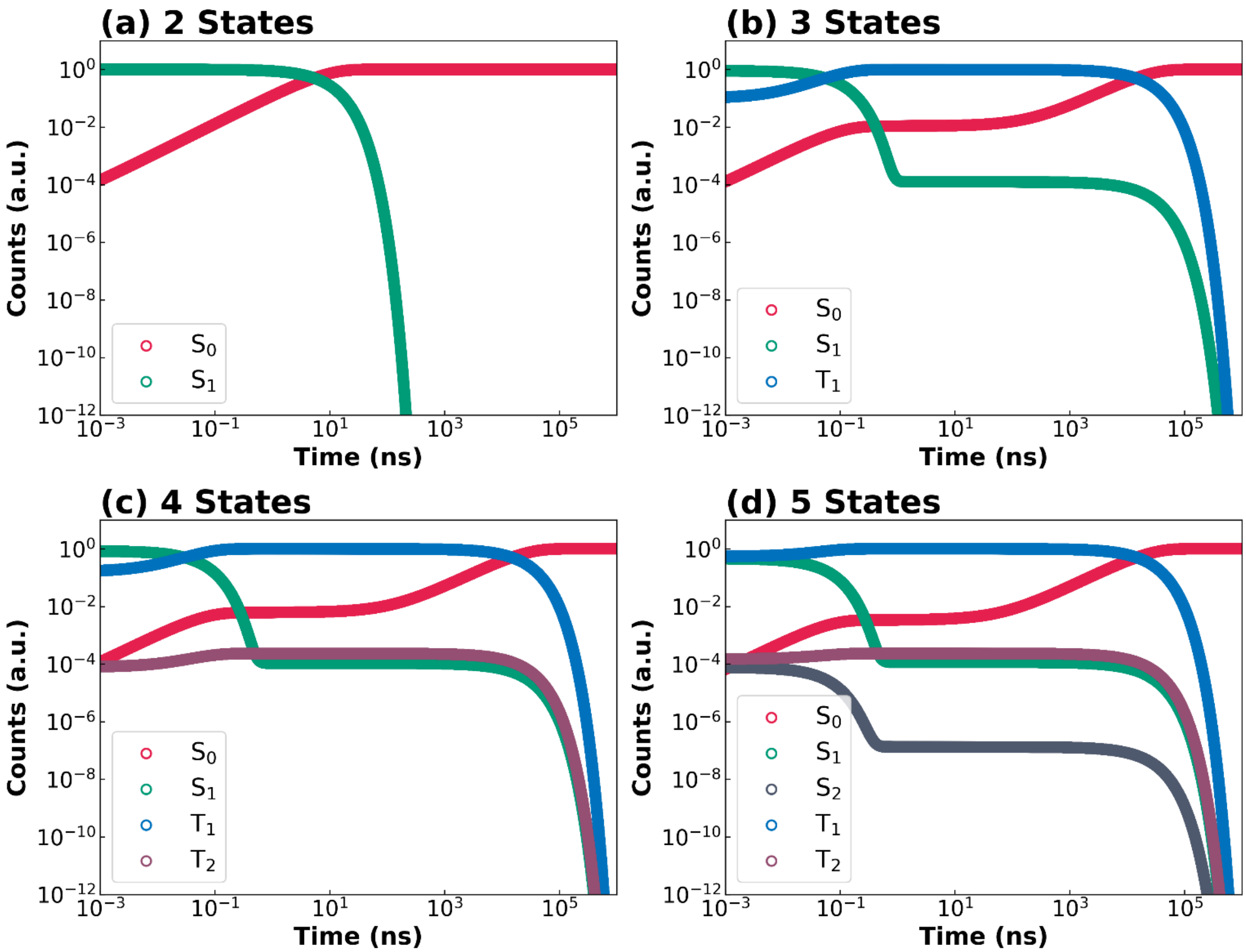

**Figure 5.** Simulated population decay kinetics of DiKTa over a 1 ms timescale based on (a) two-state ($S_0$, $S_1$), (b) three-state ($S_0$, $S_1$, $T_1$), (c) four-state ($S_0$, $S_1$, $T_1$, $T_2$), and (d) five-state ($S_0$, $S_1$, $S_2$, $T_1$, $T_2$) kinetic models. The number of absorbed photons is set to 1, with the excitation pulse width of 10 ps. Both time (x-axis) and population (y-axis) are plotted on logarithmic scales.

DiKTa (Figure 5) exhibits significantly faster (r)ISC rate constants than DOBNA (Figure 3a-b), leading to markedly distinct excited-state decay kinetics. In the three-state model (Figure 5b), the $T_1$ population becomes significantly larger than that of DOBNA (Figure 4b) because of the much larger $S_1$ to $T_1$ ISC rate constant (~$1.1\times10^{10}$ $s^{-1}$) than that of DOBNA (~$1.2\times10^{6}$ $s^{-1}$). Moreover, fast $S_1 \rightarrow T_1$ ISC suppresses direct decay from $S_1$ to $S_0$, causing a pronounced lag in $S_0$ population growth. As $S_1$ is gradually repopulated through rISC from $T_1$, the $S_0$ population increases again, producing a characteristic sigmoidal profile which was not observed for DOBNA. When moving to the four-state model (Figure 5c), the inclusion of a higher-lying triplet state $T_2$ does not introduce an efficient delayed fluorescence pathway analogous to $T_1 \rightarrow T_2 \rightarrow S_1 \rightarrow S_0$ observed for DOBNA. This is because IC rate constant for $T_2 \rightarrow T_1$ (~$1.4\times10^{14}$ $s^{-1}$) is substantially faster than the rIC $T_1 \rightarrow T_2$ rate constant (~$3.3\times10^{10}$ $s^{-1}$). Instead, $T_2$ primarily provides a competing non-radiative pathway ($S_1 \rightarrow T_2 \rightarrow T_1 \rightarrow S_0$) that effectively reduces the PLQY. In the five-state model, the higher-lying state $S_2$ introduces an additional delayed fluorescence pathway ($T_1 \rightarrow S_2 \rightarrow S_1 \rightarrow S_0$). However, this turns out to be a minor contribution that slightly increases the PLQY value, as the rISC rate constant for $T_1 \rightarrow S_2$ (~$2.9\times10^{5}$ $s^{-1}$) is smaller than that for $T_1 \rightarrow S_1$ rISC rate constant (~$1.4\times10^{6}$ $s^{-1}$).

Unlike DOBNA, the higher-lying states in DiKTa play a quantitatively important role in the excited-state kinetics. The calculated fluorescence PLQY decreases progressively from 83.1% in the two-state model to 28.6% in the three-state model and 25.4% and 26.5% in the four- and five-state model, the latter two models yielding PLQY values in excellent agreement with the experimental value of 26% (see the green bars in Figure 6). Mechanistically, the computed decrease in the PLQY value when going from the three-state to the four-state and five-state models originates from the introduction of $T_2$ as an additional non-radiative decay pathway, whereas $S_2$ partially compensates for this loss by providing an extra delayed fluorescence pathway. Overall, while the predominant delayed fluorescence pathway in DiKTa remains that of the three-state model, the inclusion of the higher-lying excited states improves the quantitative agreement with respect to the experiment. In addition, the predicted delayed fluorescence lifetime increases from 20.57 μs in the three-state model to 21.50 μs and 21.02 μs in the four and five state models, the latter two in excellent agreement with the experimental value of 23 μs. All in all, these results show that in DiKTa the inclusion of $T_2$ is not only mechanistically important but also essential for quantitatively reproducing the experimental PLQY, owing to fast (r)ISC and substantial triplet state populations. In contrast, $S_2$ primarily improves the mechanistic description of TADF pathways but it is not required to accurately predict the experimentally observable.

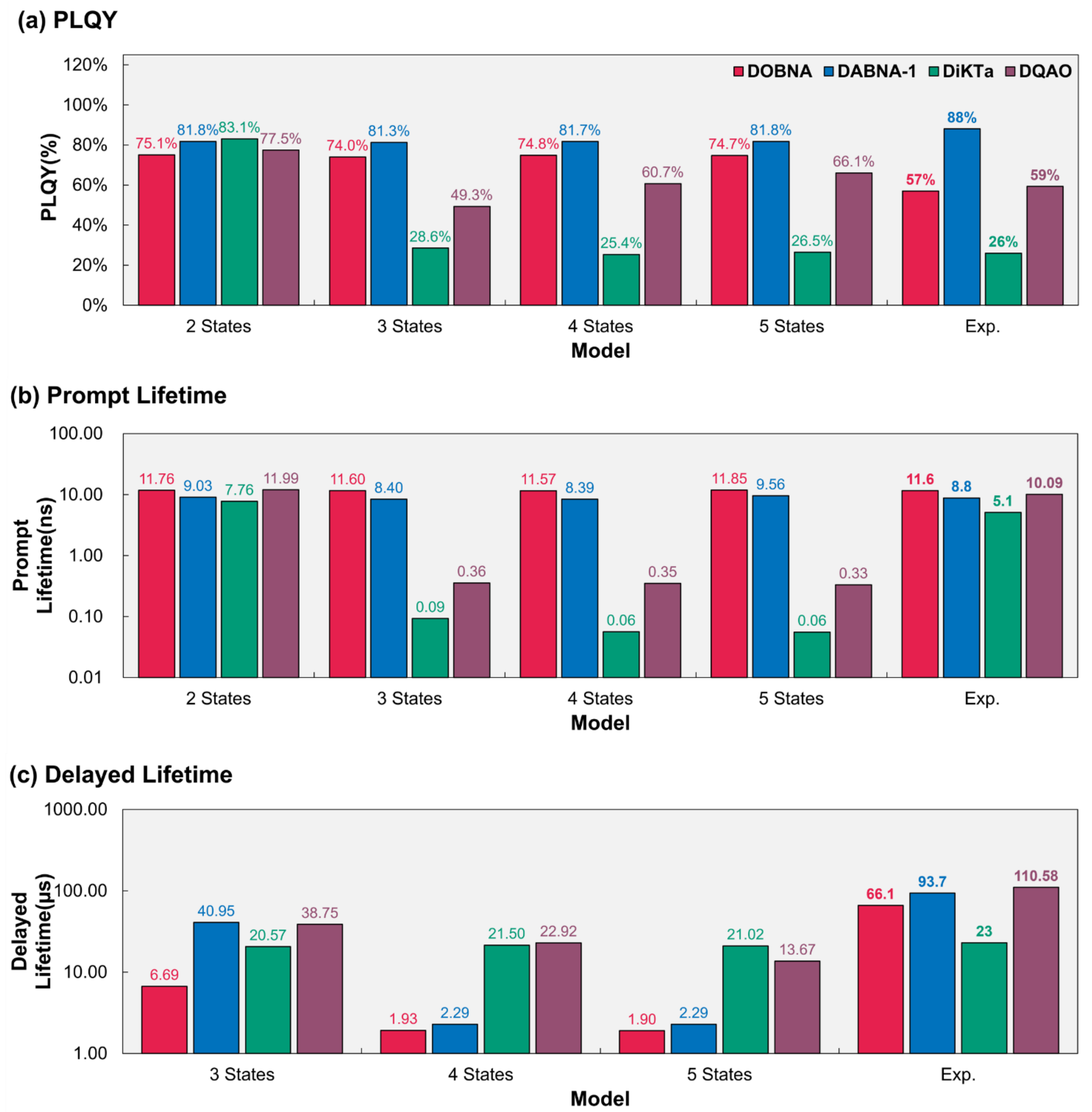

**Figure 6.** Calculated (a) photoluminescence quantum yield (PLQY), (b) prompt fluorescence lifetime, and (c) delayed fluorescence lifetime for DOBNA (red), DABNA-1 (blue), DiKTa (green), and DQAO (purple) compared with experimental results.

As a derivative of DiKTa, DQAO exhibits similar excited-state dynamics (Figure S12). The calculated fluorescence PLQY decreases progressively from 77.5% in the two-state model to 66.1% in the five-state model. The comparison of the PLQY values for the three-state (49.3%) and the four-state model (60.7%) reveals that only after the inclusion of $T_2$ in the kinetic models can we recover the experimental value of 59% (see the purple bars in Figure 6). In addition, the delayed fluorescence lifetime decreases from ~38.75 μs in the three-state model to ~22.92 μs in the four-state model, and to ~13.67 μs in the five-state model. Consistent with the results obtained for DiKTa, the inclusion of $T_2$ in DQAO is therefore essential for quantitatively reproducing the experimental PLQY, reflecting the fast (r)ISC processes and the dominant triplet state populations. Unlike in DiKTa, $T_2$ in DQAO additionally mediates a minor alternative delayed fluorescence pathway ($T_1 \rightarrow T_2 \rightarrow S_1 \rightarrow S_0$), as reflected in the changes in delayed fluorescence lifetime, rather than acting solely as a competing non-radiative decay pathway. The inclusion of $S_2$ further refines the mechanistic description of the TADF process but has only a minor impact on the predicted experimentally observables. Overall, these results demonstrate that the adequacy of four-state kinetic models is likewise determined by the magnitude of (r)ISC rate constants relative to fluorescence and IC and by whether higher-lying triplet states provide kinetically competitive rISC pathways. In systems such as DiKTa and DQAO, where fast (r)ISC leads to substantial triplet state populations, the explicit inclusion of $T_2$ is required to quantitatively reproduce experimentally observable quantities.

**2.4. Excited-State Kinetics Simulation beyond TADF emitters.** As discussed above, higher-lying excited states may not only modify delayed fluorescence pathways but can also introduce additional nonradiative decay channels. Consequently, for certain non-TADF organic emitters, a three-state kinetic model is likewise insufficient. This behavior is illustrated below using DBT (Figure S13) and PPDs-1 (Figure S14).

For DBT, the three-state model already predicts a substantial population of $T_1$ (Figure S13b) from fast ISC $S_1 \rightarrow T_1$ (~$2.13 \times 10^8$ $s^{-1}$), giving rise to a nonradiative decay pathway $S_1 \rightarrow T_1 \rightarrow S_0$. Consequently, the $S_1$ population exhibits a sigmoidal temporal profile, similar to the one observed by DiKTa. When moving to the four-state model, the inclusion of $T_2$ introduces an additional major non-radiative pathway ($S_1 \rightarrow T_2 \rightarrow T_1 \rightarrow S_0$), because the ISC rate constant for $S_1 \rightarrow T_2$ is also very large (~$2.65 \times 10^8$ $s^{-1}$). In the five-state model, the higher-lying state $S_2$ provides another nonradiative pathway ($S_1 \rightarrow S_2 \rightarrow S_0$). However, its impact on the overall kinetics is of very minor relevance compared to the dominant ISC-mediated processes. As a result, the calculated PLQY progressively decreases from 73.3% in the two-state model to 17.1% in the three-state model and further to 8.8% in the four-state model and 8.5% in the five-state model (Figure 7a). The PLQY of the four-state model is thus in good agreement with the experimental value of 9%. This highlights the necessity of including a higher-lying triplet excited state ($T_2$) to attain quantitatively accurate PLQY estimates. In parallel, the calculated prompt fluorescence lifetime decreases from 15.4 ns in the two-state model to 1.8 ns in the four-state model and 2.6 ns in the five-states model, closely matching the experimental value of ca. 3 ns. All in all, the higher-lying $T_2$ plays a crucial role by introducing competing nonradiative decay pathways ($S_1 \rightarrow T_2 \rightarrow T_1 \rightarrow S_0$), which are essential for quantitatively reproducing both PLQY and prompt fluorescence lifetimes in systems dominated by fast ISC processes involving higher-lying triplets. Accordingly, the four-state model, explicit inclusion of $T_2$ represents the minimal extension needed to achieve quantitative agreement for non-TADF organic emitters in which ISC processes involving higher-lying triplets are significant. We note, however, that the present analysis does not exclude the possible involvement of higher-lying triplet states beyond $T_2$ (e.g., $T_3$) in other molecular systems. In such cases, extensions of the present framework to include additional excited states may be required to achieve quantitative accuracy, which is beyond the scope of the present work.

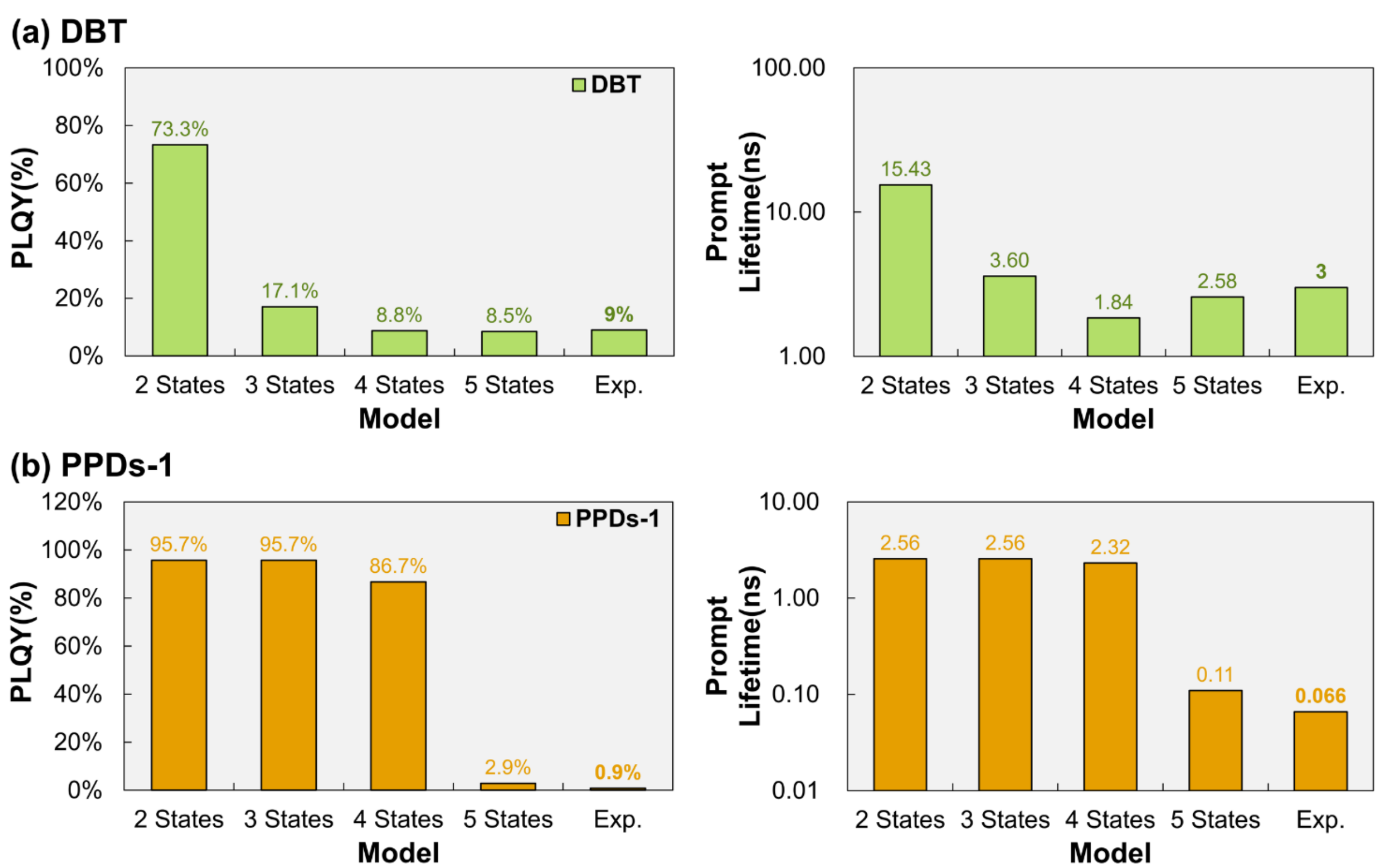

**Figure 7.** Calculated photoluminescence quantum yield (PLQY) and prompt fluorescence lifetime for (a) DBT (light green) and (b) PPDs-1 (orange), compared with experimental results.

For PPDs-1, the three-state model (Figure S14b) predicts a low $T_1$ population due to the relatively slow ISC $S_1 \rightarrow T_1$ (~$4.61\times10^3$ $s^{-1}$) relative to IC $S_1 \rightarrow S_0$ (~$1.67\times10^7$ $s^{-1}$) and fluorescence (~$3.74\times10^8$ $s^{-1}$). In the four-state model, a faster ISC from $S_1 \rightarrow T_2$ (~$4.07\times10^7$ $s^{-1}$) increases the $T_1$ population and introduces an additional nonradiative decay pathway ($S_1 \rightarrow T_2 \rightarrow T_1 \rightarrow S_0$). In the five-state model, the presence of the higher-lying singlet state $S_2$ opens up a highly efficient nonradiative decay channel, $S_1 \rightarrow S_2 \rightarrow S_0$, which dominates the excited-state dynamics. As a result, the calculated PLQY decreases progressively from 95.7% in the two-state model to 2.9% in the five-state model (Figure 7b), in good agreement with the experimental value of 0.9%. Similarly, the calculated prompt fluorescence lifetime decreases from 2.56 ns to 0.11 ns in the five-state model, closely matching the experimental value of approximately 0.066 ns. These results demonstrate that higher-lying singlet states $S_2$ can play a decisive role by introducing efficient competing nonradiative decay pathways ($S_1 \rightarrow S_2 \rightarrow S_0$), which are essential for quantitatively reproducing both PLQY and prompt fluorescence lifetimes in systems dominated by fast rIC processes. Accordingly, the five-state model, explicitly including both $T_2$ and $S_2$, represents the minimal kinetic description beyond the three-state description required to achieve quantitative agreement for non-TADF emitters dominated by fast ISC and rIC processes.

### 2.5. Accuracy and Limitations of Our Approach.

Including higher-lying excited states in the kinetic models improves the mechanistic completeness of the excited state dynamic description and can yield more reliable predictions of PLQY and excited-state lifetimes by explicitly accounting for all relevant photophysical pathways in the solution of the rate equations. Nevertheless, achieving quantitative agreement with experiment from ab initio calculations remains highly challenging, largely due to uncertainties in the calculated excited-state decay rate constants. For instance, as illustrated in Figure S15, even small variations in the calculated energy gaps $\Delta E(S_1\text{-}S_0)$ can lead to orders-of-magnitude changes in the calculated $S_1 \rightarrow S_0$ IC rate constants. Consequently, extending the kinetic model to include additional excited states and pathways inevitably increases the number of rate constants that must be estimated, which can result in the accumulation and propagation of errors in the predicted observables.

For DOBNA and its derivative DABNA-1, the calculated delayed fluorescence lifetimes obtained from the four- and five-state kinetic models are underestimated by nearly one order of magnitude with respect to the experimental values. Even for the most rigid systems, errors of this size in the computed lifetimes are not uncommon. All in all, the deviations in the predicted PLQY and delayed fluorescence lifetime are to be unambiguously traced back to inaccuracies in some specific rate constant calculations. For DOBNA (Figure S16a), gradually varying the value of the $S_1 \rightarrow S_0$ IC rate constant over a range of $10^7$-$10^8$ $s^{-1}$ while keeping all the other computed rate constants fixed leads to a pronounced change in the calculated PLQY, decreasing from approximately 80% to 40%, whereas the prompt and delayed fluorescence lifetimes remain essentially unaffected. This indicates that uncertainties in the calculated $S_1 \rightarrow S_0$ IC rate constants are the primary source of the PLQY deviation. Furthermore, Figure S16b demonstrates that variations in the $T_1 \rightarrow S_0$ ISC rate constant strongly affect the delayed fluorescence lifetime, identifying this process as the dominant contributor to the inaccuracy in the predicted delayed lifetime for DOBNA. For its derivative DABNA-1 (Figures S16c and S16d), the analysis indicates that uncertainties in the calculation of the $S_1 \rightarrow S_0$ IC rate constant predominantly affect the predicted PLQY, whereas variations in the $T_1 \rightarrow S_0$ ISC rate constant largely govern the accuracy of the delayed fluorescence lifetime.

For DiKTa and its derivative DQAO, the three-, four-, and five-state kinetic models all slightly underestimate the prompt fluorescence lifetime by nearly one order of magnitude. Among them, the four-state model yields the closest overall agreement with experiment when prompt and delayed fluorescence lifetimes and PLQY are considered simultaneously. In contrast to DOBNA and DABNA-1, the deviations observed for DiKTa and DQAO cannot be traced back to a single dominant rate constant but rather reflect the coupled and interdependent contributions of multiple excited-state decay processes. For the non-TADF emitters DBT and PPDs-1, the calculated lifetimes and PLQY show deviations from the experimental data within an acceptable range.

All in all, the above findings highlight a limitation of our fully *ab initio* kinetic modeling approach: the quantitative accuracy is ultimately determined by the accuracy in the state-of-the-art approaches to calculate excited state decay rate constants. Nevertheless, and despite the inherent uncertainties in the calculated rate constants, our investigations unambiguously highlight that a three-state model may be insufficient for some MR-TADF and non-TADF organic emitters, underscoring the need for more comprehensive excited-state kinetic models.

## 3. Conclusion

In this study, we developed KinLuv, a Python-based kinetic modeling package which explicitly includes higher-lying excited states ($T_2$ and $S_2$) into the excited state kinetic analysis. All rate constants were computed using Fermi's golden rule, explicitly including the HT vibronic coupling effect. Our approach was applied to prototypical MR-TADF emitters (DOBNA, DiKTa) and their derivatives (DABNA-1, DQAO), as well as to non-TADF organic emitters (DBT, PPDs-1). Key findings are summarized as follows:

a) In the investigated organic emitters, fast (r)ISC is primarily governed by HT vibronic coupling, as revealed by the relative contributions of HT and FC mechanisms to the calculated rate constants.

b) Mechanistically, inclusion of $T_2$ state in the four-state model additionally mediates an additional delayed fluorescence pathway ($T_1 \rightarrow T_2 \rightarrow S_1 \rightarrow S_0$) and a competing non-radiative decay pathway ($S_1 \rightarrow T_2 \rightarrow T_1 \rightarrow S_0$). Inclusion of $S_2$ state in the five-state model enables an additional delayed fluorescence pathway ($T_1 \rightarrow S_2 \rightarrow S_1 \rightarrow S_0$) and introduces further efficient competing nonradiative decay pathways ($S_1 \rightarrow S_2 \rightarrow S_0$).

c) Quantitively, when (r)ISC processes are slow relative to fluorescence and IC, higher-lying excited states are not significantly populated and do not affect the overall decay dynamics; in this regime, a three-state description already reproduces the experimental observables (e.g., DOBNA and DABNA-1).

d) When (r)ISC processes are fast leading to substantial triplet populations, and when the population of higher-lying triplet states become kinetically relevant; in this regime, explicit inclusion of at least $T_2$ is required to achieve quantitative agreement (e.g., DiKTa, DQAO and DBT).

e) When rIC processes dominate the excited-state dynamics, higher-lying singlet states become essential nonradiative decay channels; in this regime, explicit inclusion of $S_2$ is required to reproduce the experimental observables (e.g., PPDs-1).

Overall, this unified framework clarifies that higher-lying excited states may play a mechanistically important role by introducing additional decay pathways and that their inclusion in the kinetic models may become necessary to attain quantitative accuracy in the calculated values of PLQY and fluorescence lifetimes. At the same time, increased mechanistic completeness does not inherently guarantee improved quantitative accuracy owing to the amplification of errors and uncertainties in the underlying calculation of *ab initio* rate constants. Further developments in FGR-based theories, including anharmonic effects, will be necessary to improve the quantitative accuracy of PLQY and fluorescence lifetimes. Taken together, these insights establish a rational criterion for selecting minimal kinetic models that balance physical insight and numerical robustness, with direct implications for *in silico* design of high-performance organic emitters.

## 4. COMPUTATIONAL METHODS

**PLQY and Lifetimes Predictions.** We developed KinLuv, a Python-based kinetic simulation tool designed to predict prompt and delayed fluorescence lifetimes as well as PLQY using fully calculated *ab initio* rate constants as input. KinLuv is suited to modeling intricate multistate photophysical processes including fluorescence, phosphorescence, (reverse) internal conversion (IC), (r)ISC by solving coupled systems of ordinary differential equations (ODEs). The simulation framework consists of three main steps: excitation, decay, and PLQY evaluation. Required inputs include necessary rate constants and initial conditions, such as the number of absorbed photons (e.g., 1) and the excitation pulse width (e.g., 10 ps). For two- and three-state models, KinLuv provides analytical solutions, while for more sophisticated four- and five-state systems, it employs numerical simulations with separate timescales for excitation (e.g., 1 ns) and decay (e.g., 1 ms) to ensure accurate and stable results. By comprehensively accounting for all relevant transitions, KinLuv facilitates direct comparison with experimental data and provides clear insights into the underlying excited-state decay mechanisms. The accuracy of the simulated photophysical properties is inherently limited by uncertainties in the computed rate constants, which may arise from methodological approximations (e.g., validity of the harmonic oscillator approximation). These limitations are well described for the state-of-the art excited state decay rate constant theories[40] and they are further elaborated for the specific cases studied here in Section 2.5.

**Geometry Optimizations.** The geometries of ground, singlet and triplet excited states were optimized using density functional theory (DFT)[45,46], time-dependent DFT (TD-DFT)[47,48], and Tamm-Dancoff approximation DFT (TDA-DFT)[49], respectively. The CAM-B3LYP functional[50] along with the 6-311G(d,p) basis set were employed for the TADF emitters DOBNA, DiKTa, and DQAO. The 6-31G(d,p) basis set was applied for the larger TADF molecule DABNA-1 and non-TADF systems DBT and PPDs-1, which is sufficiently accurate for the electronic states considered within the scope of this study. This computational protocol was applied to all subsequent electronic structure calculations, except for the adiabatic energy gap calculations. Grimme's D3 dispersion correction[51] was included for all the investigated molecules. Solvent effects were described using the solvation model based on density (SMD)[52] with toluene for DiKTa, cyclohexane for DBT, and dichloromethane for PPDs-1, to mimic the corresponding experimental conditions. All optimizations were performed using Gaussian (16A03) with default convergence criteria and grids[53], followed by frequency calculations to obtain the Hessian matrix and to confirm the absence of imaginary frequencies.

**Adiabatic Energy Gap Calculations.** The protocol described above was refined to improve the accuracy of the computed adiabatic energy gaps. Towards this end, single-point energy calculations on the respective optimized geometries were carried out with the spin-component-scaled (SCS)[54] second-order algebraic diagrammatic construction (ADC(2)) method[55] and the def2-TZVP basis set[56]. This level of theory has been proven highly accurate (within 0.05 eV) for the adiabatic energy gap of the first singlet and triplet excited states of organic emitters, including e.g., MR-TADF dyes[57–59]. The resolution of identity (RI) approximation was employed with no frozen core approximation[60]. These calculations were performed with the Turbomole (7.7)[61].

**Excited State Decay Rate Constants.** Transition dipole moments (TDMs) were computed using TD(A)-CAM-B3LYP at the $S_1$ and $T_1$ optimized geometries. Nonadiabatic coupling matrix elements (NACMEs) were determined at TD-CAM-B3LYP level of theory and evaluated at the optimized geometries of corresponding electronic states using the electron-translation factors (ETF) correction, as implemented in Q-CHEM (5.3)[62]. Spin orbit coupling matrix elements (SOCMEs) were evaluated at the TDA-CAM-B3LYP level of theory with ORCA (5.0.4, 6.0)[63,64], specifying a higher grid (defgrid3) and four-center integrals for both Coulomb and exchange terms (SOCFlags = 1,4,4,1). Excited-state decay rate constants were computed using the thermal vibration-correlation function (TVCF) formalism[65], also known as the time-dependent approach. The adiabatic Hessian (AH) vibronic model[66] was chosen, which incorporates Duschinsky rotation effects. We employed the previously optimized geometries, gradients, Hessian matrices, TDMs, NACMEs, and SOCMEs, together with adiabatic energy gaps obtained from SCS-ADC(2). Calculations were carried out at room temperature (~300 K) in the vibronic framework, and a Lorentzian broadening with a half width at half maximum (HWHM) of 10 $cm^{-1}$ was applied. Fluorescence and IC rate constants were evaluated with FCClasses (3.0.1)[67], while phosphorescence and (r)ISC rates were computed using ORCA (5.0.4, 6.0). HT vibronic coupling was included in the radiative and ISC calculations, whereas IC rate constants can only be treated under the Franck-Condon (FC) approximation.

## ASSOCIATED CONTENT

**Supporting Information**.

The Supporting Information is available free of charge.

Includes theoretical background; ODEs and algebraic equations under SSA; natural transition orbitals; optimized geometries of ground and excited states; adiabatic excitation energies and rate constant calculations; error diagnosis of computational rate constants; simulated excitation and decay kinetics results generated by KinLuv.

**Code Availability**.
The source code, KinLuv, used in this study is publicly available on GitHub at: https://github.com/stevenuoa/KinLuv

# AUTHOR INFORMATION

## Corresponding Author

Daniel Escudero (D.E.) – Quantum Chemistry and Physical Chemistry Division, Department of Chemistry, KU Leuven, Celestijnenlaan 200F, 3001 Leuven, Belgium; http://orcid.org/0000-0002-1777-8578; Email: daniel.escudero@kuleuven.be

## Author

Yue He (Y.H) – Quantum Chemistry and Physical Chemistry Division, Department of Chemistry, KU Leuven, Celestijnenlaan 200F, 3001 Leuven, Belgium

## Author Contributions

Y.H. performed the theoretical calculations, and D.E. supervised the project. All authors participated in the manuscript preparation and approved the final version for publication.

## Notes

The authors declare no competing interests.

# ACKNOWLEDGMENT

D.E. acknowledges FWO (project numbers G079122N, G022324N) and KU Leuven Internal Funds (NovelTADF, 3E250385). The quantum chemical calculations were performed on the VSC (Flemish Supercomputer Center), funded by the Research Foundation - Flanders (FWO) and the Flemish Government. We acknowledge the support of OpenAI's ChatGPT in polishing the language of this manuscript and assisting with code refinement.

**SUPPORTING INFORMATION**

# Quantifying the Role of Higher-Lying Excited States in Organic Emitters via Multistate *Ab Initio* Kinetic Modeling

Yue He, Daniel Escudero*

Quantum Chemistry and Physical Chemistry Division, Department of Chemistry, KU Leuven, Celestijnenlaan 200F, 3001 Leuven, Belgium

*E-mail: Daniel.Escudero@kuleuven.be

**CONTENTS**

## 1. Theoretical Background for Rate Constants Calculations

**Fermi's Golden Rule (FGR).** The transition rate constant ($k_{if}$) between the initial state and the final state is given by FGR[1]:

$$k_{if} = \frac{2\pi}{\hbar}\left|\langle\Psi_f|\hat{\mathcal{H}}|\Psi_i\rangle\right|^2\rho \tag{S1.1}$$

where $\Psi_i$ and $\Psi_f$ are the total wave functions (electronic and vibrational) of the initial and final states, $\rho$ is the density of states and $\left|\langle\Psi_f|\hat{\mathcal{H}}|\Psi_i\rangle\right|$ is the matrix element of perturbation $\hat{\mathcal{H}}$ between the initial and final states. Moreover, the total wave function ($\Psi$) can be approximated by the product of its electronic ($\Phi$) and vibrational parts ($\Theta$) following the Born-Oppenheimer (BO) approximation[2]. By further applying the Condon approximation[3], eq S1.1 can be rewritten as:

$$k_{if} = \frac{2\pi}{\hbar}\sum_{\nu_i,\nu_f} P_{i,\nu_i}(T)\left|\left\langle\Theta_{f,\nu_f}\middle|\langle\Phi_f|\hat{\mathcal{H}}|\Phi_i\rangle\middle|\Theta_{i,\nu_i}\right\rangle\right|^2\delta\left(E_{f,\nu_f}-E_{i,\nu_i}\right) \tag{S1.2}$$

where the summation is carried out over all vibrational states ($\nu_i$, $\nu_f$) of the initial ($i, \nu_i$) and final ($f,\ \nu_f$) electronic states, $P_{i,\nu_i}(T)$ is the Boltzmann population of the initial vibrational level at a temperature $T$, $\hbar$ is the reduced Planck's constant, and $\delta$ is the Dirac delta function which selects the energy corresponding to the transition. The three classes of rate constants are computed via FGR with different perturbing operators[4,5]: the radiative rate constant $k_{\mathrm{r}}$ is obtained by using the electric dipole (ELD) operator as the perturbation matrix; the rate constant of intersystem crossing (ISC) $k_{\mathrm{ISC}}$ is calculated by employing the spin orbit coupling (SOC) operator; the rate constant of internal conversion (IC) $k_{\mathrm{IC}}$ is derived with the nonadiabatic coupling (NAC) operator. An essential component of all rate constant calculations is the overlap between vibrational wave functions of the two electronic states involved, which are typically evaluated with a harmonic oscillator approximation. Adiabatic Hessian (AH) is one of the popular vibronic models, where each state's vibrational wave functions are evaluated at its own optimized geometry (the Hessian at its minimum)[6]. Vibrational modes are then mapped onto one another via the Duschinsky rotation[7]. AH can break down if the two states' geometries differ excessively.

**Radiative Rate Constants.** When the perturbation ($\hat{\mathcal{H}}$) is replaced by the ELD ($\hat{\mu}$), the radiative rate constant can be obtained with eq S1.2. The transition dipole moments (TDMs) $\vec{\mu}_{if}$ depending on the vibrational normal coordinate ($Q$) can be expanded as:

$$\vec{\mu}_{if}(Q) = \left(\langle\Phi_f|\hat{\mu}|\Phi_i\rangle\right)_0 + \sum_k\left(\frac{\partial\langle\Phi_f|\hat{\mu}|\Phi_i\rangle}{\partial Q_k}\right)_0 Q_k \tag{S1.3}$$

where $\left(\langle\Phi_f|\hat{\mu}|\Phi_i\rangle\right)_0$ is the zeroth-order TMD in the Franck-Condon (FC) approximation, and the second term refers to the so-called Herzberg-Teller (HT) effects[8], which arises from the first-order dependence of the TDMs on the nuclear displacements. The summation runs over all normal vibrational modes ($Q_k$).

**ISC Rate Constants.** When the perturbation ($\hat{\mathcal{H}}$) becomes the SOC operator ($\hat{\mathcal{H}}_{\mathrm{SOC}}$), the ISC rate constant can be obtained with eq S1.2. Analogously to that for radiative rate constants, an HT-like expression for ISC also exists. The spin orbit coupling matrix elements (SOCMEs) include not only the zeroth-order term $\left(\langle\Phi_f|\hat{\mathcal{H}}_{\mathrm{SOC}}|\Phi_i\rangle\right)_0$ but also the first-order expansion with respect to the normal coordinates ($Q$), which accounts for the vibronic spin-orbit interactions[9]:

$$SOC(Q) = \left(\langle\Phi_f|\hat{\mathcal{H}}_{\mathrm{SOC}}|\Phi_i\rangle\right)_0 + \sum_k\left(\frac{\partial\langle\Phi_f|\hat{\mathcal{H}}_{\mathrm{SOC}}|\Phi_i\rangle}{\partial Q_k}\right)_0 Q_k \tag{S1.4}$$

**IC Rate Constants.** The IC rate constant, expressed in terms of NAC operator, can be obtained as follows[10]:

$$k_{\mathrm{IC}} = \sum_{k,l}\frac{R_{kl}}{\hbar^2}\int_{-\infty}^{\infty} e^{i\omega_{if}t} Z_{iv}^{-1}\rho_{\mathrm{IC},kl}(t,T)\,dt \tag{S1.5}$$

where $\rho_{\mathrm{IC},kl}(t,T)$ denotes the thermal vibration correlation function (TVCF) for IC, $Z_{iv}$ is the partition function, and $R_{kl}$ is the electronic coupling part of the NAC. Similar to TDMs, nonadiabatic coupling matrix elements (NACMEs) can be obtained from quantum chemical calculations. However, HT-like expansion for the IC rate constant is not yet widely implemented in the available codes.

## 2. Ordinary Differential Equations (ODEs) for Different Models of the Excitation Period

Two-state Model ($S_0$, $S_1$):

$$-[\mathrm{S_0}]k_{\mathrm{ABS}}^{\mathrm{S0\to S1}} + [\mathrm{S_1}]\left(k_{\mathrm{FL}}^{\mathrm{S1\to S0}} + k_{\mathrm{IC}}^{\mathrm{S1\to S0}}\right) = \frac{\mathrm{d}[\mathrm{S_0}]}{\mathrm{d}t} \tag{S2.1}$$

$$[\mathrm{S_0}]k_{\mathrm{ABS}}^{\mathrm{S0\to S1}} - [\mathrm{S_1}]\left(k_{\mathrm{FL}}^{\mathrm{S1\to S0}} + k_{\mathrm{IC}}^{\mathrm{S1\to S0}}\right) = \frac{\mathrm{d}[\mathrm{S_1}]}{\mathrm{d}t} \tag{S2.2}$$

Three-state Model ($S_0$, $S_1$, $T_1$):

$$-[\mathrm{S_0}]k_{\mathrm{ABS}}^{\mathrm{S0\to S1}} + [\mathrm{S_1}]\left(k_{\mathrm{FL}}^{\mathrm{S1\to S0}} + k_{\mathrm{IC}}^{\mathrm{S1\to S0}}\right) + [\mathrm{T_1}]\left(k_{\mathrm{PH}}^{\mathrm{T1\to S0}} + k_{\mathrm{ISC}}^{\mathrm{T1\to S0}}\right) = \frac{\mathrm{d}[\mathrm{S_0}]}{\mathrm{d}t} \tag{S2.3}$$

$$[\mathrm{S_0}]k_{\mathrm{ABS}}^{\mathrm{S0\to S1}} - [\mathrm{S_1}]\left(k_{\mathrm{ISC}}^{\mathrm{S1\to T1}} + k_{\mathrm{FL}}^{\mathrm{S1\to S0}} + k_{\mathrm{IC}}^{\mathrm{S1\to S0}}\right) + [\mathrm{T_1}]k_{\mathrm{rISC}}^{\mathrm{T1\to S1}} = \frac{\mathrm{d}[\mathrm{S_1}]}{\mathrm{d}t} \tag{S2.4}$$

$$[\mathrm{S_1}]k_{\mathrm{ISC}}^{\mathrm{S1\to T1}} - [\mathrm{T_1}]\left(k_{\mathrm{PH}}^{\mathrm{T1\to S0}} + k_{\mathrm{ISC}}^{\mathrm{T1\to S0}} + k_{\mathrm{rISC}}^{\mathrm{T1\to S1}}\right) = \frac{\mathrm{d}[\mathrm{T_1}]}{\mathrm{d}t} \tag{S2.5}$$

Four-state Model ($S_0$, $S_1$, $T_1$, $T_2$):

$$-[\mathrm{S_0}]k_{\mathrm{ABS}}^{\mathrm{S0\to S1}} + [\mathrm{S_1}]\left(k_{\mathrm{FL}}^{\mathrm{S1\to S0}} + k_{\mathrm{IC}}^{\mathrm{S1\to S0}}\right) + [\mathrm{T_1}]\left(k_{\mathrm{PH}}^{\mathrm{T1\to S0}} + k_{\mathrm{ISC}}^{\mathrm{T1\to S0}}\right) = \frac{\mathrm{d}[\mathrm{S_0}]}{\mathrm{d}t} \tag{S2.6}$$

$$[\mathrm{S_0}]k_{\mathrm{ABS}}^{\mathrm{S0\to S1}} - [\mathrm{S_1}]\left(k_{\mathrm{ISC}}^{\mathrm{S1\to T1}} + k_{\mathrm{ISC}}^{\mathrm{S1\to T2}} + k_{\mathrm{FL}}^{\mathrm{S1\to S0}} + k_{\mathrm{IC}}^{\mathrm{S1\to S0}}\right) + [\mathrm{T_1}]k_{\mathrm{rISC}}^{\mathrm{T1\to S1}} + [\mathrm{T_2}]k_{\mathrm{rISC}}^{\mathrm{T2\to S1}} = \frac{\mathrm{d}[\mathrm{S_1}]}{\mathrm{d}t} \tag{S2.7}$$

$$[\mathrm{S_1}]k_{\mathrm{ISC}}^{\mathrm{S1\to T1}} - [\mathrm{T_1}]\left(k_{\mathrm{PH}}^{\mathrm{T1\to S0}} + k_{\mathrm{ISC}}^{\mathrm{T1\to S0}} + k_{\mathrm{rISC}}^{\mathrm{T1\to S1}} + k_{\mathrm{rIC}}^{\mathrm{T1\to T2}}\right) + [\mathrm{T_2}]k_{\mathrm{IC}}^{\mathrm{T2\to T1}} = \frac{\mathrm{d}[\mathrm{T_1}]}{\mathrm{d}t} \tag{S2.8}$$

$$[\mathrm{S_1}]k_{\mathrm{ISC}}^{\mathrm{S1\to T2}} + [\mathrm{T_1}]k_{\mathrm{rIC}}^{\mathrm{T1\to T2}} - [\mathrm{T_2}]\left(k_{\mathrm{rISC}}^{\mathrm{T2\to S1}} + k_{\mathrm{IC}}^{\mathrm{T2\to T1}}\right) = \frac{\mathrm{d}[\mathrm{T_2}]}{\mathrm{d}t} \tag{S2.9}$$

Five-state Model ($S_0$, $S_1$, $S_2$, $T_1$, $T_2$):

$$-[\mathrm{S_0}]k_{\mathrm{ABS}}^{\mathrm{S0\to S2}} + [\mathrm{S_1}]\left(k_{\mathrm{FL}}^{\mathrm{S1\to S0}} + k_{\mathrm{IC}}^{\mathrm{S1\to S0}}\right) + [\mathrm{S_2}]\left(k_{\mathrm{FL}}^{\mathrm{S2\to S0}} + k_{\mathrm{IC}}^{\mathrm{S2\to S0}}\right) + [\mathrm{T_1}]\left(k_{\mathrm{PH}}^{\mathrm{T1\to S0}} + k_{\mathrm{ISC}}^{\mathrm{T1\to S0}}\right) = \frac{\mathrm{d}[\mathrm{S_0}]}{\mathrm{d}t} \tag{S2.10}$$

$$-[\mathrm{S_1}]\left(k_{\mathrm{ISC}}^{\mathrm{S1\to T1}} + k_{\mathrm{ISC}}^{\mathrm{S1\to T2}} + k_{\mathrm{FL}}^{\mathrm{S1\to S0}} + k_{\mathrm{IC}}^{\mathrm{S1\to S0}} + k_{\mathrm{rIC}}^{\mathrm{S1\to S2}}\right) + [\mathrm{S_2}]k_{\mathrm{IC}}^{\mathrm{S2\to S1}} + [\mathrm{T_1}]k_{\mathrm{rISC}}^{\mathrm{T1\to S1}} + [\mathrm{T_2}]k_{\mathrm{rISC}}^{\mathrm{T2\to S1}} = \frac{\mathrm{d}[\mathrm{S_1}]}{\mathrm{d}t} \tag{S2.11}$$

$$[\mathrm{S_0}]k_{\mathrm{ABS}}^{\mathrm{S0\to S2}} + [\mathrm{S_1}]k_{\mathrm{rIC}}^{\mathrm{S1\to S2}} - [\mathrm{S_2}]\left(k_{\mathrm{IC}}^{\mathrm{S2\to S1}} + k_{\mathrm{ISC}}^{\mathrm{S2\to T1}} + k_{\mathrm{ISC}}^{\mathrm{S2\to T2}} + k_{\mathrm{FL}}^{\mathrm{S2\to S0}} + k_{\mathrm{IC}}^{\mathrm{S2\to S0}}\right) + [\mathrm{T_1}]k_{\mathrm{rISC}}^{\mathrm{T1\to S2}} + [\mathrm{T_2}]k_{\mathrm{rISC}}^{\mathrm{T2\to S2}} = \frac{\mathrm{d}[\mathrm{S_2}]}{\mathrm{d}t} \tag{S2.12}$$

$$[\mathrm{S_1}]k_{\mathrm{ISC}}^{\mathrm{S1\to T1}} + [\mathrm{S_2}]k_{\mathrm{ISC}}^{\mathrm{S2\to T1}} - [\mathrm{T_1}]\left(k_{\mathrm{PH}}^{\mathrm{T1\to S0}} + k_{\mathrm{ISC}}^{\mathrm{T1\to S0}} + k_{\mathrm{rISC}}^{\mathrm{T1\to S1}} + k_{\mathrm{rISC}}^{\mathrm{T1\to S2}} + k_{\mathrm{rIC}}^{\mathrm{T1\to T2}}\right) + [\mathrm{T_2}]k_{\mathrm{IC}}^{\mathrm{T2\to T1}} = \frac{\mathrm{d}[\mathrm{T_1}]}{\mathrm{d}t} \tag{S2.13}$$

$$[\mathrm{S_1}]k_{\mathrm{ISC}}^{\mathrm{S1\to T2}} + [\mathrm{S_2}]k_{\mathrm{ISC}}^{\mathrm{S2\to T2}} + [\mathrm{T_1}]k_{\mathrm{rIC}}^{\mathrm{T1\to T2}} - [\mathrm{T_2}]\left(k_{\mathrm{rISC}}^{\mathrm{T2\to S1}} + k_{\mathrm{rISC}}^{\mathrm{T2\to S2}} + k_{\mathrm{IC}}^{\mathrm{T2\to T1}}\right) = \frac{\mathrm{d}[\mathrm{T_2}]}{\mathrm{d}t} \tag{S2.14}$$

The transient photoluminescence decay curve is typically measured following a short excitation pulse. Equations (eq S2.1-S2.14) above describe the excited-state population kinetics during this excitation period. Consequently, the excited-state populations at the end of the excitation pulse serve as the initial conditions for the subsequent decay period. To solve the ODEs, the absorption rate ($k_{\mathrm{ABS}}^{\mathrm{S0\to S1}}$, $k_{\mathrm{ABS}}^{\mathrm{S0\to S2}}$) and the duration of the excitation pulse $t_{pulse}$ must be specified. These parameters will not affect the prompt or delayed fluorescence lifetimes during the decay period. In this case study, $k_{\mathrm{ABS}}^{\mathrm{S0\to S1}} = 1\times10^{13}$ ($\mathrm{s}^{-1}$), $k_{\mathrm{ABS}}^{\mathrm{S0\to S2}} = 1\times10^{13}$ ($\mathrm{s}^{-1}$), $t_{pulse} = 10$ (ps).

## 3. ODEs in Vector-Matrix Form for Different Models of the Decay Period ($k_{\mathrm{ABS}}^{\mathrm{S0}\rightarrow\mathrm{S1}}=0$, $k_{\mathrm{ABS}}^{\mathrm{S0}\rightarrow\mathrm{S2}}=0$)

Two-state Model ($S_0$, $S_1$):

$$\frac{\mathrm{d}}{\mathrm{d}t}\begin{pmatrix}[\mathrm{S}_0]\\ [\mathrm{S}_1]\end{pmatrix}=\begin{pmatrix}0 & \left(k_{\mathrm{FL}}^{\mathrm{S1}\rightarrow\mathrm{S0}}+k_{\mathrm{IC}}^{\mathrm{S1}\rightarrow\mathrm{S0}}\right)\\ 0 & -\left(k_{\mathrm{FL}}^{\mathrm{S1}\rightarrow\mathrm{S0}}+k_{\mathrm{IC}}^{\mathrm{S1}\rightarrow\mathrm{S0}}\right)\end{pmatrix}\begin{pmatrix}[\mathrm{S}_0]\\ [\mathrm{S}_1]\end{pmatrix} \tag{S3.1}$$

Three-state Model ($S_0$, $S_1$, $T_1$):

$$\frac{\mathrm{d}}{\mathrm{d}t}\begin{pmatrix}[\mathrm{S}_0]\\ [\mathrm{S}_1]\\ [\mathrm{T}_1]\end{pmatrix}=\begin{pmatrix}0 & \left(k_{\mathrm{FL}}^{\mathrm{S1}\rightarrow\mathrm{S0}}+k_{\mathrm{IC}}^{\mathrm{S1}\rightarrow\mathrm{S0}}\right) & \left(k_{\mathrm{PH}}^{\mathrm{T1}\rightarrow\mathrm{S0}}+k_{\mathrm{ISC}}^{\mathrm{T1}\rightarrow\mathrm{S0}}\right)\\ 0 & -\left(k_{\mathrm{ISC}}^{\mathrm{S1}\rightarrow\mathrm{T1}}+k_{\mathrm{FL}}^{\mathrm{S1}\rightarrow\mathrm{S0}}+k_{\mathrm{IC}}^{\mathrm{S1}\rightarrow\mathrm{S0}}\right) & k_{\mathrm{rISC}}^{\mathrm{T1}\rightarrow\mathrm{S1}}\\ 0 & k_{\mathrm{ISC}}^{\mathrm{S1}\rightarrow\mathrm{T1}} & -\left(k_{\mathrm{PH}}^{\mathrm{T1}\rightarrow\mathrm{S0}}+k_{\mathrm{ISC}}^{\mathrm{T1}\rightarrow\mathrm{S0}}+k_{\mathrm{rISC}}^{\mathrm{T1}\rightarrow\mathrm{S1}}\right)\end{pmatrix}\begin{pmatrix}[\mathrm{S}_0]\\ [\mathrm{S}_1]\\ [\mathrm{T}_1]\end{pmatrix} \tag{S3.2}$$

Four-state Model ($S_0$, $S_1$, $T_1$, $T_2$):

$$\frac{\mathrm{d}}{\mathrm{d}t}\begin{pmatrix}[\mathrm{S}_0]\\ [\mathrm{S}_1]\\ [\mathrm{T}_1]\\ [\mathrm{T}_2]\end{pmatrix}=\begin{pmatrix}0 & \left(k_{\mathrm{FL}}^{\mathrm{S1}\rightarrow\mathrm{S0}}+k_{\mathrm{IC}}^{\mathrm{S1}\rightarrow\mathrm{S0}}\right) & \left(k_{\mathrm{PH}}^{\mathrm{T1}\rightarrow\mathrm{S0}}+k_{\mathrm{ISC}}^{\mathrm{T1}\rightarrow\mathrm{S0}}\right) & 0\\ 0 & -\left(k_{\mathrm{ISC}}^{\mathrm{S1}\rightarrow\mathrm{T1}}+k_{\mathrm{ISC}}^{\mathrm{S1}\rightarrow\mathrm{T2}}+k_{\mathrm{FL}}^{\mathrm{S1}\rightarrow\mathrm{S0}}+k_{\mathrm{IC}}^{\mathrm{S1}\rightarrow\mathrm{S0}}\right) & k_{\mathrm{rISC}}^{\mathrm{T1}\rightarrow\mathrm{S1}} & k_{\mathrm{rISC}}^{\mathrm{T2}\rightarrow\mathrm{S1}}\\ 0 & k_{\mathrm{ISC}}^{\mathrm{S1}\rightarrow\mathrm{T1}} & -\left(k_{\mathrm{PH}}^{\mathrm{T1}\rightarrow\mathrm{S0}}+k_{\mathrm{ISC}}^{\mathrm{T1}\rightarrow\mathrm{S0}}+k_{\mathrm{rISC}}^{\mathrm{T1}\rightarrow\mathrm{S1}}+k_{\mathrm{rIC}}^{\mathrm{T1}\rightarrow\mathrm{T2}}\right) & k_{\mathrm{IC}}^{\mathrm{T2}\rightarrow\mathrm{T1}}\\ 0 & k_{\mathrm{ISC}}^{\mathrm{S1}\rightarrow\mathrm{T2}} & k_{\mathrm{rIC}}^{\mathrm{T1}\rightarrow\mathrm{T2}} & -\left(k_{\mathrm{rISC}}^{\mathrm{T2}\rightarrow\mathrm{S1}}+k_{\mathrm{IC}}^{\mathrm{T2}\rightarrow\mathrm{T1}}\right)\end{pmatrix}\begin{pmatrix}[\mathrm{S}_0]\\ [\mathrm{S}_1]\\ [\mathrm{T}_1]\\ [\mathrm{T}_2]\end{pmatrix} \tag{S3.3}$$

Five-state Model ($S_0$, $S_1$, $S_2$, $T_1$, $T_2$):

$$\frac{\mathrm{d}}{\mathrm{d}t}\begin{pmatrix}[\mathrm{S}_0]\\ [\mathrm{S}_1]\\ [\mathrm{S}_2]\\ [\mathrm{T}_1]\\ [\mathrm{T}_2]\end{pmatrix}=K\begin{pmatrix}[\mathrm{S}_0]\\ [\mathrm{S}_1]\\ [\mathrm{S}_2]\\ [\mathrm{T}_1]\\ [\mathrm{T}_2]\end{pmatrix} \tag{S3.4}$$

$$K=\begin{pmatrix}0 & \left(k_{\mathrm{FL}}^{\mathrm{S1}\rightarrow\mathrm{S0}}+k_{\mathrm{IC}}^{\mathrm{S1}\rightarrow\mathrm{S0}}\right) & \left(k_{\mathrm{FL}}^{\mathrm{S2}\rightarrow\mathrm{S0}}+k_{\mathrm{IC}}^{\mathrm{S2}\rightarrow\mathrm{S0}}\right) & \left(k_{\mathrm{PH}}^{\mathrm{T1}\rightarrow\mathrm{S0}}+k_{\mathrm{ISC}}^{\mathrm{T1}\rightarrow\mathrm{S0}}\right) & 0\\ 0 & \sum \mathrm{S}_1 & k_{\mathrm{IC}}^{\mathrm{S2}\rightarrow\mathrm{S1}} & k_{\mathrm{rISC}}^{\mathrm{T1}\rightarrow\mathrm{S1}} & k_{\mathrm{rISC}}^{\mathrm{T2}\rightarrow\mathrm{S1}}\\ 0 & k_{\mathrm{rIC}}^{\mathrm{S1}\rightarrow\mathrm{S2}} & \sum \mathrm{S}_2 & k_{\mathrm{rISC}}^{\mathrm{T1}\rightarrow\mathrm{S2}} & k_{\mathrm{rISC}}^{\mathrm{T2}\rightarrow\mathrm{S2}}\\ 0 & k_{\mathrm{ISC}}^{\mathrm{S1}\rightarrow\mathrm{T1}} & k_{\mathrm{ISC}}^{\mathrm{S2}\rightarrow\mathrm{T1}} & \sum \mathrm{T}_1 & k_{\mathrm{IC}}^{\mathrm{T2}\rightarrow\mathrm{T1}}\\ 0 & k_{\mathrm{ISC}}^{\mathrm{S1}\rightarrow\mathrm{T2}} & k_{\mathrm{ISC}}^{\mathrm{S2}\rightarrow\mathrm{T2}} & k_{\mathrm{rIC}}^{\mathrm{T1}\rightarrow\mathrm{T2}} & \sum \mathrm{T}_2\end{pmatrix} \tag{S3.5}$$

$$\sum \mathrm{S}_1=-\left(k_{\mathrm{ISC}}^{\mathrm{S1}\rightarrow\mathrm{T1}}+k_{\mathrm{ISC}}^{\mathrm{S1}\rightarrow\mathrm{T2}}+k_{\mathrm{FL}}^{\mathrm{S1}\rightarrow\mathrm{S0}}+k_{\mathrm{IC}}^{\mathrm{S1}\rightarrow\mathrm{S0}}+k_{\mathrm{rIC}}^{\mathrm{S1}\rightarrow\mathrm{S2}}\right) \tag{S3.6}$$

$$\sum \mathrm{S}_2=-\left(k_{\mathrm{IC}}^{\mathrm{S2}\rightarrow\mathrm{S1}}+k_{\mathrm{ISC}}^{\mathrm{S2}\rightarrow\mathrm{T1}}+k_{\mathrm{ISC}}^{\mathrm{S2}\rightarrow\mathrm{T2}}+k_{\mathrm{FL}}^{\mathrm{S2}\rightarrow\mathrm{S0}}+k_{\mathrm{IC}}^{\mathrm{S2}\rightarrow\mathrm{S0}}\right) \tag{S3.7}$$

$$\sum \mathrm{T}_1=-\left(k_{\mathrm{PH}}^{\mathrm{T1}\rightarrow\mathrm{S0}}+k_{\mathrm{ISC}}^{\mathrm{T1}\rightarrow\mathrm{S0}}+k_{\mathrm{rISC}}^{\mathrm{T1}\rightarrow\mathrm{S1}}+k_{\mathrm{rISC}}^{\mathrm{T1}\rightarrow\mathrm{S2}}+k_{\mathrm{rIC}}^{\mathrm{T1}\rightarrow\mathrm{T2}}\right) \tag{S3.8}$$

$$\sum \mathrm{T}_2=-\left(k_{\mathrm{rISC}}^{\mathrm{T2}\rightarrow\mathrm{S1}}+k_{\mathrm{rISC}}^{\mathrm{T2}\rightarrow\mathrm{S2}}+k_{\mathrm{IC}}^{\mathrm{T2}\rightarrow\mathrm{T1}}\right) \tag{S3.9}$$

The general solution of ODEs is given by

$$\vec{N}(t)=\sum_{i=1}^{n} c_i\vec{v_i}e^{\lambda_i t} \tag{S3.10}$$

where $\vec{N}(t)$ is the state population vector, $\vec{v_i}$ denotes the eigenvector, $\lambda_i$ is the corresponding eigenvalue and $c_i$ is the coefficient associated with the matrix. For example, in the case of the three-state model ($S_0$, $S_1$, $T_1$), These parameters can be determined

$$\begin{pmatrix}\lambda_1\\ \lambda_2\\ \lambda_3\end{pmatrix}=\begin{pmatrix}0\\ -1/2\left(k_{\mathrm{sum}}^{\mathrm{S1}}-k_{\mathrm{sum}}^{\mathrm{T1}}+\sqrt{(k_{\mathrm{sum}}^{\mathrm{S1}}-k_{\mathrm{sum}}^{\mathrm{T1}})^2+4k_{\mathrm{ISC}}^{\mathrm{S1}\rightarrow\mathrm{T1}}k_{\mathrm{rISC}}^{\mathrm{T1}\rightarrow\mathrm{S1}}}\right)\\ -1/2\left(k_{\mathrm{sum}}^{\mathrm{S1}}-k_{\mathrm{sum}}^{\mathrm{T1}}-\sqrt{(k_{\mathrm{sum}}^{\mathrm{S1}}-k_{\mathrm{sum}}^{\mathrm{T1}})^2+4k_{\mathrm{ISC}}^{\mathrm{S1}\rightarrow\mathrm{T1}}k_{\mathrm{rISC}}^{\mathrm{T1}\rightarrow\mathrm{S1}}}\right)\end{pmatrix} \tag{S3.11}$$

$$\vec{v_1}=\begin{pmatrix}1\\ 0\\ 0\end{pmatrix},\vec{v_2}=\begin{pmatrix}-\left(\lambda_2+k_{\mathrm{sum}}^{\mathrm{S1}}+k_{\mathrm{rISC}}^{\mathrm{T1}\rightarrow\mathrm{S1}}\right)\\ k_{\mathrm{rISC}}^{\mathrm{T1}\rightarrow\mathrm{S1}}\\ \lambda_2+k_{\mathrm{sum}}^{\mathrm{S1}}\end{pmatrix},\vec{v_3}=\begin{pmatrix}-\left(\lambda_3+k_{\mathrm{sum}}^{\mathrm{S1}}+k_{\mathrm{rISC}}^{\mathrm{T1}\rightarrow\mathrm{S1}}\right)\\ k_{\mathrm{rISC}}^{\mathrm{T1}\rightarrow\mathrm{S1}}\\ \lambda_3+k_{\mathrm{sum}}^{\mathrm{S1}}\end{pmatrix} \tag{S3.12}$$

where $k_{\mathrm{sum}}^{\mathrm{S1}}=k_{\mathrm{ISC}}^{\mathrm{S1}\rightarrow\mathrm{T1}}+k_{\mathrm{FL}}^{\mathrm{S1}\rightarrow\mathrm{S0}}+k_{\mathrm{IC}}^{\mathrm{S1}\rightarrow\mathrm{S0}}$, $k_{\mathrm{sum}}^{\mathrm{T1}}=k_{\mathrm{PH}}^{\mathrm{T1}\rightarrow\mathrm{S0}}+k_{\mathrm{ISC}}^{\mathrm{T1}\rightarrow\mathrm{S0}}+k_{\mathrm{rISC}}^{\mathrm{T1}\rightarrow\mathrm{S1}}$. Consequently, the excited state populations can be described as follows:

$$\begin{pmatrix}[\mathrm{S_0}]\\ [\mathrm{S_1}]\\ [\mathrm{T_1}]\end{pmatrix}=\begin{pmatrix}c_1-c_2\left(k_{\mathrm{sum}}^{\mathrm{S1}}+k_{\mathrm{rISC}}^{\mathrm{T1\to S1}}-k_{\mathrm{PF}}\right)e^{-k_{\mathrm{PF}}t}-c_3\left(k_{\mathrm{sum}}^{\mathrm{S1}}+k_{\mathrm{rISC}}^{\mathrm{T1\to S1}}-k_{\mathrm{DF}}\right)e^{-k_{\mathrm{DF}}t}\\ c_2k_{\mathrm{rISC}}^{\mathrm{T1\to S1}}e^{-k_{\mathrm{PF}}t}+c_3k_{\mathrm{rISC}}^{\mathrm{T1\to S1}}e^{-k_{\mathrm{DF}}t}\\ c_2(k_{\mathrm{sum}}^{\mathrm{S1}}-k_{\mathrm{PF}})e^{-k_{\mathrm{PF}}t}+c_3(k_{\mathrm{sum}}^{\mathrm{S1}}-k_{\mathrm{DF}})e^{-k_{\mathrm{DF}}t}\end{pmatrix} \tag{S3.13}$$

where $k_{\mathrm{PF}}=-\lambda_2$, $k_{\mathrm{DF}}=-\lambda_3$, and coefficients $c_i$ can be obtained from eq S3.14.

$$\begin{pmatrix}1 & -\left(\lambda_2+k_{\mathrm{sum}}^{\mathrm{S1}}+k_{\mathrm{rISC}}^{\mathrm{T1\to S1}}\right) & -\left(\lambda_3+k_{\mathrm{sum}}^{\mathrm{S1}}+k_{\mathrm{rISC}}^{\mathrm{T1\to S1}}\right)\\ 0 & k_{\mathrm{rISC}}^{\mathrm{T1\to S1}} & k_{\mathrm{rISC}}^{\mathrm{T1\to S1}}\\ 0 & \lambda_2+k_{\mathrm{sum}}^{\mathrm{S1}} & \lambda_3+k_{\mathrm{sum}}^{\mathrm{S1}}\end{pmatrix}\begin{pmatrix}c_1\\ c_2\\ c_3\end{pmatrix}=\begin{pmatrix}[\mathrm{S_0}]_{t=0}\\ [\mathrm{S_1}]_{t=0}\\ [\mathrm{T_1}]_{t=0}\end{pmatrix} \tag{S3.14}$$

$$\begin{pmatrix}c_1\\ c_2\\ c_3\end{pmatrix}=\begin{pmatrix}[\mathrm{S_0}]_{t=0}+[\mathrm{S_1}]_{t=0}+[\mathrm{T_1}]_{t=0}\\ \dfrac{(k_{\mathrm{sum}}^{\mathrm{S1}}-k_{\mathrm{DF}})[\mathrm{S_1}]_{t=0}-k_{\mathrm{rISC}}^{\mathrm{T1\to S1}}[\mathrm{T_1}]_{t=0}}{k_{\mathrm{rISC}}^{\mathrm{T1\to S1}}(k_{\mathrm{PF}}-k_{\mathrm{DF}})}\\ \dfrac{-(k_{\mathrm{sum}}^{\mathrm{S1}}-k_{\mathrm{PF}})[\mathrm{S_1}]_{t=0}+k_{\mathrm{rISC}}^{\mathrm{T1\to S1}}[\mathrm{T_1}]_{t=0}}{k_{\mathrm{rISC}}^{\mathrm{T1\to S1}}(k_{\mathrm{PF}}-k_{\mathrm{DF}})}\end{pmatrix} \tag{S3.15}$$

Especially when $[\mathrm{S_0}]_{t=0}=[\mathrm{T_1}]_{t=0}=0$, $[\mathrm{S_1}]$ and $[\mathrm{T_1}]$ can be written as below:

$$[\mathrm{S_1}]=\frac{[\mathrm{S_1}]_{t=0}}{(k_{\mathrm{PF}}-k_{\mathrm{DF}})}(k_{\mathrm{sum}}^{\mathrm{S1}}-k_{\mathrm{DF}})e^{-k_{\mathrm{PF}}t}+\frac{[\mathrm{S_1}]_{t=0}}{(k_{\mathrm{PF}}-k_{\mathrm{DF}})}e^{-k_{\mathrm{DF}}t} \tag{S3.16}$$

$$[\mathrm{T_1}]=\frac{[\mathrm{S_1}]_{t=0}k_{\mathrm{ISC}}^{\mathrm{S1\to T1}}}{(k_{\mathrm{PF}}-k_{\mathrm{DF}})}-e^{-k_{\mathrm{PF}}t}+\frac{[\mathrm{S_1}]_{t=0}k_{\mathrm{ISC}}^{\mathrm{S1\to T1}}}{(k_{\mathrm{PF}}-k_{\mathrm{DF}})}e^{-k_{\mathrm{DF}}t} \tag{S3.17}$$

## 4. Ordinary Algebraic Equations for Different Models Under Steady State Approximation (SSA)

Two-state Model ($S_0$, $S_1$):

$$\begin{pmatrix} -k_{\mathrm{ABS}}^{\mathrm{S0\to S1}} & \left(k_{\mathrm{FL}}^{\mathrm{S1\to S0}}+k_{\mathrm{IC}}^{\mathrm{S1\to S0}}\right) \\ k_{\mathrm{ABS}}^{\mathrm{S0\to S1}} & -\left(k_{\mathrm{FL}}^{\mathrm{S1\to S0}}+k_{\mathrm{IC}}^{\mathrm{S1\to S0}}\right) \end{pmatrix}\begin{pmatrix} [\mathrm{S_0}] \\ [\mathrm{S_1}] \end{pmatrix}=0 \tag{S4.1}$$

Three-state Model ($S_0$, $S_1$, $T_1$):

$$\begin{pmatrix} -k_{\mathrm{ABS}}^{\mathrm{S0\to S1}} & \left(k_{\mathrm{FL}}^{\mathrm{S1\to S0}}+k_{\mathrm{IC}}^{\mathrm{S1\to S0}}\right) & \left(k_{\mathrm{PH}}^{\mathrm{T1\to S0}}+k_{\mathrm{ISC}}^{\mathrm{T1\to S0}}\right) \\ k_{\mathrm{ABS}}^{\mathrm{S0\to S1}} & -\left(k_{\mathrm{ISC}}^{\mathrm{S1\to T1}}+k_{\mathrm{FL}}^{\mathrm{S1\to S0}}+k_{\mathrm{IC}}^{\mathrm{S1\to S0}}\right) & k_{\mathrm{rISC}}^{\mathrm{T1\to S1}} \\ 0 & k_{\mathrm{ISC}}^{\mathrm{S1\to T1}} & -\left(k_{\mathrm{PH}}^{\mathrm{T1\to S0}}+k_{\mathrm{ISC}}^{\mathrm{T1\to S0}}+k_{\mathrm{rISC}}^{\mathrm{T1\to S1}}\right) \end{pmatrix}\begin{pmatrix} [\mathrm{S_0}] \\ [\mathrm{S_1}] \\ [\mathrm{T_1}] \end{pmatrix}=0 \tag{S4.2}$$

Four-state Model ($S_0$, $S_1$, $T_1$, $T_2$):

$$\begin{pmatrix} -k_{\mathrm{ABS}}^{\mathrm{S0\to S1}} & \left(k_{\mathrm{FL}}^{\mathrm{S1\to S0}}+k_{\mathrm{IC}}^{\mathrm{S1\to S0}}\right) & \left(k_{\mathrm{PH}}^{\mathrm{T1\to S0}}+k_{\mathrm{ISC}}^{\mathrm{T1\to S0}}\right) & 0 \\ k_{\mathrm{ABS}}^{\mathrm{S0\to S1}} & -\left(k_{\mathrm{ISC}}^{\mathrm{S1\to T1}}+k_{\mathrm{ISC}}^{\mathrm{S1\to T2}}+k_{\mathrm{FL}}^{\mathrm{S1\to S0}}+k_{\mathrm{IC}}^{\mathrm{S1\to S0}}\right) & k_{\mathrm{rISC}}^{\mathrm{T1\to S1}} & k_{\mathrm{rISC}}^{\mathrm{T2\to S1}} \\ 0 & k_{\mathrm{ISC}}^{\mathrm{S1\to T1}} & -\left(k_{\mathrm{PH}}^{\mathrm{T1\to S0}}+k_{\mathrm{ISC}}^{\mathrm{T1\to S0}}+k_{\mathrm{rISC}}^{\mathrm{T1\to S1}}+k_{\mathrm{rIC}}^{\mathrm{T1\to T2}}\right) & k_{\mathrm{IC}}^{\mathrm{T2\to T1}} \\ 0 & k_{\mathrm{ISC}}^{\mathrm{S1\to T2}} & k_{\mathrm{rIC}}^{\mathrm{T1\to T2}} & -\left(k_{\mathrm{rISC}}^{\mathrm{T2\to S1}}+k_{\mathrm{IC}}^{\mathrm{T2\to T1}}\right) \end{pmatrix}\begin{pmatrix} [\mathrm{S_0}] \\ [\mathrm{S_1}] \\ [\mathrm{T_1}] \\ [\mathrm{T_2}] \end{pmatrix}=0 \tag{S4.3}$$

Five-state Model ($S_0$, $S_1$, $S_2$, $T_1$, $T_2$):

$$K\begin{pmatrix} [\mathrm{S_0}] \\ [\mathrm{S_1}] \\ [\mathrm{S_2}] \\ [\mathrm{T_1}] \\ [\mathrm{T_2}] \end{pmatrix}=0 \tag{S4.4}$$

$$K=\begin{pmatrix} -k_{\mathrm{ABS}}^{\mathrm{S0\to S2}} & \left(k_{\mathrm{FL}}^{\mathrm{S1\to S0}}+k_{\mathrm{IC}}^{\mathrm{S1\to S0}}\right) & \left(k_{\mathrm{FL}}^{\mathrm{S2\to S0}}+k_{\mathrm{IC}}^{\mathrm{S2\to S0}}\right) & \left(k_{\mathrm{PH}}^{\mathrm{T1\to S0}}+k_{\mathrm{ISC}}^{\mathrm{T1\to S0}}\right) & 0 \\ 0 & \sum \mathrm{S_1} & k_{\mathrm{IC}}^{\mathrm{S2\to S1}} & k_{\mathrm{rISC}}^{\mathrm{T1\to S1}} & k_{\mathrm{rISC}}^{\mathrm{T2\to S1}} \\ k_{\mathrm{ABS}}^{\mathrm{S0\to S2}} & k_{\mathrm{rIC}}^{\mathrm{S1\to S2}} & \sum \mathrm{S_2} & k_{\mathrm{rISC}}^{\mathrm{T1\to S2}} & k_{\mathrm{rISC}}^{\mathrm{T2\to S2}} \\ 0 & k_{\mathrm{ISC}}^{\mathrm{S1\to T1}} & k_{\mathrm{ISC}}^{\mathrm{S2\to T1}} & \sum \mathrm{T_1} & k_{\mathrm{IC}}^{\mathrm{T2\to T1}} \\ 0 & k_{\mathrm{ISC}}^{\mathrm{S1\to T2}} & k_{\mathrm{ISC}}^{\mathrm{S2\to T2}} & k_{\mathrm{rIC}}^{\mathrm{T1\to T2}} & \sum \mathrm{T_2} \end{pmatrix} \tag{S4.5}$$

$$\sum \mathrm{S_1}=-\left(k_{\mathrm{ISC}}^{\mathrm{S1\to T1}}+k_{\mathrm{ISC}}^{\mathrm{S1\to T2}}+k_{\mathrm{FL}}^{\mathrm{S1\to S0}}+k_{\mathrm{IC}}^{\mathrm{S1\to S0}}+k_{\mathrm{rIC}}^{\mathrm{S1\to S2}}\right) \tag{S4.6}$$

$$\sum \mathrm{S_2}=-\left(k_{\mathrm{IC}}^{\mathrm{S2\to S1}}+k_{\mathrm{ISC}}^{\mathrm{S2\to T1}}+k_{\mathrm{ISC}}^{\mathrm{S2\to T2}}+k_{\mathrm{FL}}^{\mathrm{S2\to S0}}+k_{\mathrm{IC}}^{\mathrm{S2\to S0}}\right) \tag{S4.7}$$

$$\sum \mathrm{T_1}=-\left(k_{\mathrm{PH}}^{\mathrm{T1\to S0}}+k_{\mathrm{ISC}}^{\mathrm{T1\to S0}}+k_{\mathrm{rISC}}^{\mathrm{T1\to S1}}+k_{\mathrm{rISC}}^{\mathrm{T1\to S2}}+k_{\mathrm{rIC}}^{\mathrm{T1\to T2}}\right) \tag{S4.8}$$

$$\sum \mathrm{T_2}=-\left(k_{\mathrm{rISC}}^{\mathrm{T2\to S1}}+k_{\mathrm{rISC}}^{\mathrm{T2\to S2}}+k_{\mathrm{IC}}^{\mathrm{T2\to T1}}\right) \tag{S4.9}$$

The quantum yield ($\Phi_i$) of a photophysical process is defined as

$$\Phi_i=\frac{\text{Number of molecules undergoing that process}}{\text{Number of photons absorbed by the reactant}} \tag{S4.10}$$

The two-state model under SSA (eq S4.1) yields the following relationship:

$$[\mathrm{S_0}]k_{\mathrm{ABS}}^{\mathrm{S0\to S1}}=[\mathrm{S_1}]\left(k_{\mathrm{FL}}^{\mathrm{S1\to S0}}+k_{\mathrm{IC}}^{\mathrm{S1\to S0}}\right) \tag{S4.11}$$

where $[\mathrm{S_0}]k_{\mathrm{ABS}}^{\mathrm{S0\to S1}}$ denotes the number of molecules excited per unit volume per unit time, equals the number of photons absorbed per unit volume per unit time under the one-photon absorption mechanism.

Accordingly, the photoluminescence quantum yield (PLQY) of fluorescence is given by:

$$\Phi_{\mathrm{FL}}=\frac{[\mathrm{S_1}]k_{\mathrm{FL}}^{\mathrm{S1\to S0}}}{[\mathrm{S_0}]k_{\mathrm{ABS}}^{\mathrm{S0\to S1}}}=\frac{[\mathrm{S_1}]k_{\mathrm{FL}}^{\mathrm{S1\to S0}}}{[\mathrm{S_1}](k_{\mathrm{FL}}^{\mathrm{S1\to S0}}+k_{\mathrm{IC}}^{\mathrm{S1\to S0}})}=\frac{k_{\mathrm{FL}}^{\mathrm{S1\to S0}}}{(k_{\mathrm{FL}}^{\mathrm{S1\to S0}}+k_{\mathrm{IC}}^{\mathrm{S1\to S0}})} \tag{S4.12}$$

Similarly, the three-state model under SSA (eq S4.2) gives:

$$[\mathrm{S_0}]k_{\mathrm{ABS}}^{\mathrm{S0\to S1}}=[\mathrm{S_1}]\left(k_{\mathrm{FL}}^{\mathrm{S1\to S0}}+k_{\mathrm{IC}}^{\mathrm{S1\to S0}}\right)+[\mathrm{T_1}]\left(k_{\mathrm{PH}}^{\mathrm{T1\to S0}}+k_{\mathrm{ISC}}^{\mathrm{T1\to S0}}\right) \tag{S4.13}$$

$$[\mathrm{T_1}]=[\mathrm{S_1}]\frac{k_{\mathrm{ISC}}^{\mathrm{S1\to T1}}}{(k_{\mathrm{PH}}^{\mathrm{T1\to S0}}+k_{\mathrm{ISC}}^{\mathrm{T1\to S0}}+k_{\mathrm{rISC}}^{\mathrm{T1\to S1}})} \tag{S4.14}$$

Therefore, PLQY of fluorescence for three-state model is given by:

$$\Phi_{\mathrm{FL}} = \frac{[\mathrm{S}_1]k_{\mathrm{FL}}^{\mathrm{S1\to S0}}}{[\mathrm{S}_0]k_{\mathrm{ABS}}^{\mathrm{S0\to S1}}} = \frac{[\mathrm{S}_1]k_{\mathrm{FL}}^{\mathrm{S1\to S0}}}{[\mathrm{S}_1](k_{\mathrm{FL}}^{\mathrm{S1\to S0}} + k_{\mathrm{IC}}^{\mathrm{S1\to S0}}) + [\mathrm{T}_1](k_{\mathrm{PH}}^{\mathrm{T1\to S0}} + k_{\mathrm{ISC}}^{\mathrm{T1\to S0}})}$$
$$= \frac{[\mathrm{S}_1]k_{\mathrm{FL}}^{\mathrm{S1\to S0}}}{[\mathrm{S}_1](k_{\mathrm{FL}}^{\mathrm{S1\to S0}} + k_{\mathrm{IC}}^{\mathrm{S1\to S0}}) + [\mathrm{S}_1]\frac{k_{\mathrm{ISC}}^{\mathrm{S1\to T1}}(k_{\mathrm{PH}}^{\mathrm{T1\to S0}} + k_{\mathrm{ISC}}^{\mathrm{T1\to S0}})}{(k_{\mathrm{PH}}^{\mathrm{T1\to S0}} + k_{\mathrm{ISC}}^{\mathrm{T1\to S0}} + k_{\mathrm{rISC}}^{\mathrm{T1\to S1}})}} = \frac{k_{\mathrm{FL}}^{\mathrm{S1\to S0}}}{(k_{\mathrm{FL}}^{\mathrm{S1\to S0}} + k_{\mathrm{IC}}^{\mathrm{S1\to S0}}) + \frac{k_{\mathrm{ISC}}^{\mathrm{S1\to T1}}(k_{\mathrm{PH}}^{\mathrm{T1\to S0}} + k_{\mathrm{ISC}}^{\mathrm{T1\to S0}})}{(k_{\mathrm{PH}}^{\mathrm{T1\to S0}} + k_{\mathrm{ISC}}^{\mathrm{T1\to S0}} + k_{\mathrm{rISC}}^{\mathrm{T1\to S1}})}} \quad (\mathrm{S4.15})$$

Four-state (eq S4.3) and five-state (eq S4.4) models can follow a similar workflow to determine the PLQY of fluorescence.

## 5. Natural Transition Orbitals of DOBNA and DiKTa

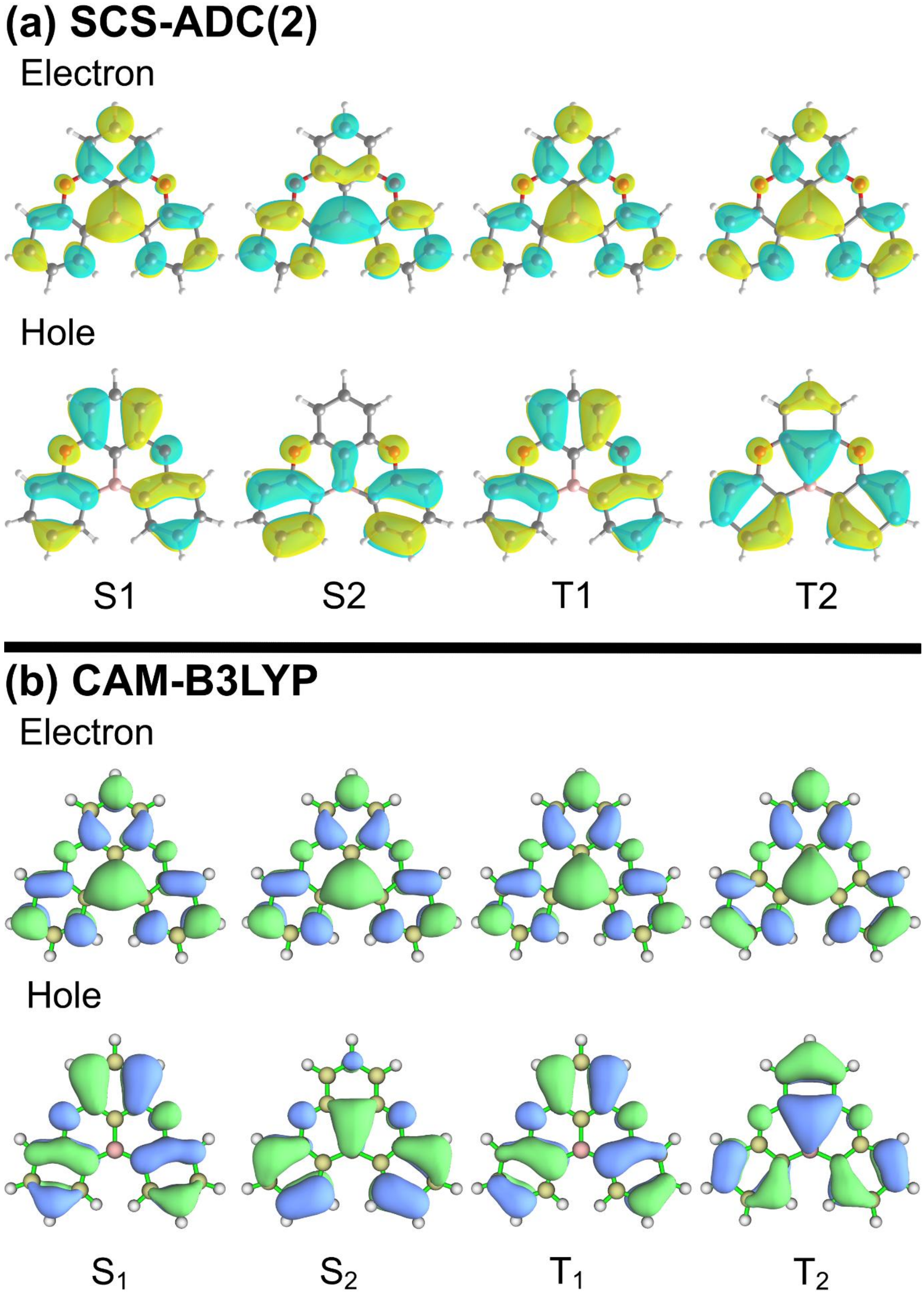

**Figure S1.** Natural transition orbitals (NTOs) of DOBNA at the ground-state geometry calculated by (a) SCS-ADC(2) and (b) TD(A)-CAM-B3LYP. Visualizations were generated using (a) VESTA[11] and (b) Multiwfn[12,13] (isovalue = 0.026).

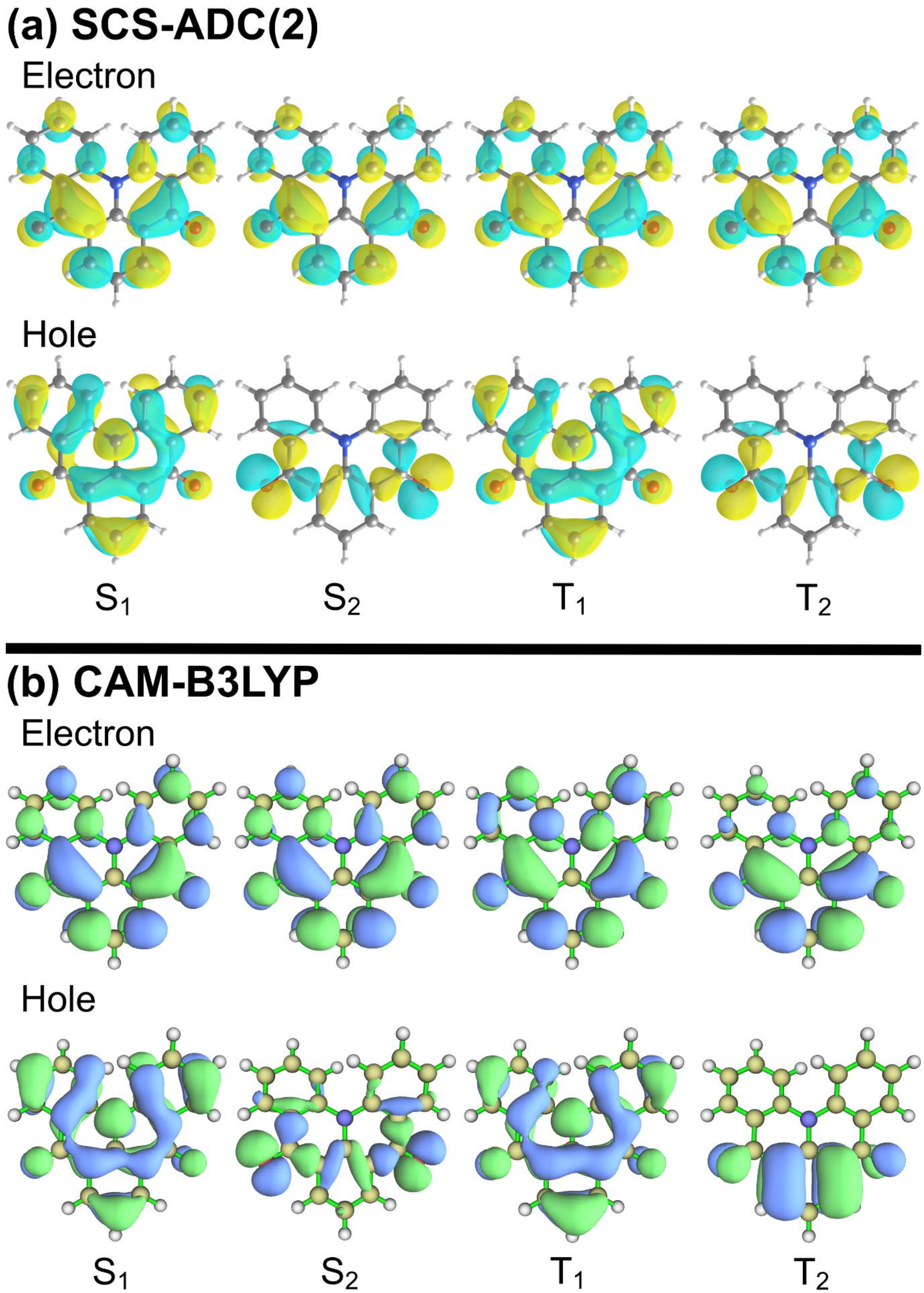

**Figure S2.** Natural transition orbitals (NTOs) of DiKTa at the ground-state geometry calculated by (a) SCS-ADC(2) and (b) TD(A)-CAM-B3LYP. Visualizations were generated using (a) VESTA and (b) Multiwfn (isovalue = 0.026).

## 6. Optimized Ground and Excited State Geometries

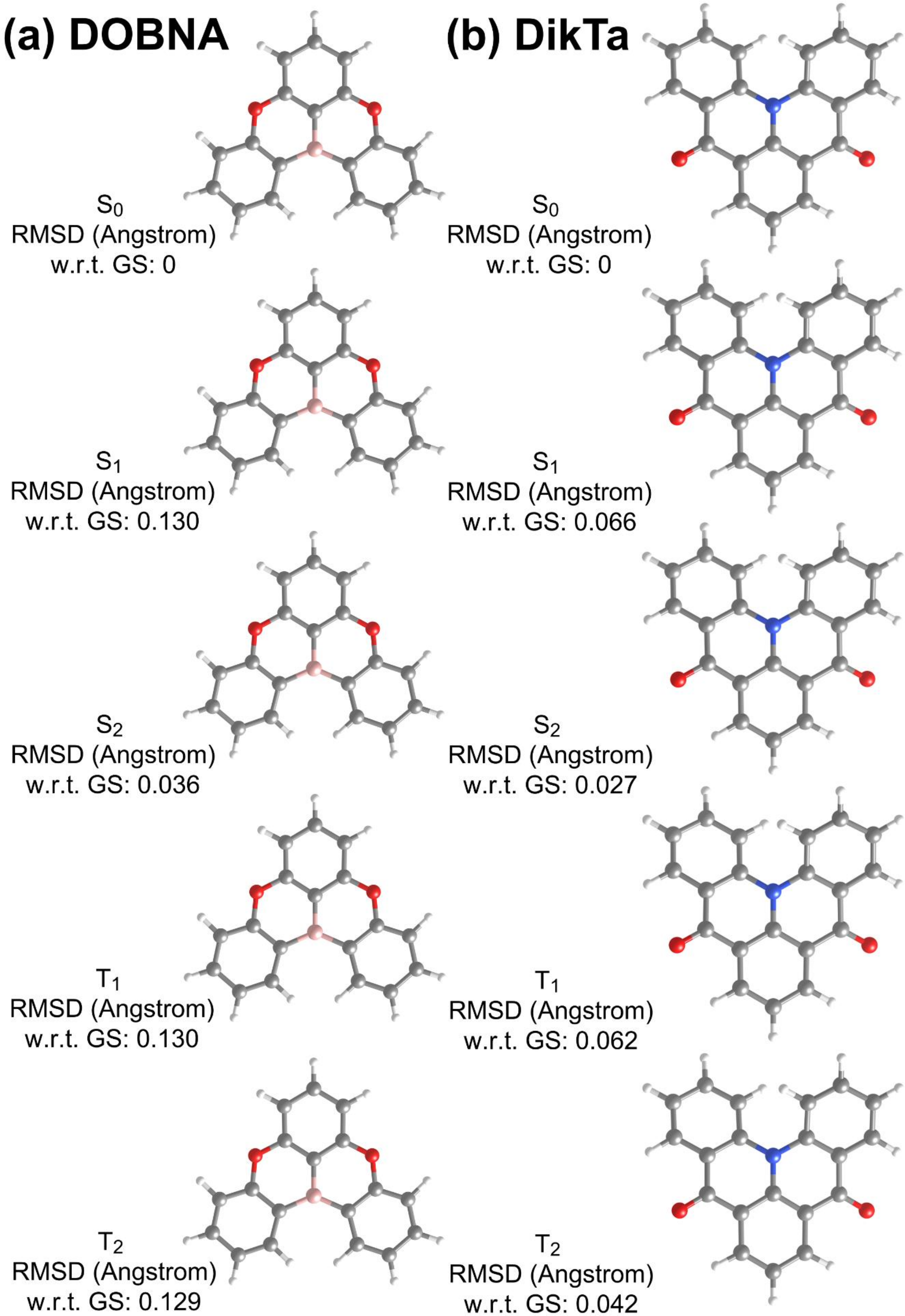

**Figure S3.** Optimized ground and excited state geometries of (a) DOBNA and (b) DiKTa at the TD(A)-CAM-B3LYP/6-311G(d,p) level. Blue: nitrogen, red: oxygen, pink: boron. Root Mean Square Deviations (RMSD) quantify the structural changes upon excitation with respect to (w.r.t) the ground state (GS).

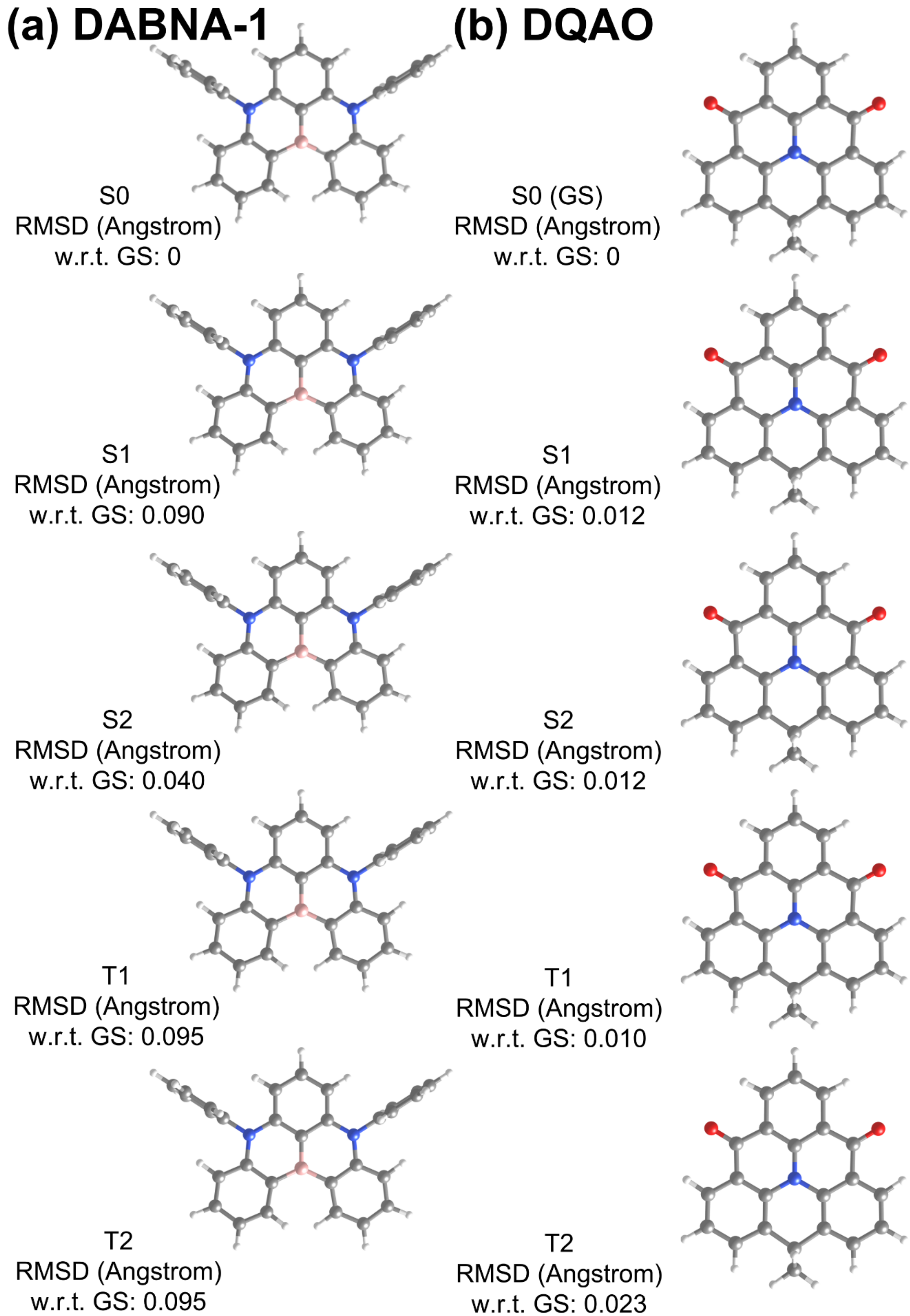

**Figure S4.** Optimized ground and excited state geometries of (a) DABNA-1 and (b) DQAO at the TD(A)-CAM-B3LYP/6-31(1)G(d,p) level. Blue: nitrogen, red: oxygen, pink: boron. Root Mean Square Deviations (RMSD) quantify the structural changes upon excitation with respect to (w.r.t) the ground state (GS).

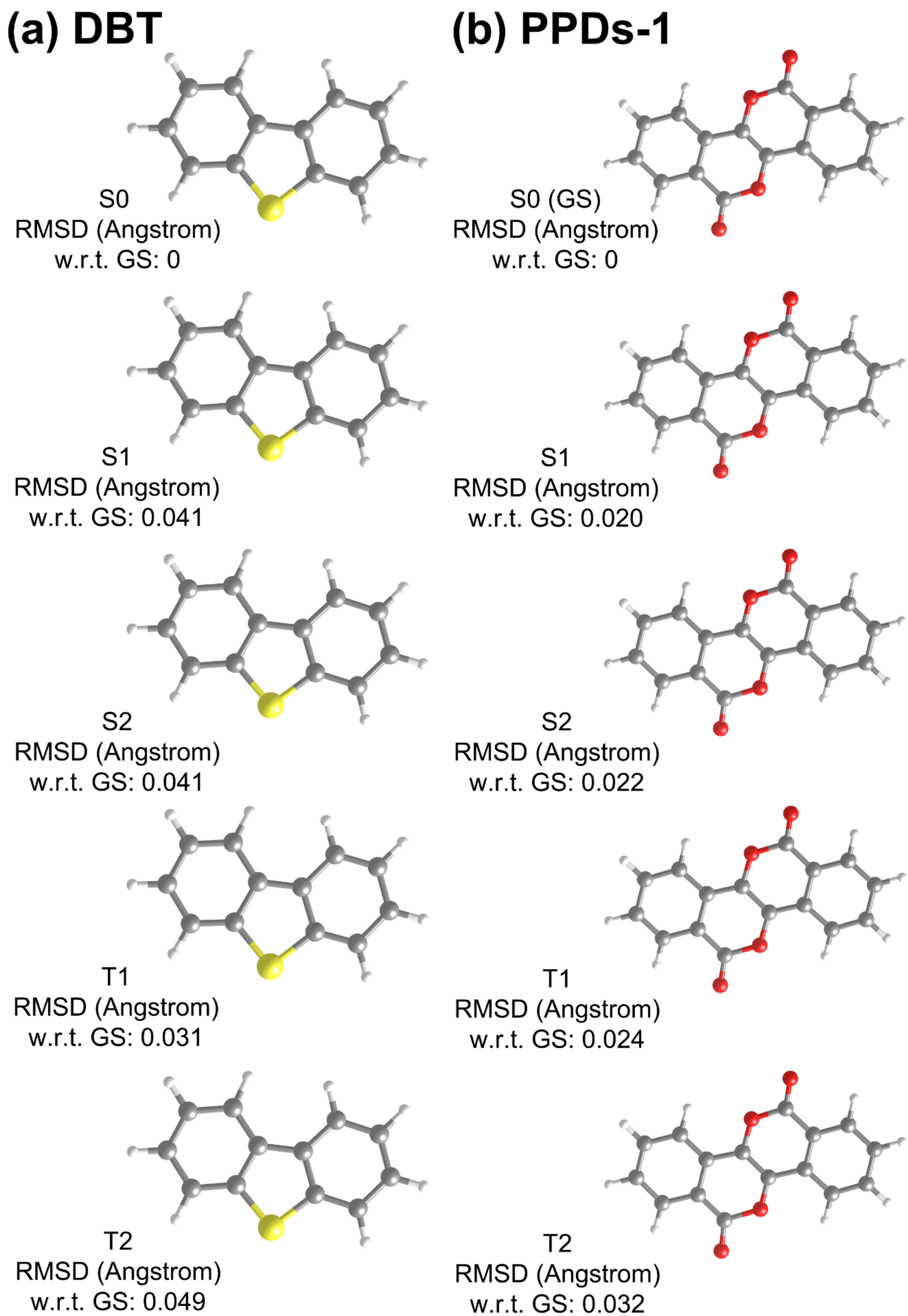

**Figure S5.** Optimized ground and excited state geometries of (a) DBT and (b) PPDs-1 at the TD(A)-CAM-B3LYP/6-31G(d,p) level. Yellow: sulfur, red: oxygen. Root Mean Square Deviations (RMSD) quantify the structural changes upon excitation with respect to (w.r.t) the ground state (GS).

## 7. Adiabatic Excitation Energies

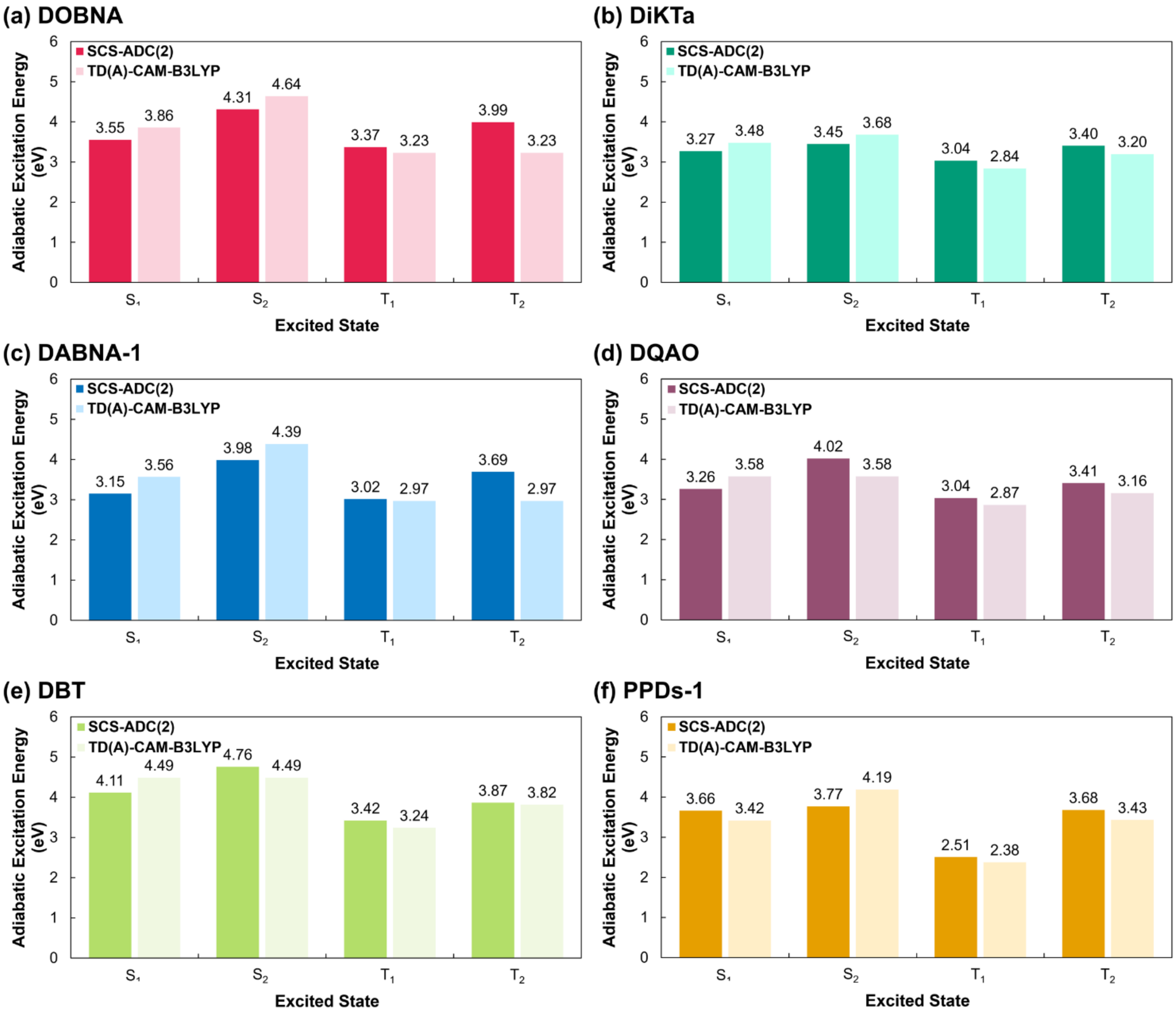

**Figure S6.** Adiabatic excitation energies (eV) for the excited states calculated at the TD(A)-CAM-B3LYP and SCS-ADC(2) levels for (a) DOBNA (red), (b) DiKTa (green), (c) DABNA-1 (blue), (d) DQAO (purple), (e) DBT (light green) and (f) PPDs-1 (orange).

## 8. Detailed Rate Constants Calculations of DOBNA and DiKTa

**Table S1. ISC rate constants for the individual sublevels (Ms = 0, ±1) of DOBNA[a]**

| ISC Transition | Sublevel (Ms) | Rate (1/s) |
|---|---|---|
| $S_1 \rightarrow T_1$ | 0 | $2.24\times10^{4}$ |
| $S_1 \rightarrow T_1$ | ±1 | $5.94\times10^{5}$ |
| $S_1 \rightarrow T_2$ | 0 | $7.77\times10^{2}$ |
| $S_1 \rightarrow T_2$ | ±1 | $1.08\times10^{5}$ |
| $S_2 \rightarrow T_1$ | 0 | $1.33\times10^{4}$ |
| $S_2 \rightarrow T_1$ | ±1 | $5.38\times10^{6}$ |
| $S_2 \rightarrow T_2$ | 0 | $6.78\times10^{5}$ |
| $S_2 \rightarrow T_2$ | ±1 | $2.44\times10^{7}$ |
| $T_1 \rightarrow S_0$ | 0 | $8.71\times10^{3}$ |
| $T_1 \rightarrow S_0$ | ±1 | $2.15\times10^{5}$ |
| $T_1 \rightarrow S_1$ | 0 | $7.70\times10^{1}$ |
| $T_1 \rightarrow S_1$ | ±1 | $5.32\times10^{3}$ |
| $T_1 \rightarrow S_2$ | 0 | $3.83\times10^{2}$ |
| $T_1 \rightarrow S_2$ | ±1 | $4.77\times10^{3}$ |
| $T_2 \rightarrow S_1$ | 0 | $1.26\times10^{5}$ |
| $T_2 \rightarrow S_1$ | ±1 | $2.15\times10^{6}$ |
| $T_2 \rightarrow S_2$ | 0 | $7.23\times10^{1}$ |
| $T_2 \rightarrow S_2$ | ±1 | $1.90\times10^{4}$ |

[a]The total ISC rate constant for one singlet-to-triplet transition is obtained by summing the rate constants of the three triplet sublevels, whereas the total rISC rate constant for one triplet-to-singlet transition is calculated as the average over these sublevels.

**Table S2. ISC rate constants for the individual sublevels (Ms = 0, ±1) of DiKTa[a]**

| ISC Transition | Sublevel | Rate (1/s) |
|---|---|---|
| $S_1 \rightarrow T_1$ | 0 | $3.39\times10^{8}$ |
| $S_1 \rightarrow T_1$ | ±1 | $5.15\times10^{9}$ |
| $S_1 \rightarrow T_2$ | 0 | $8.84\times10^{8}$ |
| $S_1 \rightarrow T_2$ | ±1 | $3.05\times10^{9}$ |
| $S_2 \rightarrow T_1$ | 0 | $2.76\times10^{10}$ |
| $S_2 \rightarrow T_1$ | ±1 | $1.28\times10^{11}$ |
| $S_2 \rightarrow T_2$ | 0 | $1.30\times10^{11}$ |
| $S_2 \rightarrow T_2$ | ±1 | $3.95\times10^{11}$ |
| $T_1 \rightarrow S_0$ | 0 | $1.22\times10^{4}$ |
| $T_1 \rightarrow S_0$ | ±1 | $4.23\times10^{4}$ |
| $T_1 \rightarrow S_1$ | 0 | $2.69\times10^{5}$ |
| $T_1 \rightarrow S_1$ | ±1 | $1.91\times10^{6}$ |
| $T_1 \rightarrow S_2$ | 0 | $6.82\times10^{4}$ |
| $T_1 \rightarrow S_2$ | ±1 | $4.05\times10^{5}$ |
| $T_2 \rightarrow S_1$ | 0 | $4.63\times10^{8}$ |
| $T_2 \rightarrow S_1$ | ±1 | $3.30\times10^{9}$ |
| $T_2 \rightarrow S_2$ | 0 | $1.21\times10^{7}$ |
| $T_2 \rightarrow S_2$ | ±1 | $7.22\times10^{7}$ |

[a]The total ISC rate constant for one singlet-to-triplet transition is obtained by summing the rate constants of the three triplet sublevels, whereas the total rISC rate constant for one triplet-to-singlet transition is calculated as the average over these sublevels.

**Table S3. Phosphorescence rate constants for the individual sublevels (Ms = 0, ±1)[a]**

| Molecule | Sublevel | Rate (1/s) |
|---|---|---|
| DOBNA | 0 | $9.02\times10^{-2}$ |
| DOBNA | 1 | $4.97\times10^{-2}$ |
| DOBNA | -1 | $1.55\times10^{-1}$ |
| DiKTa | 0 | $8.00\times10^{-1}$ |
| DiKTa | 1 | $4.02\times10^{0}$ |
| DiKTa | -1 | $1.17\times10^{-2}$ |

[a]The total phosphorescence rate constant is calculated as the average over these sublevels.

**Table S4. Reorganization energies for ISC transitions in DOBNA**

| ISC Transition | Reorganization Energy ($cm^{-1}$) | Reorganization Energy (eV) |
|---|---|---|
| $S_1 \rightarrow T_1$ | 145.48 | 0.02 |
| $S_1 \rightarrow T_2$ | 145.31 | 0.02 |
| $S_2 \rightarrow T_1$ | 1185.07 | 0.15 |
| $S_2 \rightarrow T_2$ | 1178.94 | 0.15 |
| $T_1 \rightarrow S_0$ | 697.74 | 0.09 |
| $T_1 \rightarrow S_1$ | 172.32 | 0.02 |
| $T_1 \rightarrow S_2$ | 1012.63 | 0.13 |
| $T_2 \rightarrow S_1$ | 172.18 | 0.02 |
| $T_2 \rightarrow S_2$ | 1015.26 | 0.13 |

**Table S5. Reorganization energies for ISC transitions in DiKTa**

| ISC Transition | Reorganization Energy (1/cm) | Reorganization Energy (eV) |
|---|---|---|
| $S_1 \rightarrow T_1$ | 186.73 | 0.02 |
| $S_1 \rightarrow T_2$ | 2085.2 | 0.26 |
| $S_2 \rightarrow T_1$ | 5628.08 | 0.70 |
| $S_2 \rightarrow T_2$ | 4876.38 | 0.60 |
| $T_1 \rightarrow S_0$ | 847.07 | 0.11 |
| $T_1 \rightarrow S_1$ | 194.99 | 0.02 |
| $T_1 \rightarrow S_2$ | 953.82 | 0.12 |
| $T_2 \rightarrow S_1$ | 1526.58 | 0.19 |
| $T_2 \rightarrow S_2$ | 2629.44 | 0.33 |

**Table S6. SOCMEs at the Franck-Condon (FC) geometry for ISC transitions in DOBNA**

| ISC Transition | Sublevel | FC SOCME ($cm^{-1}$) |
|---|---|---|
| $S_1 \rightarrow T_1$ | 0 | $1.07\times10^{-6}$ |
| $S_1 \rightarrow T_1$ | ±1 | $1.74\times10^{-4}$ |
| $S_1 \rightarrow T_2$ | 0 | $2.99\times10^{-2}$ |
| $S_1 \rightarrow T_2$ | ±1 | $5.32\times10^{-4}$ |
| $S_2 \rightarrow T_1$ | 0 | $2.25\times10^{-2}$ |
| $S_2 \rightarrow T_1$ | ±1 | $6.81\times10^{-4}$ |
| $S_2 \rightarrow T_2$ | 0 | $2.07\times10^{-5}$ |
| $S_2 \rightarrow T_2$ | ±1 | $1.30\times10^{-4}$ |
| $T_1 \rightarrow S_0$ | 0 | $1.60\times10^{-2}$ |
| $T_1 \rightarrow S_0$ | ±1 | $1.60\times10^{-1}$ |
| $T_1 \rightarrow S_1$ | 0 | $1.35\times10^{-6}$ |
| $T_1 \rightarrow S_1$ | ±1 | $1.77\times10^{-4}$ |
| $T_1 \rightarrow S_2$ | 0 | $5.89\times10^{-2}$ |
| $T_1 \rightarrow S_2$ | ±1 | $1.63\times10^{-1}$ |
| $T_2 \rightarrow S_1$ | 0 | $3.42\times10^{-2}$ |
| $T_2 \rightarrow S_1$ | ±1 | $7.32\times10^{-4}$ |
| $T_2 \rightarrow S_2$ | 0 | $6.43\times10^{-7}$ |
| $T_2 \rightarrow S_2$ | ±1 | $1.98\times10^{-1}$ |

**Table S7. SOCMEs at the Franck-Condon (FC) geometry for ISC transitions in DiKTa**

| ISC Transition | Sublevel | FC SOCME ($cm^{-1}$) |
|---|---|---|
| $S_1 \rightarrow T_1$ | 0 | $7.84\times10^{-5}$ |
| $S_1 \rightarrow T_1$ | ±1 | $1.16\times10^{0}$ |
| $S_1 \rightarrow T_2$ | 0 | $5.65\times10^{0}$ |
| $S_1 \rightarrow T_2$ | ±1 | $9.88\times10^{0}$ |
| $S_2 \rightarrow T_1$ | 0 | $2.90\times10^{-3}$ |
| $S_2 \rightarrow T_1$ | ±1 | $8.16\times10^{0}$ |
| $S_2 \rightarrow T_2$ | 0 | $5.85\times10^{0}$ |
| $S_2 \rightarrow T_2$ | ±1 | $1.07\times10^{1}$ |
| $T_1 \rightarrow S_0$ | 0 | $7.83\times10^{-1}$ |
| $T_1 \rightarrow S_0$ | ±1 | $4.74\times10^{-1}$ |
| $T_1 \rightarrow S_1$ | 0 | $9.83\times10^{-5}$ |
| $T_1 \rightarrow S_1$ | ±1 | $5.13\times10^{-1}$ |
| $T_1 \rightarrow S_2$ | 0 | $4.64\times10^{-1}$ |
| $T_1 \rightarrow S_2$ | ±1 | $9.39\times10^{-1}$ |
| $T_2 \rightarrow S_1$ | 0 | $3.65\times10^{-1}$ |
| $T_2 \rightarrow S_1$ | ±1 | $1.01\times10^{0}$ |
| $T_2 \rightarrow S_2$ | 0 | $2.23\times10^{0}$ |
| $T_2 \rightarrow S_2$ | ±1 | $8.03\times10^{0}$ |

**Table S8. Franck-Condon (FC) and Herzberg-Teller (HT) contributions to the ISC transitions of DOBNA**

| ISC Transition | Sublevel | FC Component (%) | HT Component (%) |
|---|---|---|---|
| $S_1 \rightarrow T_1$ | 0 | 0 | 100 |
| $S_1 \rightarrow T_1$ | ±1 | 0 | 100 |
| $S_1 \rightarrow T_2$ | 0 | 34.17 | 65.83 |
| $S_1 \rightarrow T_2$ | ±1 | 0 | 100 |
| $S_2 \rightarrow T_1$ | 0 | 57.31 | 42.69 |
| $S_2 \rightarrow T_1$ | ±1 | 0 | 100 |
| $S_2 \rightarrow T_2$ | 0 | 0 | 100 |
| $S_2 \rightarrow T_2$ | ±1 | 0 | 100 |
| $T_1 \rightarrow S_0$ | 0 | 0.14 | 99.86 |
| $T_1 \rightarrow S_0$ | ±1 | 0.58 | 99.42 |
| $T_1 \rightarrow S_1$ | 0 | 0 | 100 |
| $T_1 \rightarrow S_1$ | ±1 | 0 | 100 |
| $T_1 \rightarrow S_2$ | 0 | 34.14 | 65.86 |
| $T_1 \rightarrow S_2$ | ±1 | 20.92 | 79.08 |
| $T_2 \rightarrow S_1$ | 0 | 3.3 | 96.7 |
| $T_2 \rightarrow S_1$ | ±1 | 0 | 100 |
| $T_2 \rightarrow S_2$ | 0 | 0 | 100 |
| $T_2 \rightarrow S_2$ | ±1 | 42.29 | 57.71 |

**Table S9. Franck-Condon (FC) and Herzberg-Teller (HT) contributions to the ISC transitions of DiKTa**

| ISC Transition | Sublevel | FC Component (%) | HT Component (%) |
|---|---|---|---|
| $S_1 \rightarrow T_1$ | 0 | 0 | 100 |
| $S_1 \rightarrow T_1$ | ±1 | 0.91 | 99.09 |
| $S_1 \rightarrow T_2$ | 0 | 4.54 | 95.46 |
| $S_1 \rightarrow T_2$ | ±1 | 4.03 | 95.97 |
| $S_2 \rightarrow T_1$ | 0 | 0 | 100 |
| $S_2 \rightarrow T_1$ | ±1 | 1.58 | 98.42 |
| $S_2 \rightarrow T_2$ | 0 | 5.58 | 94.42 |
| $S_2 \rightarrow T_2$ | ±1 | 6.19 | 93.81 |
| $T_1 \rightarrow S_0$ | 0 | 40.01 | 59.99 |
| $T_1 \rightarrow S_0$ | ±1 | 4.22 | 95.78 |
| $T_1 \rightarrow S_1$ | 0 | 0 | 100 |
| $T_1 \rightarrow S_1$ | ±1 | 8.77 | 91.23 |
| $T_1 \rightarrow S_2$ | 0 | 22.03 | 77.97 |
| $T_1 \rightarrow S_2$ | ±1 | 15.17 | 84.83 |
| $T_2 \rightarrow S_1$ | 0 | 7.48 | 92.52 |
| $T_2 \rightarrow S_1$ | ±1 | 7.95 | 92.05 |
| $T_2 \rightarrow S_2$ | 0 | 51.74 | 48.26 |
| $T_2 \rightarrow S_2$ | ±1 | 112.32 | -12.32 |

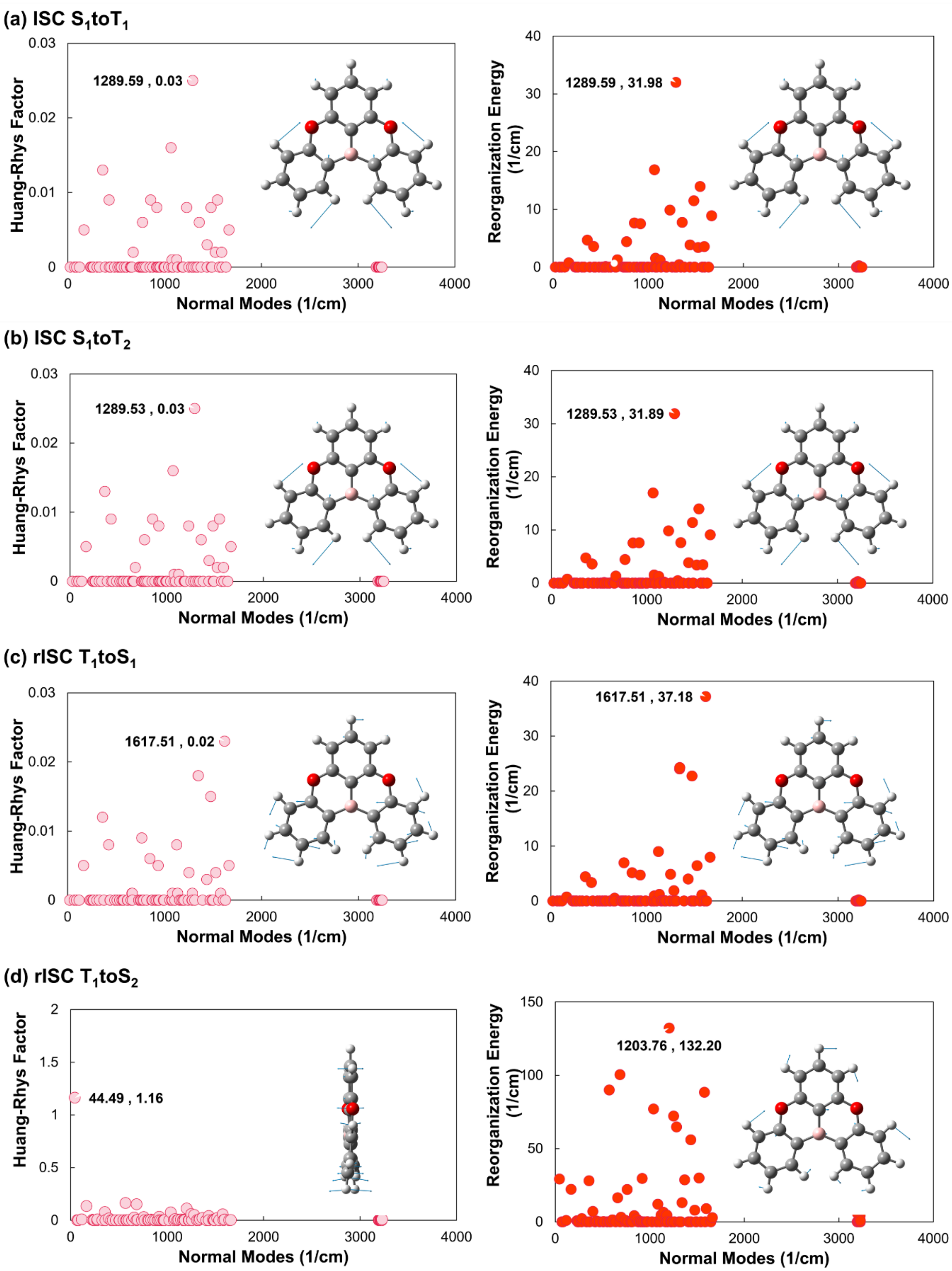

**Figure S7.** Huang-Rhys (HR) factors (left) and reorganization energies (right) as functions of vibrational normal modes for DOBNA: (a) ISC ($S_1 \rightarrow T1$), (b) ISC ($S_1 \rightarrow T_2$), (c) rISC ($T_1 \rightarrow S_1$), and (d) rISC ($T_1 \rightarrow S_2$). The highest HR factor and reorganization energy with their corresponding frequencies are indicated. The inset shows the displacement vectors of the normal modes corresponding to the highest HR factor (left) and reorganization energy (right).

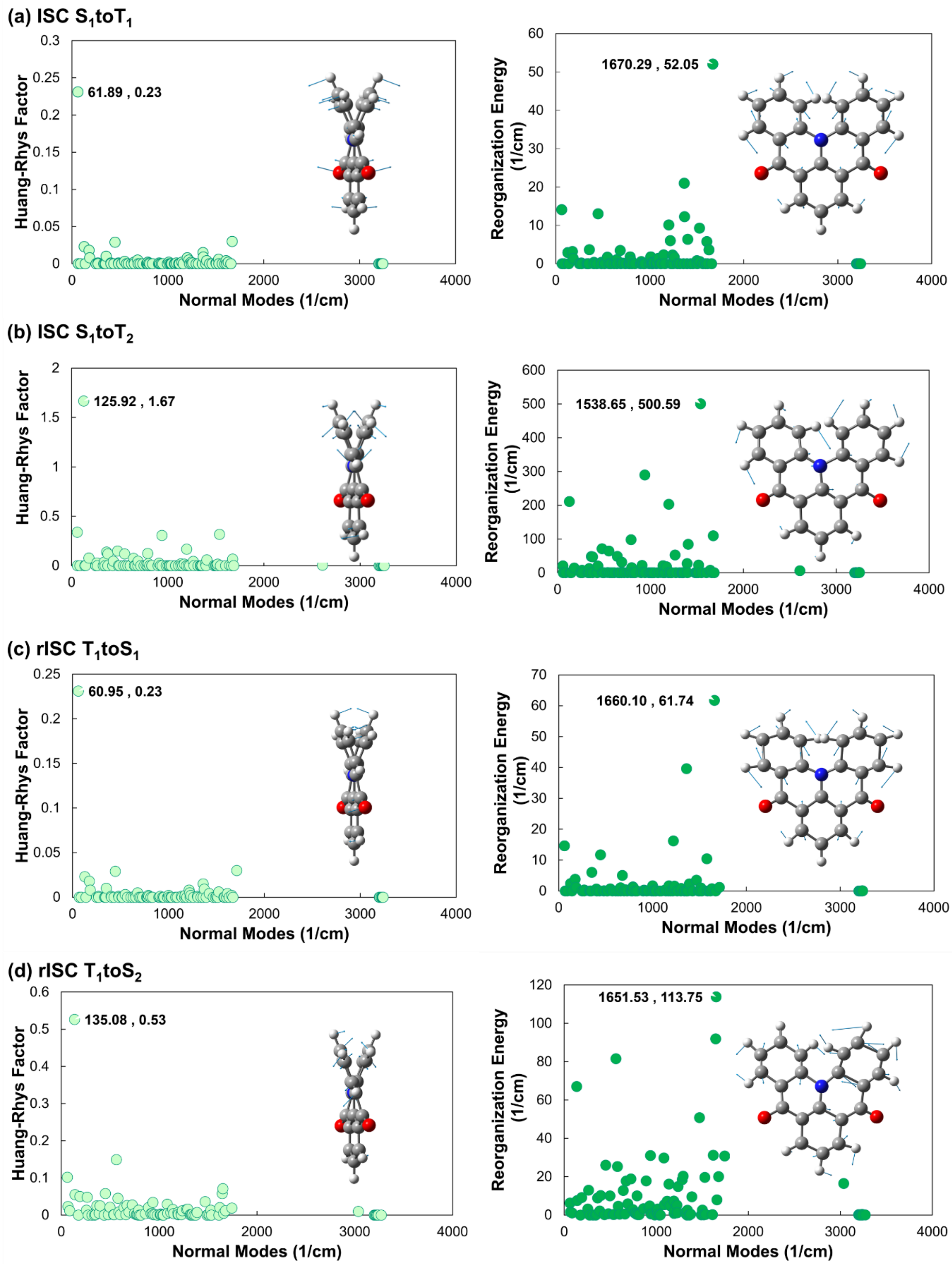

**Figure S8.** Huang-Rhys (HR) factors and reorganization energies as functions of vibrational normal modes for DiKTa: (a) ISC ($S_1 \rightarrow T_1$), (b) ISC ($S_1 \rightarrow T_2$), (c) rISC ($T_1 \rightarrow S_1$), and (d) rISC ($T_1 \rightarrow S_2$). The highest HR factor and reorganization energy with their corresponding frequencies are indicated. The inset shows the displacement vectors of the normal modes corresponding to the highest HR factor (left) and reorganization energy (right).

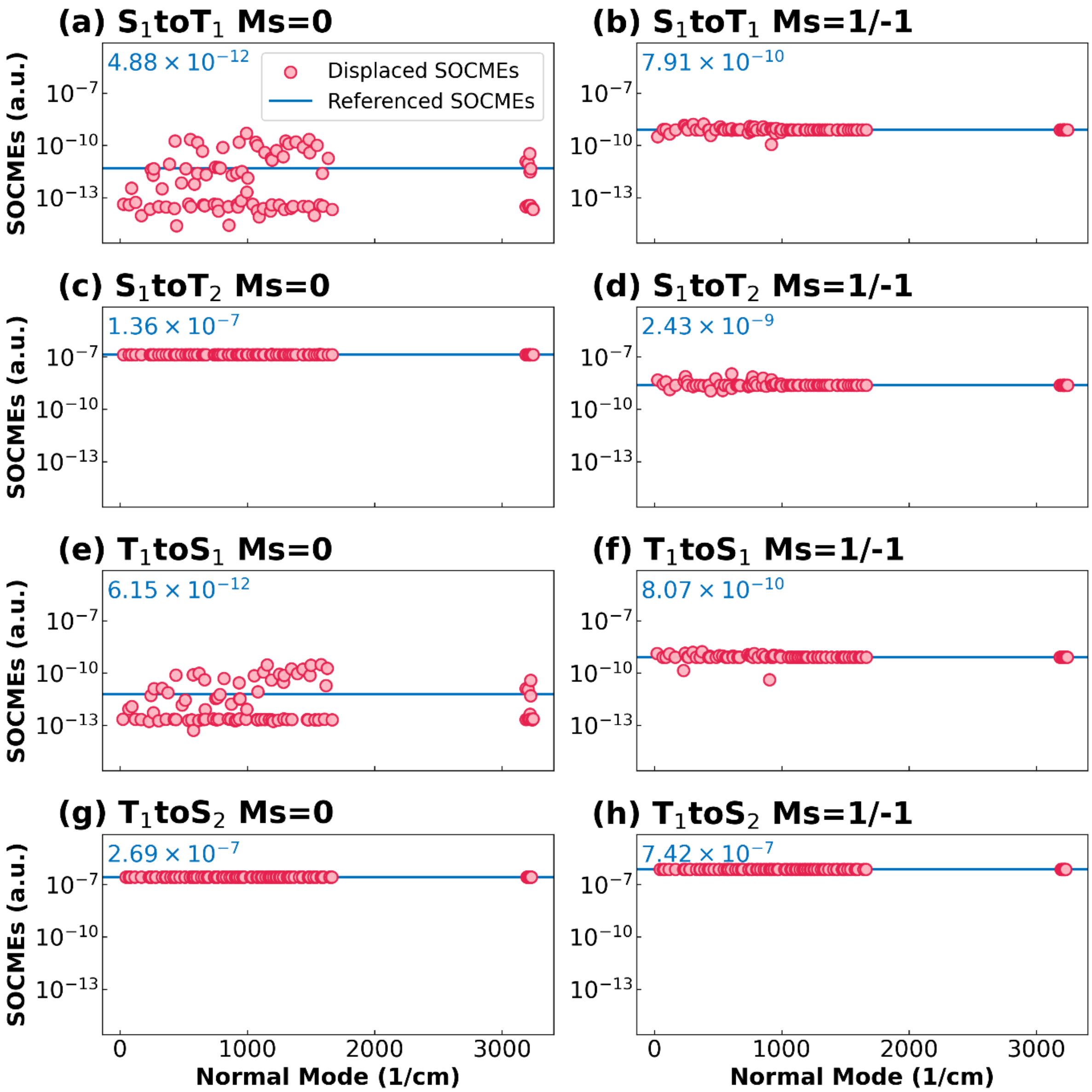

**Figure S9.** SOCMEs at the reference Franck-Condon (FC) geometry (blue) and displaced geometry (red) as functions of vibrational normal modes for DOBNA: (a-b) ISC $S_1 \rightarrow T_1$ (Ms=0) (Ms=1/-1), (c-d) ISC $S_1 \rightarrow T_2$ (Ms=0) (Ms=1/-1), (e-f) rISC $T_1 \rightarrow S_1$ (Ms=0) (Ms=1/-1), and (g-h) rISC $T_1 \rightarrow S_2$ (Ms=0) (Ms=1/-1). The SOCMEs values at the reference FC geometry are indicated (blue).

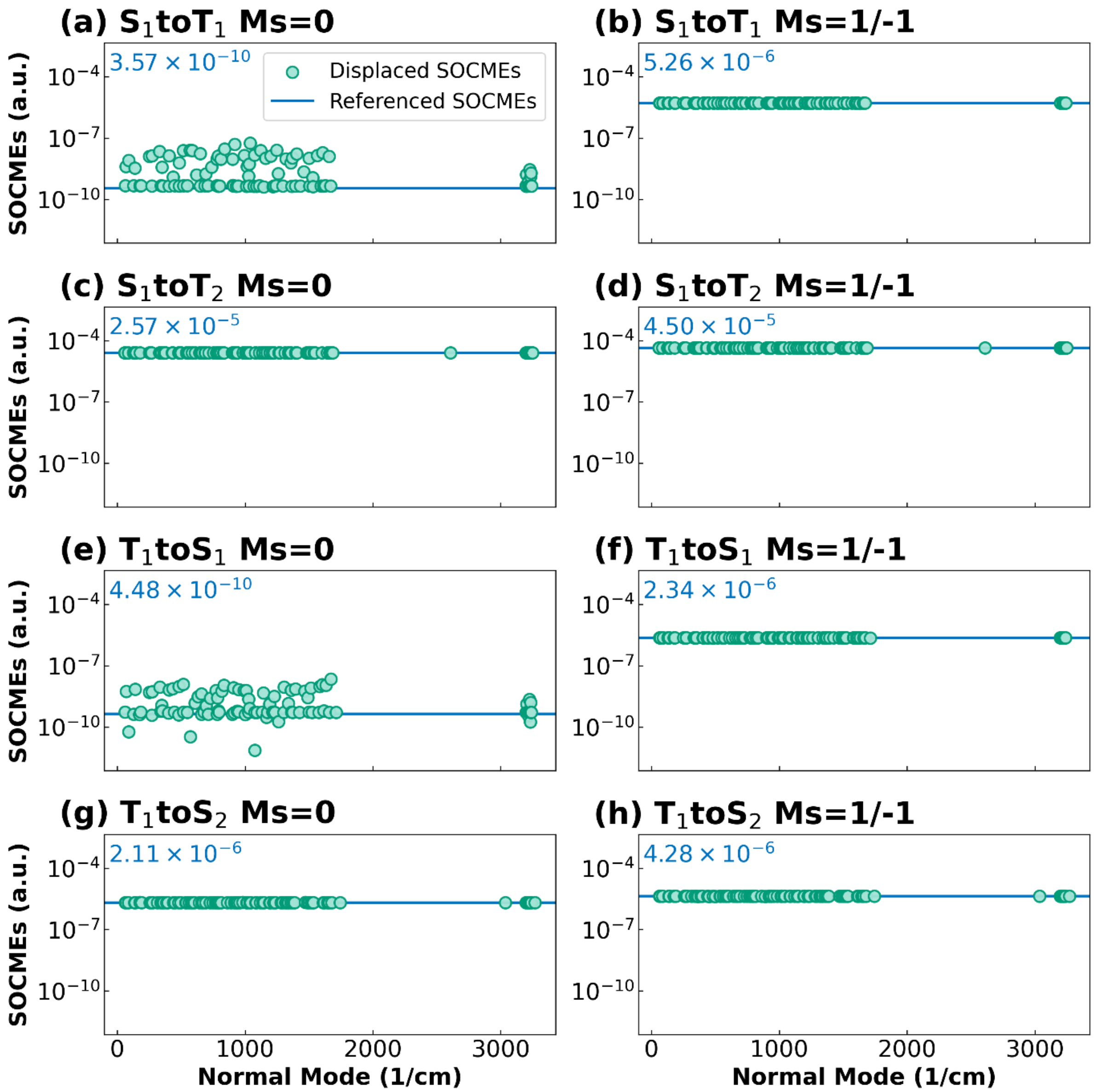

**Figure S10.** SOCMEs at the reference Franck-Condon (FC) geometry (blue) and displaced geometries (green) as functions of vibrational normal modes for DiKTa: (a-b) ISC S1→T1 (Ms=0) (Ms=1/-1), (c-d) ISC S1→T2 (Ms=0) (Ms=1/-1), (e-f) rISC T1→S1 (Ms=0) (Ms=1/-1), and (g-h) rISC T1→S2 (Ms=0) (Ms=1/-1). The SOCMEs values at the reference FC geometry are indicated (blue).

## 9. Rate Constants Calculations of DABNA-1, DQAO, DBT and PPDs-1

**Table S10. Calculated rate constant ($s^{-1}$) for all transitions considered in DABNA-1, DQAO, DBT and PPDs-1**

| Transition | Type | DABNA-1 | DQAO | DBT | PPDs-1 |
|---|---|---|---|---|---|
| $S_1 \rightarrow T_1$ | ISC | $8.30\times10^{6}$ | $2.72\times10^{9}$ | $2.13\times10^{8}$ | $4.61\times10^{3}$ |
| $S_1 \rightarrow T_2$ | ISC | $8.27\times10^{4}$ | $4.52\times10^{7}$ | $2.65\times10^{8}$ | $4.07\times10^{7}$ |
| $S_2 \rightarrow T_1$ | ISC | $3.13\times10^{6}$ | $3.96\times10^{9}$ | $5.79\times10^{4}$ | $4.87\times10^{7}$ |
| $S_2 \rightarrow T_2$ | ISC | $8.23\times10^{7}$ | $2.54\times10^{11}$ | $3.69\times10^{7}$ | $1.25\times10^{8}$ |
| $T_1 \rightarrow S_0$ | ISC | $2.00\times10^{3}$ | $9.42\times10^{3}$ | $3.83\times10^{4}$ | $4.92\times10^{5}$ |
| $T_1 \rightarrow S_1$ | rISC | $2.41\times10^{4}$ | $5.43\times10^{5}$ | - | - |
| $T_1 \rightarrow S_2$ | rISC | $3.87\times10^{3}$ | $2.50\times10^{8}$ | - | - |
| $T_2 \rightarrow S_1$ | rISC | $1.25\times10^{6}$ | $1.61\times10^{9}$ | - | - |
| $T_2 \rightarrow S_2$ | rISC | $6.15\times10^{3}$ | $1.42\times10^{8}$ | - | - |
| $S_1 \rightarrow S_0$ | EM | $9.05\times10^{7}$ | $6.46\times10^{7}$ | $4.75\times10^{7}$ | $3.74\times10^{8}$ |
| $T_1 \rightarrow S_0$ | EM | $1.58\times10^{-1}$ | $2.36\times10^{2}$ | $7.57\times10^{-1}$ | $5.55\times10^{-1}$ |
| $S_1 \rightarrow S_0$ | IC | $2.02\times10^{7}$ | $1.88\times10^{7}$ | $1.73\times10^{7}$ | $1.67\times10^{7}$ |
| $S_2 \rightarrow S_0$ | IC | $5.88\times10^{6}$ | $1.12\times10^{6}$ | $1.87\times10^{6}$ | $4.46\times10^{10}$ |
| $S_2 \rightarrow S_1$ | IC | $1.85\times10^{10}$ | $1.17\times10^{9}$ | $1.35\times10^{10}$ | $2.39\times10^{11}$ |
| $T_2 \rightarrow T_1$ | IC | $1.92\times10^{10}$ | $1.05\times10^{14}$ | $8.80\times10^{12}$ | $5.75\times10^{11}$ |
| $S_1 \rightarrow S_2$ | rIC | $2.30\times10^{8}$ | $2.17\times10^{8}$ | $5.11\times10^{9}$ | $6.69\times10^{10}$ |
| $T_1 \rightarrow T_2$ | rIC | $1.07\times10^{10}$ | $3.97\times10^{10}$ | $7.04\times10^{8}$ | $2.14\times10^{8}$ |

**Table S11. ISC rate constants for the individual sublevels (Ms = 0, ±1) in DABNA-1, DQAO, DBT and PPDs-1[a]**

| ISC Transition | Sublevel | DABNA-1 | DQAO | DBT | PPDs-1 |
|---|---|---|---|---|---|
| $S_1 \rightarrow T_1$ | 0 | $7.20\times10^{5}$ | $1.27\times10^{9}$ | $3.90\times10^{5}$ | $5.28\times10^{2}$ |
| $S_1 \rightarrow T_1$ | ±1 | $3.79\times10^{6}$ | $7.20\times10^{8}$ | $1.06\times10^{8}$ | $2.04\times10^{3}$ |
| $S_1 \rightarrow T_2$ | 0 | $1.89\times10^{4}$ | $4.39\times10^{7}$ | $9.20\times10^{4}$ | $4.50\times10^{6}$ |
| $S_1 \rightarrow T_2$ | ±1 | $3.19\times10^{4}$ | $6.53\times10^{5}$ | $1.32\times10^{8}$ | $1.81\times10^{7}$ |
| $S_2 \rightarrow T_1$ | 0 | $9.72\times10^{5}$ | $3.92\times10^{9}$ | $1.23\times10^{3}$ | $1.23\times10^{7}$ |
| $S_2 \rightarrow T_1$ | ±1 | $1.08\times10^{6}$ | $1.76\times10^{7}$ | $2.83\times10^{4}$ | $1.82\times10^{7}$ |
| $S_2 \rightarrow T_2$ | 0 | $4.48\times10^{7}$ | $4.00\times10^{10}$ | $1.28\times10^{5}$ | $2.30\times10^{7}$ |
| $S_2 \rightarrow T_2$ | ±1 | $1.87\times10^{7}$ | $1.07\times10^{11}$ | $1.84\times10^{7}$ | $5.10\times10^{7}$ |
| $T_1 \rightarrow S_0$ | 0 | $2.95\times10^{3}$ | $1.42\times10^{4}$ | $1.63\times10^{2}$ | $4.67\times10^{5}$ |
| $T_1 \rightarrow S_0$ | ±1 | $1.52\times10^{3}$ | $7.04\times10^{3}$ | $5.74\times10^{4}$ | $5.04\times10^{5}$ |
| $T_1 \rightarrow S_1$ | 0 | $1.51\times10^{4}$ | $9.57\times10^{5}$ | - | - |
| $T_1 \rightarrow S_1$ | ±1 | $2.86\times10^{4}$ | $3.36\times10^{5}$ | - | - |
| $T_1 \rightarrow S_2$ | 0 | $2.09\times10^{3}$ | $7.38\times10^{8}$ | - | - |
| $T_1 \rightarrow S_2$ | ±1 | $4.77\times10^{3}$ | $6.38\times10^{6}$ | - | - |
| $T_2 \rightarrow S_1$ | 0 | $2.38\times10^{5}$ | $3.73\times10^{9}$ | - | - |
| $T_2 \rightarrow S_1$ | ±1 | $1.75\times10^{6}$ | $5.52\times10^{8}$ | - | - |
| $T_2 \rightarrow S_2$ | 0 | $9.05\times10^{3}$ | $1.41\times10^{8}$ | - | - |
| $T_2 \rightarrow S_2$ | ±1 | $4.70\times10^{3}$ | $1.42\times10^{8}$ | - | - |

[a]The total ISC rate constant for one singlet-to-triplet transition is obtained by summing the rate constants of the three triplet sublevels, whereas the total rISC rate constant for one triplet-to-singlet transition is calculated as the average over these sublevels.

**Table S12. Phosphorescence rate constants for the individual sublevels (Ms = 0, ±1) in DABNA-1, DQAO, DBT and PPDs-1[a]**

| Sublevel | DABNA-1 | DQAO | DBT | PPDs-1 |
|---|---|---|---|---|
| 0 | $4.18\times10^{-1}$ | $7.05\times10^{2}$ | $1.80\times10^{0}$ | $7.45\times10^{-2}$ |
| 1 | $1.10\times10^{-2}$ | $2.76\times10^{0}$ | $4.90\times10^{-3}$ | $1.04\times10^{-1}$ |
| -1 | $4.40\times10^{-2}$ | $1.56\times10^{-3}$ | $4.61\times10^{-1}$ | $1.49\times10^{0}$ |

[a]The total phosphorescence rate constant is calculated as the average over these sublevels.

**Table S13. Reorganization energies ($cm^{-1}$) for ISC transitions in DABNA-1, DQAO, DBT and PPDs-1**

| ISC Transition | DABNA-1 | DQAO | DBT | PPDs-1 |
|---|---|---|---|---|
| $S_1\rightarrow T_1$ | 108.84 | 223.08 | 1984.15 | 235.45 |
| $S_1\rightarrow T_2$ | 108.79 | 1736.50 | 782.22 | 2306.24 |
| $S_2\rightarrow T_1$ | 821.67 | 232.66 | 1985.98 | 2375.36 |
| $S_2\rightarrow T_2$ | 821.67 | 1737.40 | 782.60 | 503.30 |
| $T_1\rightarrow S_0$ | 542.04 | 658.16 | 2874.72 | 3174.33 |
| $T_1\rightarrow S_1$ | 114.62 | 228.35 | - | - |
| $T_1\rightarrow S_2$ | 826.90 | 236.97 | - | - |
| $T_2\rightarrow S_1$ | 114.58 | 1291.00 | - | - |
| $T_2\rightarrow S_2$ | 826.76 | 1295.43 | - | - |

**Table S14. SOCMEs ($cm^{-1}$) at the Franck-Condon (FC) geometry for ISC transitions in DABNA-1, DQAO, DBT and PPDs-1**

| ISC Transition | Sublevel | DABNA-1 | DQAO | DBT | PPDs-1 |
|---|---|---|---|---|---|
| $S_1\rightarrow T_1$ | 0 | $4.68\times10^{-2}$ | $1.73\times10^{-4}$ | $3.54\times10^{-2}$ | $1.38\times10^{-2}$ |
| $S_1\rightarrow T_1$ | ±1 | $1.24\times10^{-5}$ | $3.12\times10^{-4}$ | $5.32\times10^{-5}$ | $1.01\times10^{-2}$ |
| $S_1\rightarrow T_2$ | 0 | $6.65\times10^{-5}$ | $4.03\times10^{-3}$ | $1.50\times10^{-4}$ | $4.96\times10^{-2}$ |
| $S_1\rightarrow T_2$ | ±1 | $2.96\times10^{-1}$ | $1.36\times10^{-2}$ | $2.70\times10^{-6}$ | $4.18\times10^{-2}$ |
| $S_2\rightarrow T_1$ | 0 | $1.98\times10^{-1}$ | $1.95\times10^{0}$ | $7.01\times10^{-5}$ | $2.52\times10^{-4}$ |
| $S_2\rightarrow T_1$ | ±1 | $9.21\times10^{-5}$ | $9.01\times10^{0}$ | $3.55\times10^{-6}$ | $4.68\times10^{-4}$ |
| $S_2\rightarrow T_2$ | 0 | $7.76\times10^{-5}$ | $2.24\times10^{1}$ | $7.12\times10^{-3}$ | $1.83\times10^{-2}$ |
| $S_2\rightarrow T_2$ | ±1 | $9.55\times10^{-2}$ | $6.54\times10^{-2}$ | $2.07\times10^{-5}$ | $1.65\times10^{-2}$ |
| $T_1\rightarrow S_0$ | 0 | $6.35\times10^{-5}$ | $1.53\times10^{-4}$ | $1.83\times10^{-2}$ | $1.41\times10^{-3}$ |
| $T_1\rightarrow S_0$ | ±1 | $1.26\times10^{-1}$ | $3.38\times10^{-2}$ | $2.65\times10^{-5}$ | $1.05\times10^{-3}$ |
| $T_1\rightarrow S_1$ | 0 | $4.95\times10^{-2}$ | $1.71\times10^{-4}$ | - | - |
| $T_1\rightarrow S_1$ | ±1 | $4.04\times10^{-6}$ | $9.93\times10^{-5}$ | - | - |
| $T_1\rightarrow S_2$ | 0 | $3.28\times10^{-5}$ | $1.62\times10^{-2}$ | - | - |
| $T_1\rightarrow S_2$ | ±1 | $2.64\times10^{-1}$ | $8.22\times10^{0}$ | - | - |
| $T_2\rightarrow S_1$ | 0 | $3.12\times10^{-5}$ | $3.08\times10^{-4}$ | - | - |
| $T_2\rightarrow S_1$ | ±1 | $3.15\times10^{-1}$ | $4.80\times10^{-2}$ | - | - |
| $T_2\rightarrow S_2$ | 0 | $1.13\times10^{-1}$ | $2.33\times10^{1}$ | - | - |
| $T_2\rightarrow S_2$ | ±1 | $1.26\times10^{-5}$ | $6.40\times10^{-2}$ | - | - |

**Table S15. Herzberg-Teller (HT) contributions (%) to the ISC transitions in DABNA-1, DQAO, DBT and PPDs-1**

| ISC Transition | Sublevel | DABNA-1 | DQAO | DBT | PPDs-1 |
|---|---|---|---|---|---|
| $S_1\rightarrow T_1$ | 0 | 76.08 | 100 | 93.49 | 98.32 |
| $S_1\rightarrow T_1$ | ±1 | 100 | 100 | 100 | 99.77 |
| $S_1 \rightarrow T_2$ | 0 | 100 | 100 | 100 | 94.02 |
| $S_1 \rightarrow T_2$ | ±1 | 47.71 | 99.98 | 100 | 98.94 |
| $S_2 \rightarrow T_1$ | 0 | 95.02 | 99.99 | 100 | 100 |
| $S_2 \rightarrow T_1$ | ±1 | 100 | 65.05 | 100 | 100 |
| $S_2 \rightarrow T_2$ | 0 | 100 | 53.59 | 99.78 | 99.62 |
| $S_2 \rightarrow T_2$ | ±1 | 90.39 | 100 | 100 | 99.86 |
| $T_1 \rightarrow S_0$ | 0 | 100 | 100 | 98.41 | 100 |
| $T_1 \rightarrow S_0$ | ±1 | 92.03 | 99.85 | 100 | 100 |
| $T_1 \rightarrow S_1$ | 0 | 48.44 | 100 | - | - |
| $T_1 \rightarrow S_1$ | ±1 | 100 | 100 | - | - |
| $T_1 \rightarrow S_2$ | 0 | 100 | 100 | - | - |
| $T_1 \rightarrow S_2$ | ±1 | 21.67 | 47.26 | - | - |
| $T_2 \rightarrow S_1$ | 0 | 100 | 100 | - | - |
| $T_2 \rightarrow S_1$ | ±1 | 95.32 | 99.89 | - | - |
| $T_2 \rightarrow S_2$ | 0 | 24.06 | 57.75 | - | - |
| $T_2 \rightarrow S_2$ | ±1 | 100 | 100 | - | - |

**10. Simulated Decay Kinetics of DABNA-1, DQAO, DBT and PPDs-1**

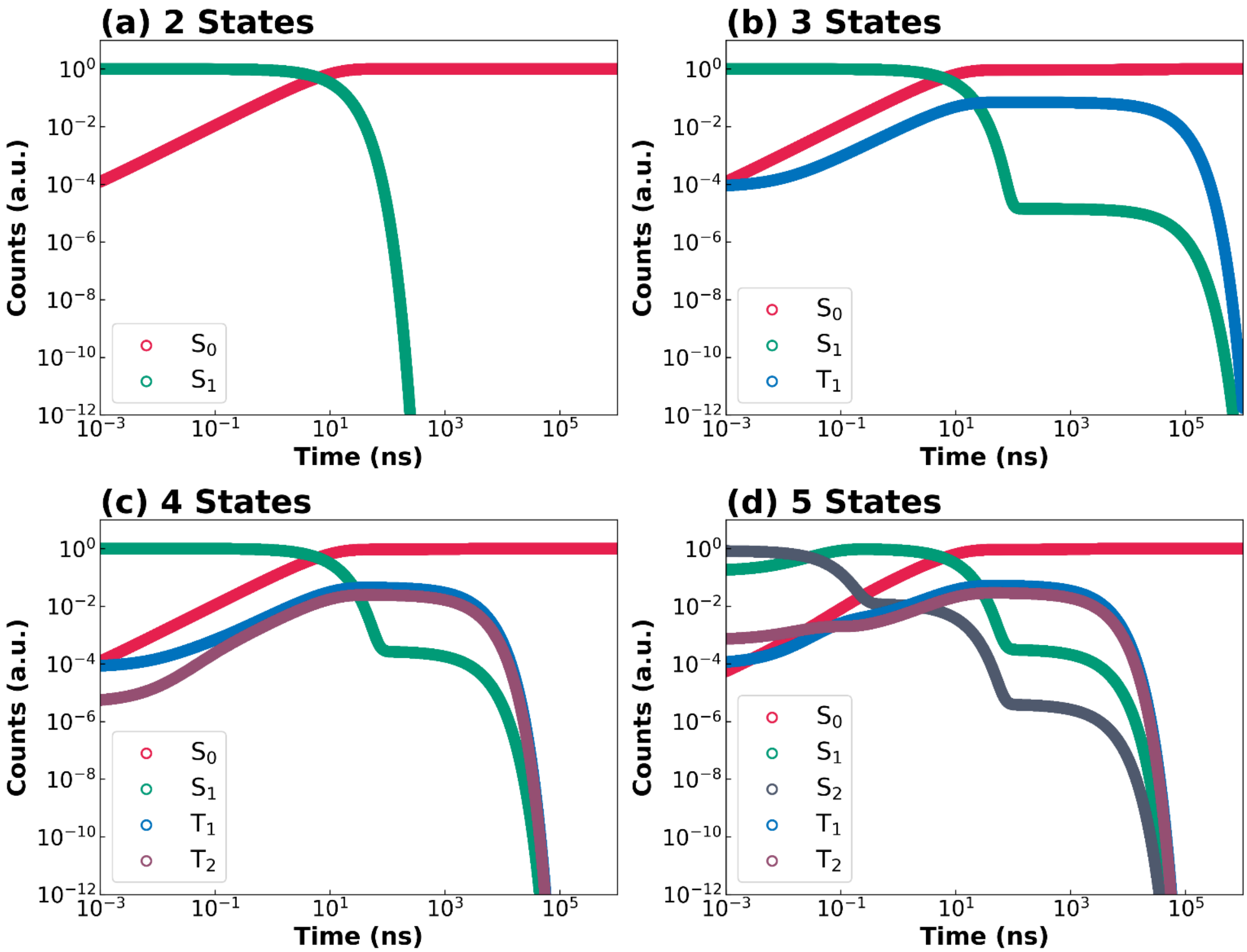

**Figure S11.** Simulated population decay kinetics of DABNA-1 over a 1 ms timescale based on (a) two-state ($S_0$, $S_1$), (b) three-state ($S_0$, $S_1$, $T_1$), (c) four-state ($S_0$, $S_1$, $T_1$, $T_2$), and (d) five-state ($S_0$, $S_1$, $S_2$, $T_1$, $T_2$) kinetic models. The number of absorbed photons is set to 1, with the excitation pulse width of 10 ps. Both time (x-axis) and population (y-axis) are plotted on logarithmic scales.

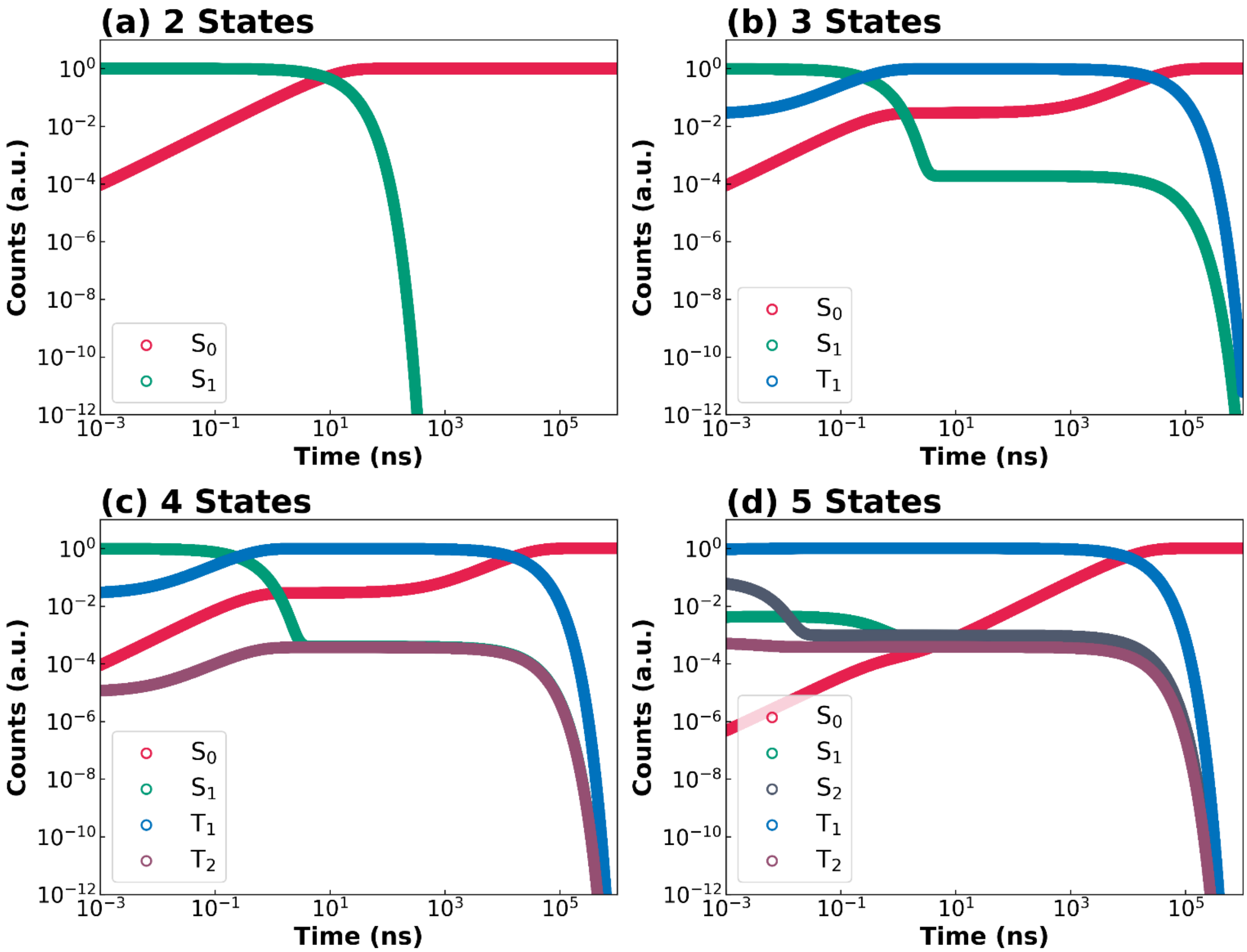

**Figure S12.** Simulated population decay kinetics of DQAO over a 1 ms timescale based on (a) two-state ($S_0$, $S_1$), (b) three-state ($S_0$, $S_1$, $T_1$), (c) four-state ($S_0$, $S_1$, $T_1$, $T_2$), and (d) five-state ($S_0$, $S_1$, $S_2$, $T_1$, $T_2$) kinetic models. The number of absorbed photons is set to 1, with the excitation pulse width of 10 ps. Both time (x-axis) and population (y-axis) are plotted on logarithmic scales.

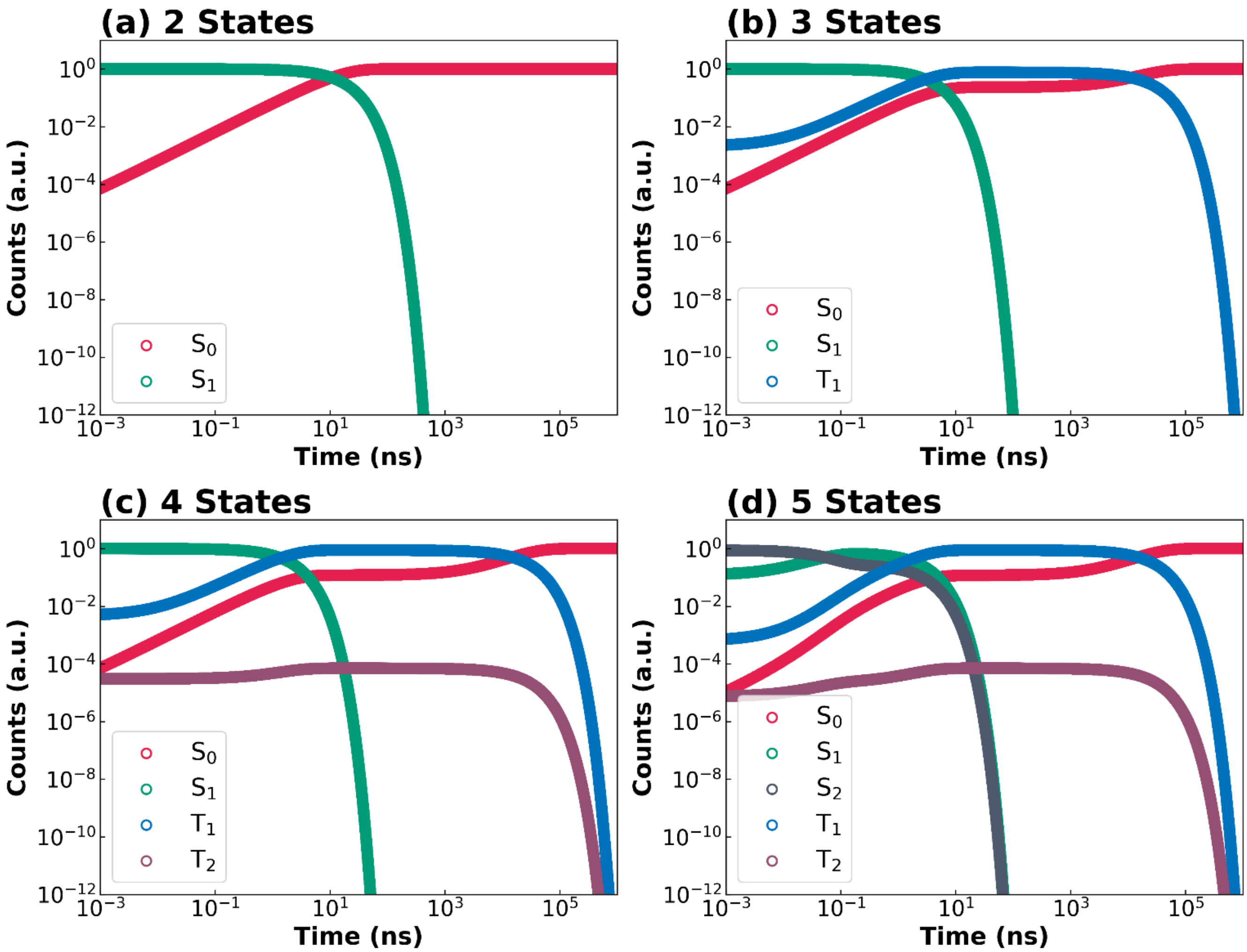

**Figure S13.** Simulated population decay kinetics of DBT over a 1 ms timescale based on (a) two-state ($S_0$, $S_1$), (b) three-state ($S_0$, $S_1$, $T_1$), (c) four-state ($S_0$, $S_1$, $T_1$, $T_2$), and (d) five-state ($S_0$, $S_1$, $S_2$, $T_1$, $T_2$) kinetic models. The number of absorbed photons is set to 1, with the excitation pulse width of 10 ps. Both time (x-axis) and population (y-axis) are plotted on logarithmic scale.

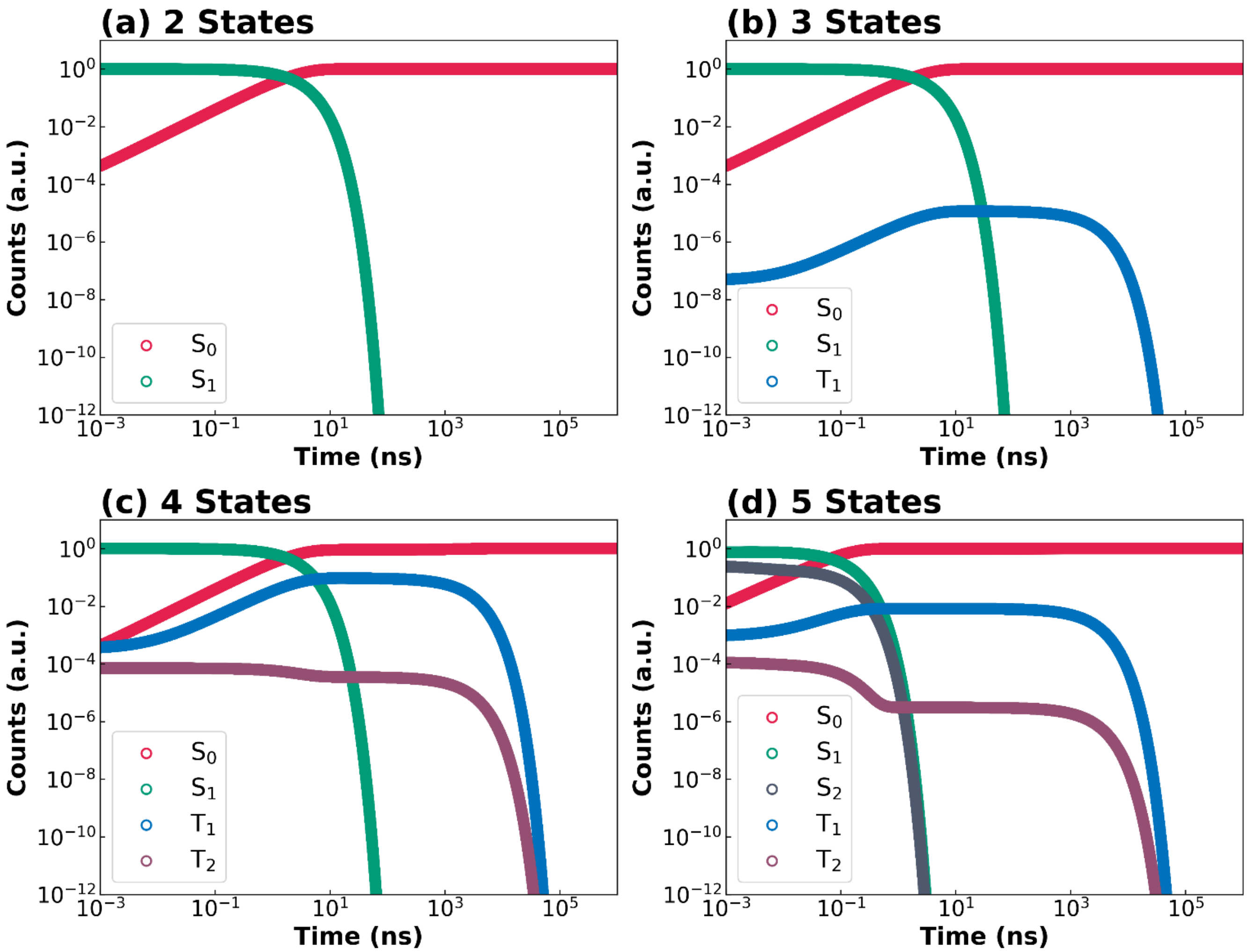

**Figure S14.** Simulated population decay kinetics of PPDs-1 over a 1 ms timescale based on (a) two-state ($S_0$, $S_1$), (b) three-state ($S_0$, $S_1$, $T_1$), (c) four-state ($S_0$, $S_1$, $T_1$, $T_2$), and (d) five-state ($S_0$, $S_1$, $S_2$, $T_1$, $T_2$) kinetic models. The number of absorbed photons is set to 1, with the excitation pulse width of 10 ps. Both time (x-axis) and population (y-axis) are plotted on logarithmic scales.

## 11. Error Diagnosis of Computational Rate Constants

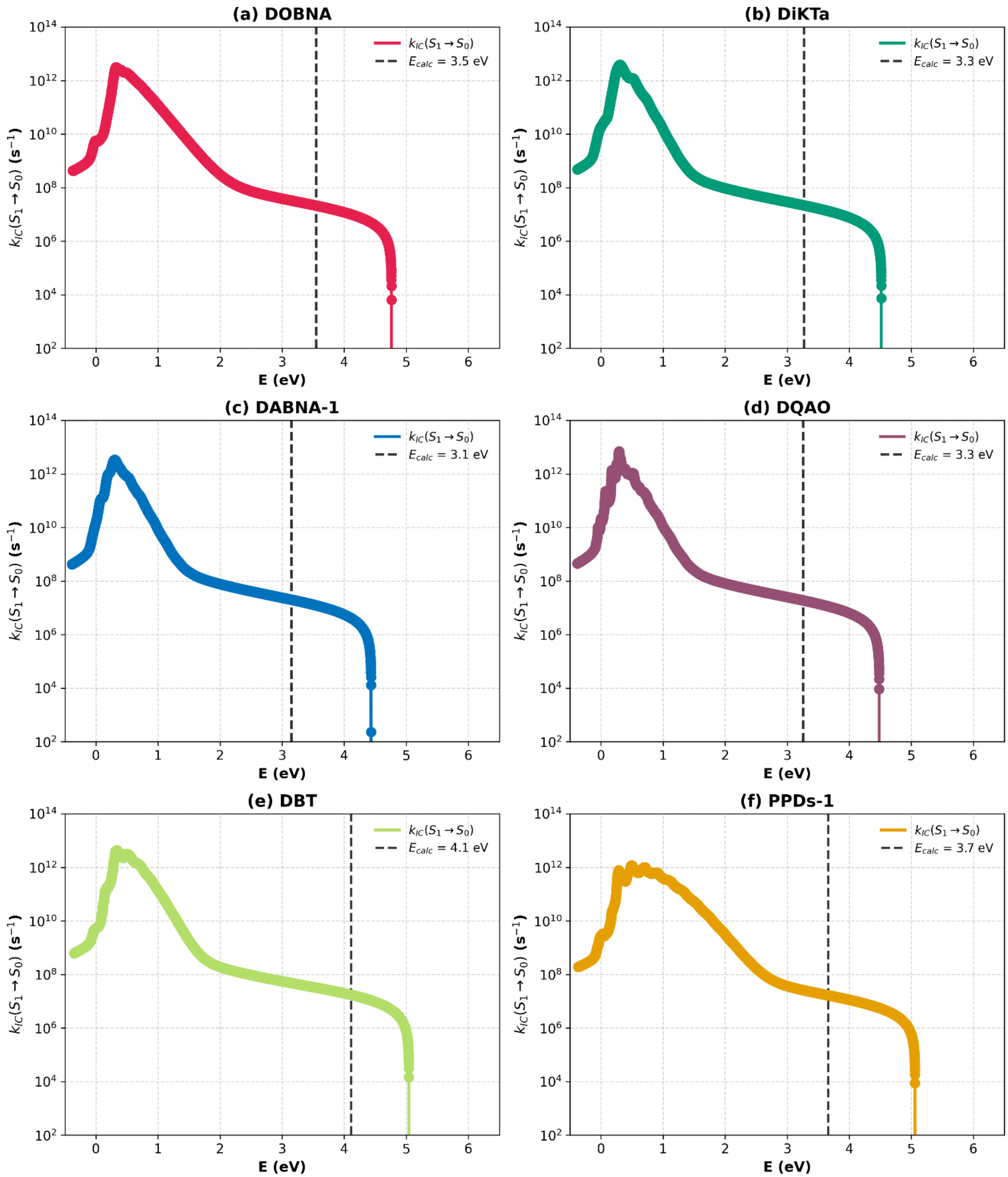

**Figure S15.** Dependencies of calculated rate constant IC $S_1 \rightarrow S_0$ on adiabatic energy gap for (a) DOBNA (red), (b) DiKTa (green), (c) DABNA-1 (blue) and (d) DQAO (purple), (e) DBT (light green), (f) PPDs-1 (orange).

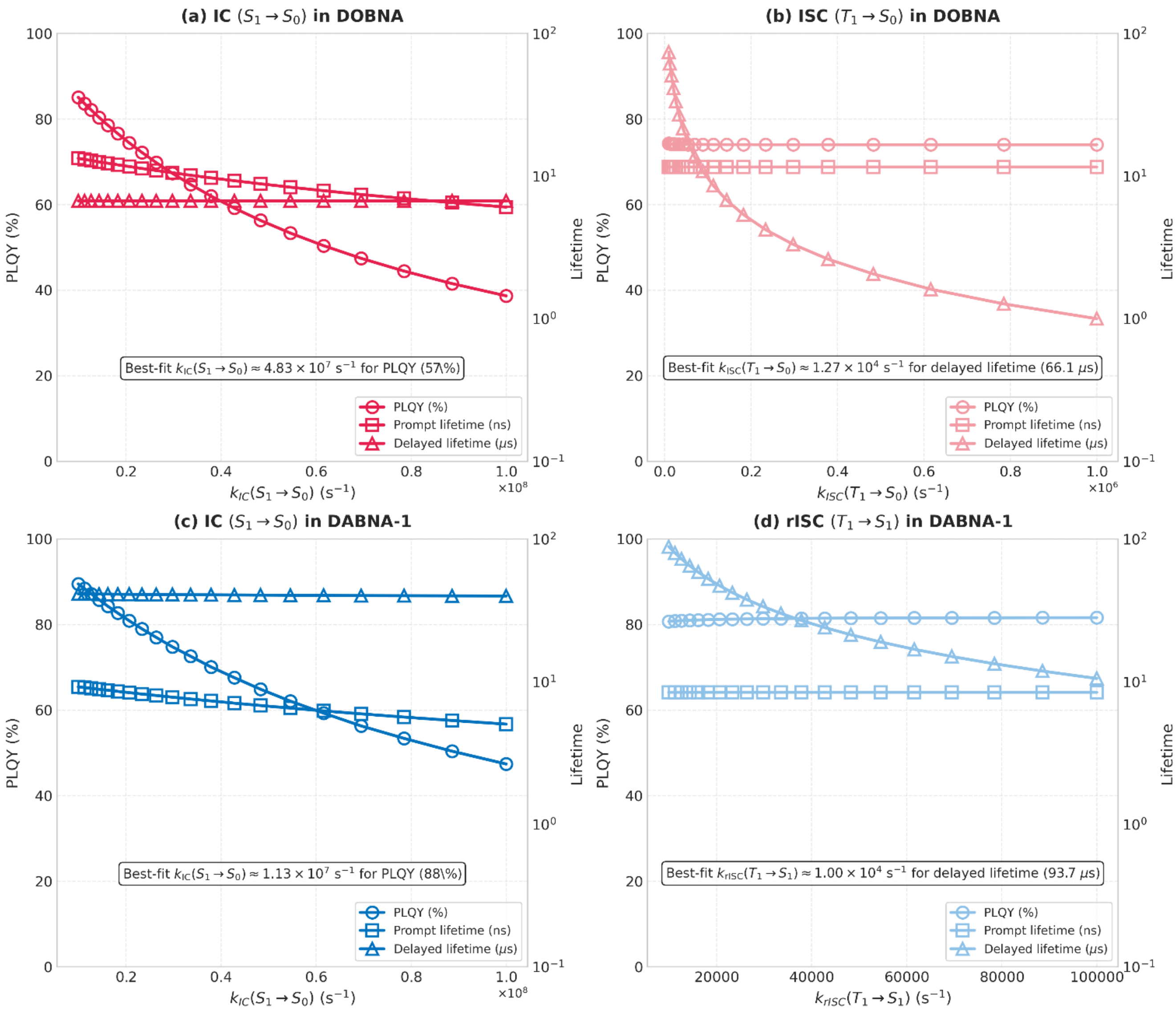

**Figure S16.** Influences of calculated rate constant IC $S_1 \rightarrow S_0$ in DOBNA (red) (a), ISC $T_1 \rightarrow S_0$ in DOBNA (light red) (b), IC $S_1 \rightarrow S_0$ in DABNA-1 (blue) (c) and rISC $T_1 \rightarrow S_0$ in DABNA-1 (light blue) (d) with three-state model.

## 12. Simulated Excitation and Decay Kinetics Results Directly Generated by KinLuv

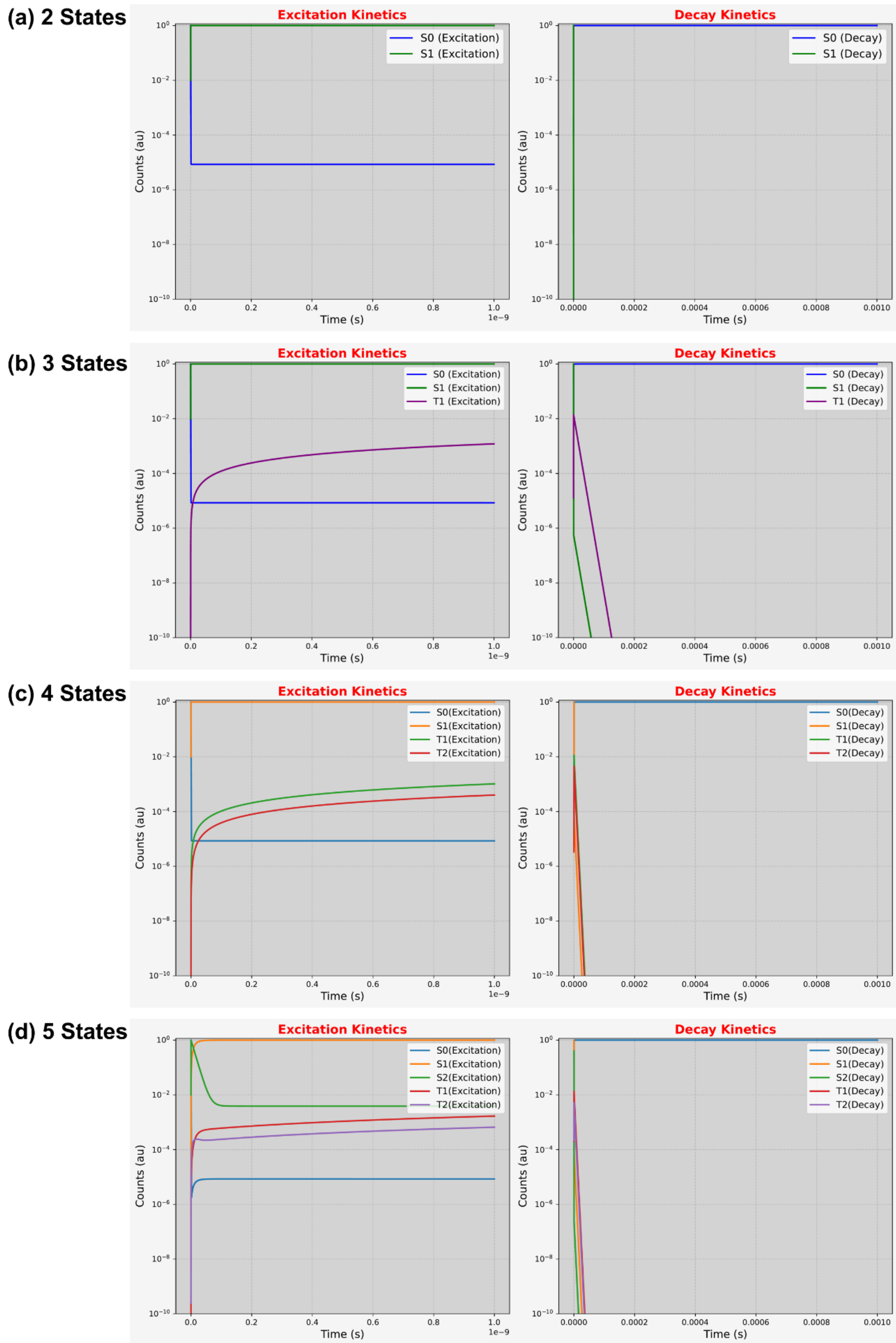

.

**Figure S17.** Excitation and decay plots generated by KinLuv for DOBNA based on (a) two-state, (b) three-state, (c) four-state, and (d) five-state kinetic models.

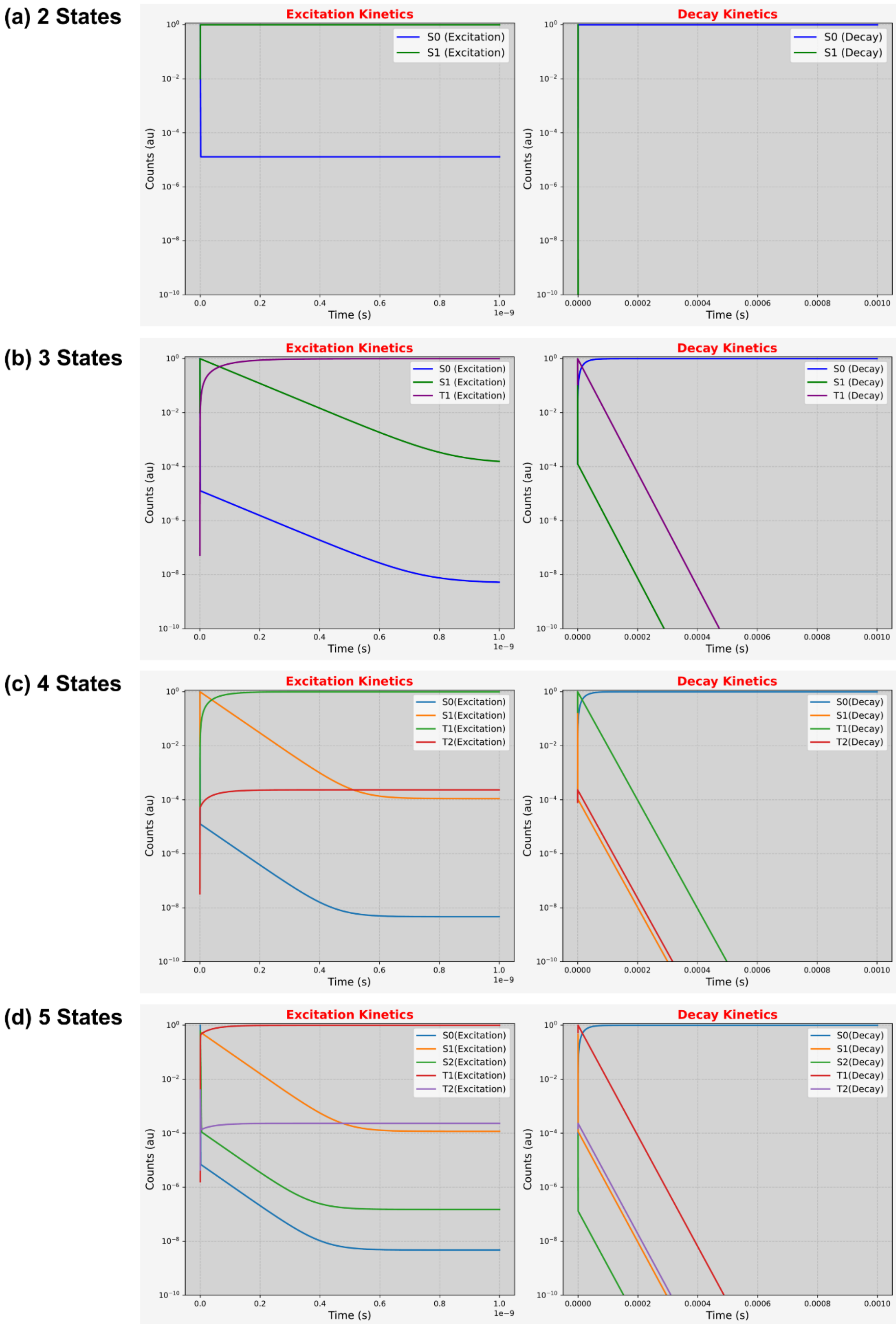

**Figure S18.** Excitation and decay plots generated by KinLuv for DiKTa based on (a) two-state, (b) three-state, (c) four-state, and (d) five-state kinetic models.

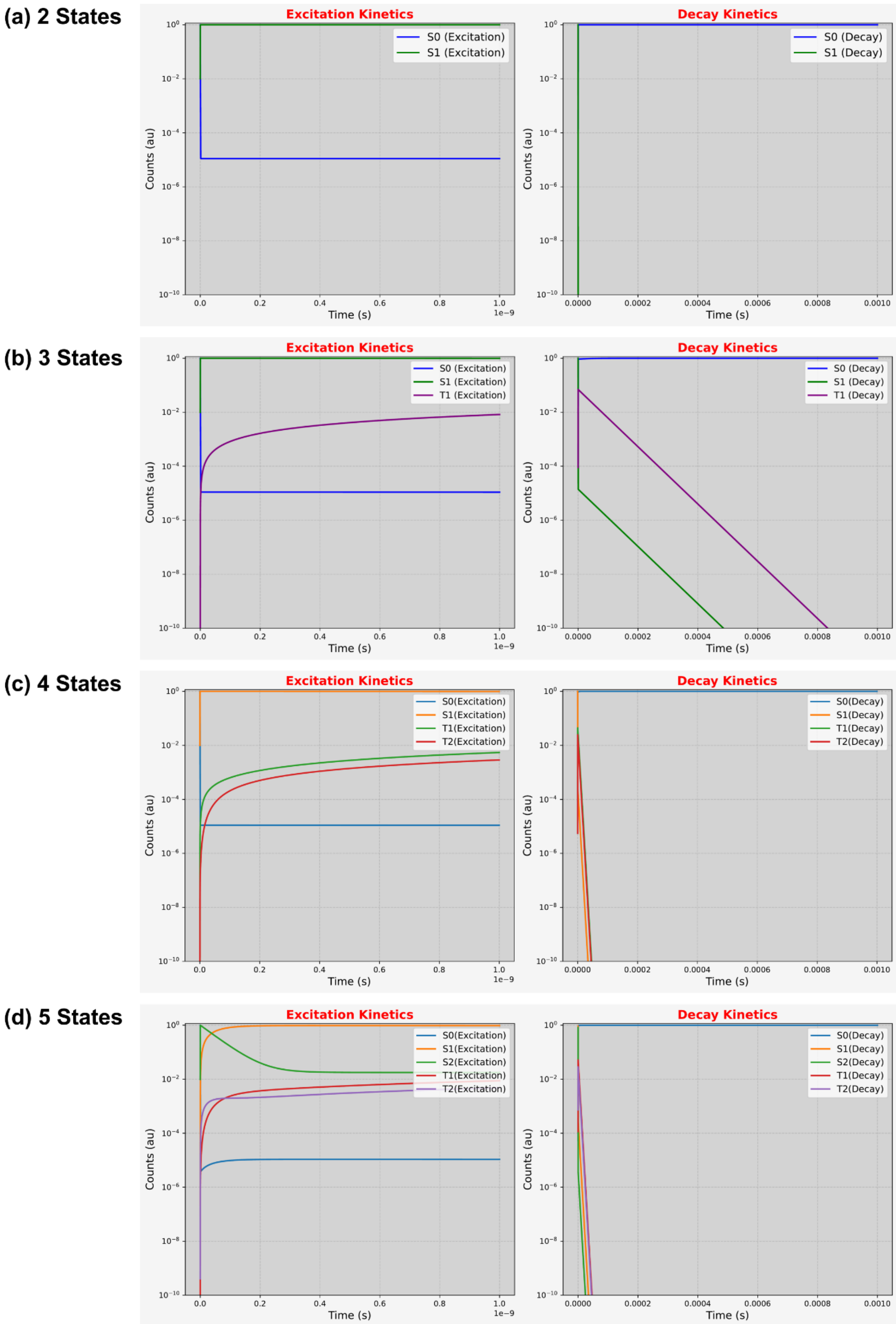

**Figure S19.** Excitation and decay plots generated by KinLuv for DABNA-1 based on (a) two-state, (b) three-state, (c) four-state, and (d) five-state kinetic models.

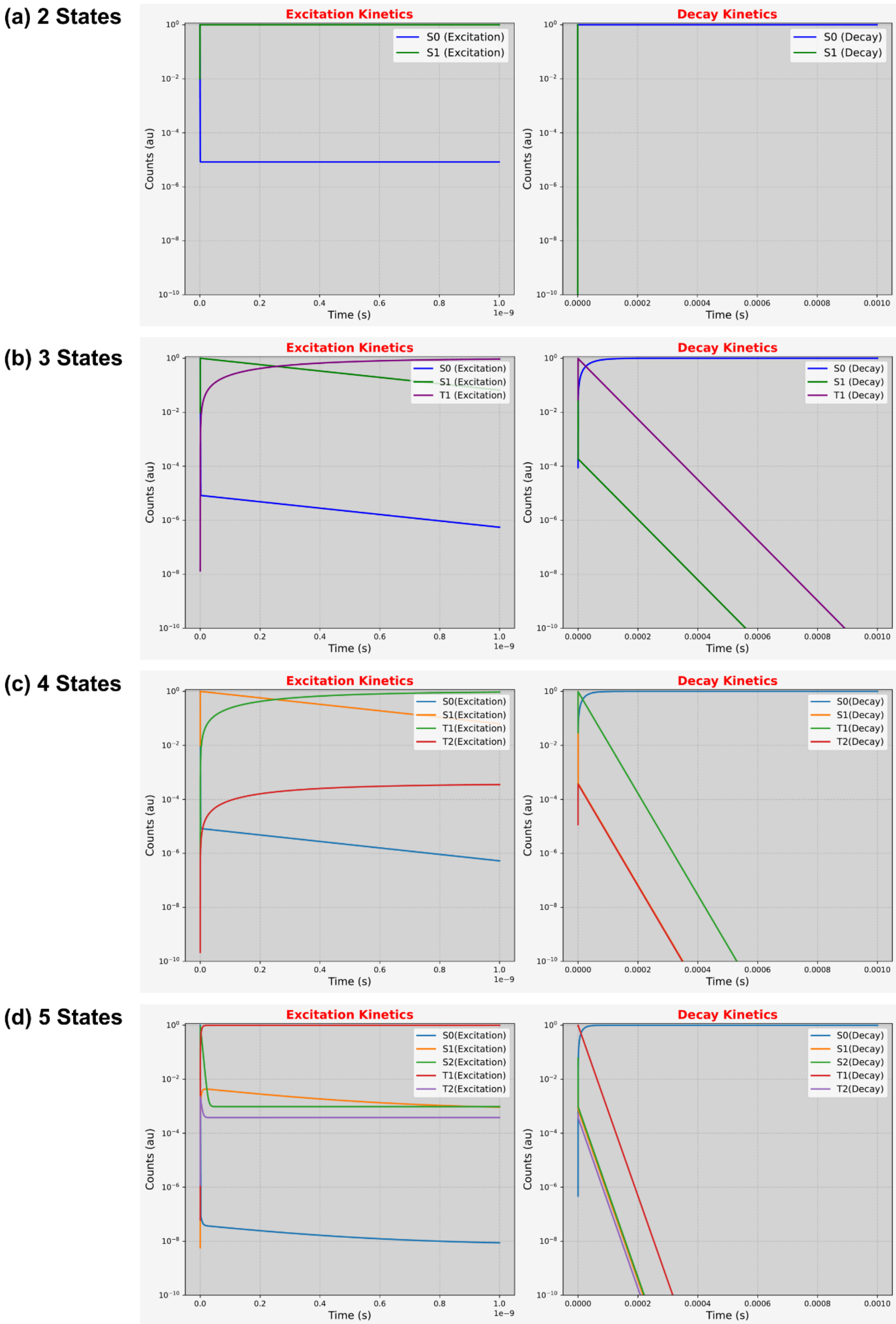

**Figure S20.** Excitation and decay plots generated by KinLuv for DQAO based on (a) two-state, (b) three-state, (c) four-state, and (d) five-state kinetic models.

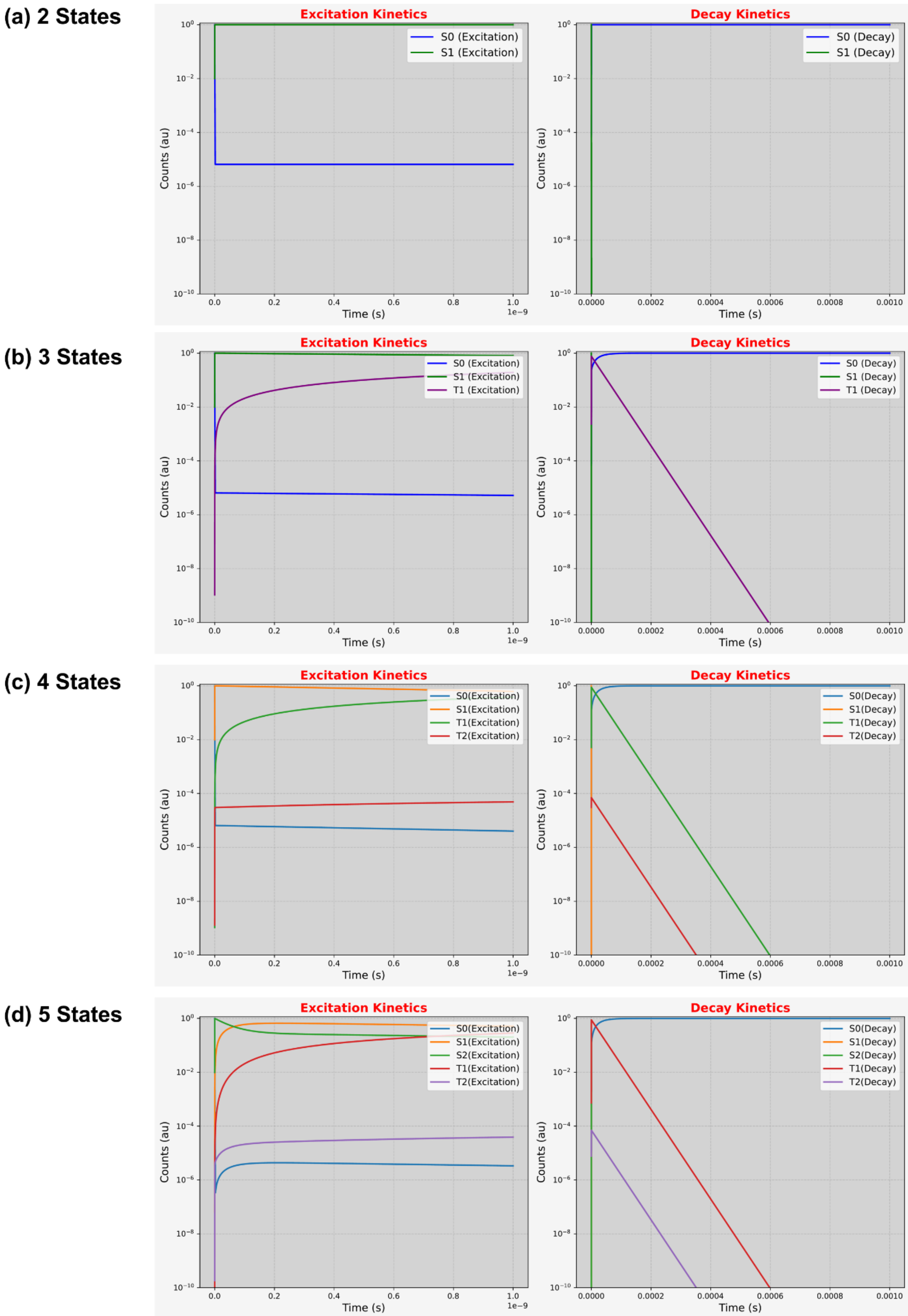

**Figure S21.** Excitation and decay plots generated by KinLuv for DBT based on (a) two-state, (b) three-state, (c) four-state, and (d) five-state kinetic models.

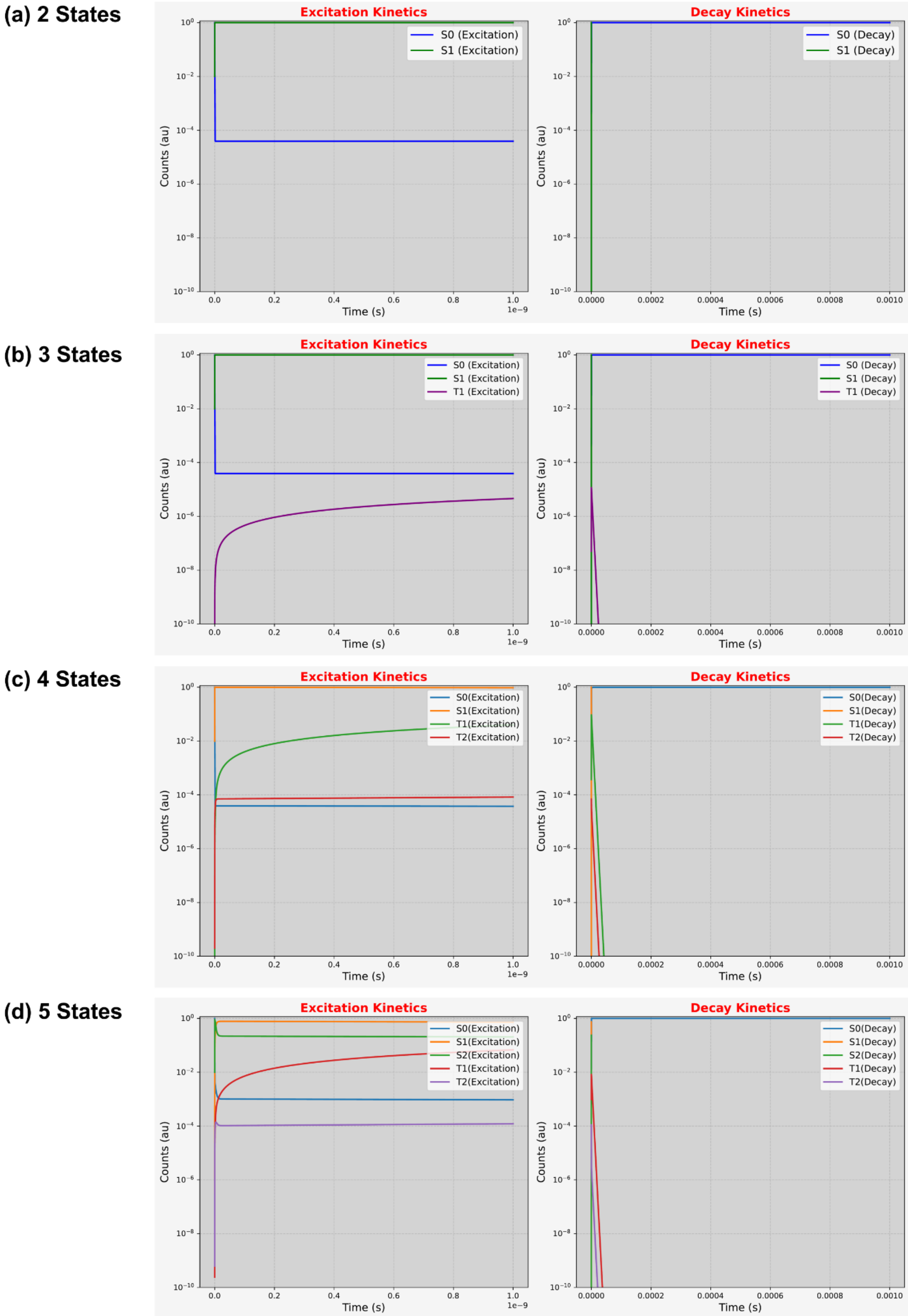

**Figure S22.** Excitation and decay plots generated by KinLuv for PPDs-1 based on (a) two-state, (b) three-state, (c) four-state, and (d) five-state kinetic models.